\documentclass{article}
\usepackage[a4paper, total={6in, 10in}]{geometry}
\usepackage{indentfirst}
\usepackage{bbm}
\usepackage{physics}
\usepackage{mdframed}
\usepackage{float}
\usepackage[caption = false]{subfig}
\usepackage{graphicx,color}
\usepackage{amsbsy,amssymb,amsmath,bm,mathrsfs,mathdots}
\usepackage{cite}
\usepackage{authblk}

\usepackage{tabularray}
\usepackage{makecell}
\usepackage[title]{appendix}
\apptocmd{\appendices}{\apptocmd{\thesection}{}{}{}}{}{}
\usepackage{hyperref}
\hypersetup{
    colorlinks=true,
    linkcolor=blue,
    filecolor=blue,      
    urlcolor=blue,
    citecolor=blue,   
    pdftitle={Overleaf Example},
    pdfpagemode=FullScreen,
    }

\def \bea {\begin{eqnarray}}
\def \eea {\end{eqnarray}}

\makeatother

\begin{document}
\title{Crystalline-equivalent topological phases of many-body fermionic systems in one dimension}

\author[1,$\ast$]{Chen-Shen Lee}
\author[2,$\dagger$]{Ken Shiozaki}
\author[1,3,4,$\ddagger$]{Chang-Tse Hsieh}
\affil[1]{Department of Physics and Center for Theoretical Physics, National Taiwan University,
Taipei 10607, Taiwan}
\affil[2]{Center for Gravitational Physics and Quantum Information, Yukawa Institute for Theoretical Physics, Kyoto University, Kyoto 606-8502, Japan}
\affil[3]{Physics Division, National Center for Theoretical Science, National Taiwan University,
Taipei 10607, Taiwan}
\affil[4]{Center for Quantum Science and Engineering, National Taiwan University, Taipei 10617, Taiwan}
\affil[$\ast$]{yausan0523270@gmail.com}
\affil[$\dagger$]{ken.shiozaki@yukawa.kyoto-u.ac.jp}
\affil[$\ddagger$]{cthsieh@phys.ntu.edu.tw}

\date{\today }

\maketitle
\selectfont
We explore one-dimensional fermionic symmetry-protected topological (SPT) phases related by the crystalline equivalence principle. In particular, we study charge-conserving many-body topological phases of fermions protected respectively by chiral and reflection symmetries. While the classifications of the two crystalline-equivalent SPT phases are identical, their topological properties and phase structures can be very different, depending on the microscopic details. 
Specifically, we consider certain extensions of the Su–Schrieffer–Heeger model, with and without interactions, that preserve both chiral and reflection symmetries, and explicitly compute the many-body topological invariants based on the systems' ground states. The phase structures determined by these topological invariants align perfectly with the many-body spectra of deformations among the models. As expected, gapped deformations exist only when all the topological invariants remain unchanged. Moreover, we show that decomposable systems—those that can be decomposed into local and decoupled subsystems—can be topologically characterized by real-space quantum numbers directly associated with the symmetries. For reflection-symmetric systems, these quantum numbers are related to the many-body topological invariants via a bulk-center correspondence, which can be justified using the Atiyah-Hirzebruch spectral sequence in generalized homology theory. Finally, we discuss the role of transition symmetry in the many-body topologies of these SPT phases.

\section{Introduction}
Topological materials are among the most intriguing materials in condensed matter physics due to their topology-governed properties. One well-known example is the topological insulator (see \cite{RMPforTI} and references therein). These systems exhibit exotic boundary phenomena, such as gaplessness and spin-momentum locking, which are under the protection of an on-site symmetry (time-reversal symmetry in this case). In this context, a topological insulator belongs to the family of symmetry-protected topological (SPT) phases \cite{SPTphases}. SPT phases can be understood as symmetry-enriched versions of trivial phases where the ground states are (or can be adiabatically connected to) product states, and they become trivial in the absence of symmetry. When we say that two (gapped) states are in the same SPT phase, we mean that there exists a continuous path of Hamiltonians connecting these two states, with the symmetry preserved and the energy gap remaining open throughout the path.

The classification of SPT phases for free fermion systems with on-site symmetries is well-established and can be described using K-theory \cite{Ktheory, Chiu:2016aa}. Beyond on-site symmetries, SPT phases can also be protected by crystalline symmetries, and their non-interacting classification can be understood using certain extensions of K-theory \cite{TKtheory1, Shiozaki:2014aa, TKtheory2, Chiu:2016aa}. While these classification schemes rely on momentum-space formalism, which is essential for free fermion systems, they are not applicable to  interacting SPT phases. Instead, topological characteristics of many-body systems are generally formulated in real space. For systems with on-site symmetries, group (super)cohomology theory provides a systematic description \cite{cohomology1, cohomology2}. For systems with crystalline symmetries, the classification problem can be addressed through the idea of cell decomposition---decomposing crystalline SPT phases into lower-dimensional SPT phases with on-site symmetries \cite{Song:2017aa, Huang:2017aa,Lower-dconstruction3, Lower-dconstruction2, Lower-dconstruction1,2024realspaceconstruction}---or, in a more mathematical framework, by using the generalized homology approach \cite{ghomo2}.

In addition to these methods, cobordism theory has also been proposed as another approach for classifying interacting SPT phases, primarily within the framework of topological quantum field theory (TQFT) \cite{Freed_2021, cobordism1, cobordism2, cobordism3}. Although cobordism theory gives a more complete classification than the group-cohomology approach, it assumes Lorentz symmetry, or space-time rotational symmetry in Euclidean signature, which is emergent only in the low-energy limit of a condensed matter system. (For fermionic systems, this requires certain spin structures refined by the underlying symmetry.) Therefore, it remains a conjecture whether cobordism theory provides a sensible classification of SPT phases of generic non-relativistic quantum matter.

Motivated by this, we investigate SPT phases related by the crystalline equivalence principle \cite{Thorngren:2018aa}. This principle, derived from Euclidean spacetime formalism and TQFT, states that the classification of crystalline topological phases with a spatial symmetry group $G$ is equivalent to the classification of topological phases with the corresponding internal symmetry $G$. For instance, reflection symmetry corresponds to chiral symmetry, the product of charge-conjugation and time-reversal symmetries, in this framework
\footnote{This is related to the CPT (more precisely, CRT) theorem in relativistic quantum field theory.
}.
In this work, we study fermionic SPT phases in the presence of charge U(1) symmetry and either chiral or reflection symmetry. Both of them are classified by $\mathbb{Z}$ for free-fermion systems \cite{Shiozaki:2014aa, Hsieh:2014aa, Chiu:2016aa} and by $\mathbb{Z}_4$ for many-body systems\cite{crystalMTI, OnsiteMTI}. In particular, the interacting classifications come from the same cobordism group $\Omega^{\text{pin}^c}_2(pt) = \mathbb{Z}_4$ (more precisely, Hom$(\Omega^{\text{pin}^c}_2(pt), \mathrm{U}(1))=\mathbb{Z}_4$) in cobordism theory. While their classifications are identical, the topological origins of the two SPT phases are fundamentally different, and we aim to distinguish them from a microscopic perspective. Indeed, the many-body topological invariants, defined through the partial transpose and partial reflection on the systems' ground states \cite{Pollmann_and_Turner2012,PRLMTI, crystalMTI, OnsiteMTI}, are intrinsically distinct when constructed from the two symmetries. 
We explicitly evaluate these topological invariants for both non-interacting and interacting fermionic models as well as for certain deformations among these models in the presence of the symmetries. As expected, gapped deformations exist only when the topological invariants remain unchanged, i.e., when there is no phase transition. Our results thus confirm the validity of these many-body topological invariants proposed in the literature.

Additionally, we provide a more physical interpretation of these many-body topological invariants for decomposable systems. A decomposable system is one that can be divided into local and decoupled subsystems. For systems with chiral symmetry, we show that the topological invariant is related to the number of unbound fermions appearing at the ends of the systems in the semi-infinite limit, consistent with the usual bulk-boundary correspondence.  For systems with reflection symmetry, we can represent the topological invariant by two symmetry quantum numbers, number of charges and reflection parity, at the reflection center. This can be regarded as a bulk-center correspondence, which is established through the Atiyah-Hirzebruch spectral sequence (AHSS) \cite{ghomo2}.

We also discuss the effect of translation symmetry on the two crystalline-equivalent SPT phases. While the presence of translation symmetry does not influence the classification of chiral-symmetric SPT phases, it does modify the classification of reflection-symmetric SPT phases.

This paper is organized as follows. Sec.~\ref{The models} introduces the models studied throughout the paper. In Sec.~\ref{Many-body topological invariants}, we briefly review the many-body topological invariants in the operator formalism for 1$d$ fermionic SPT phases with chiral and reflection symmetries. In Sec.~\ref{Chiral-symmetric SPT phases} and \ref{Reflection-symmetric SPT phases}, we explicitly compute these invariants for the proposed models and identify their phase structures associated with each symmetry. The role of translation symmetry in both types of SPT phases is discussed in Sec.~\ref{Role of translation symmetry}. Finally, Sec.~\ref{Conclusion} summarizes the findings and conclusions of this work.

\section{The models}\label{The models}
\begin{figure}[hbt!]
\centering
\subfloat[]{\includegraphics[width=0.32\textwidth]{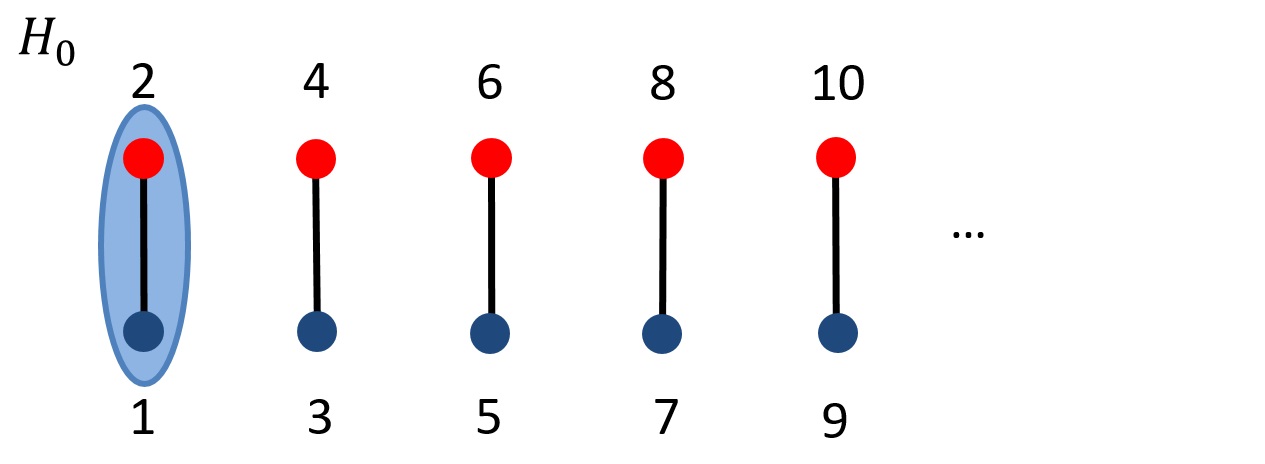}}\hskip 0.1cm
\subfloat[]{\includegraphics[width=0.32\textwidth]{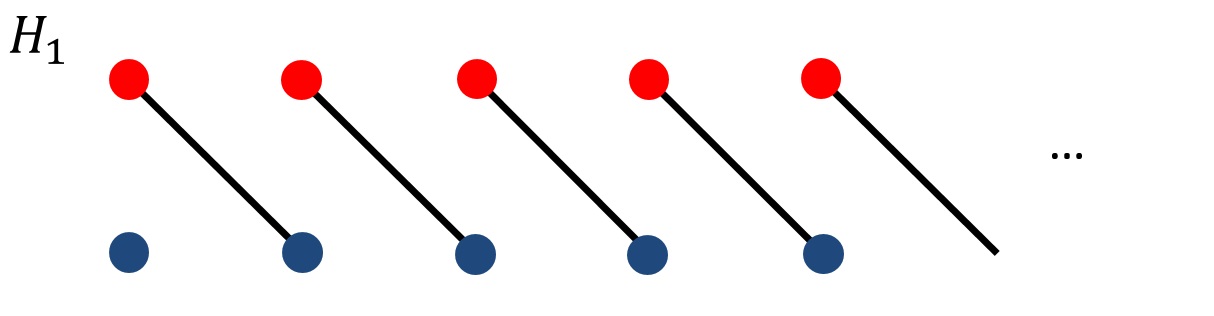}}\hskip 0.1cm
\subfloat[]{\includegraphics[width=0.32\textwidth]{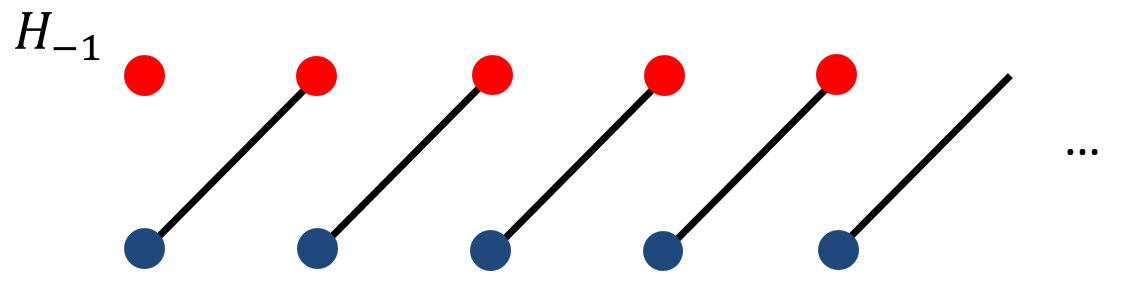}}\\
\subfloat[]{\includegraphics[width=0.32\textwidth]{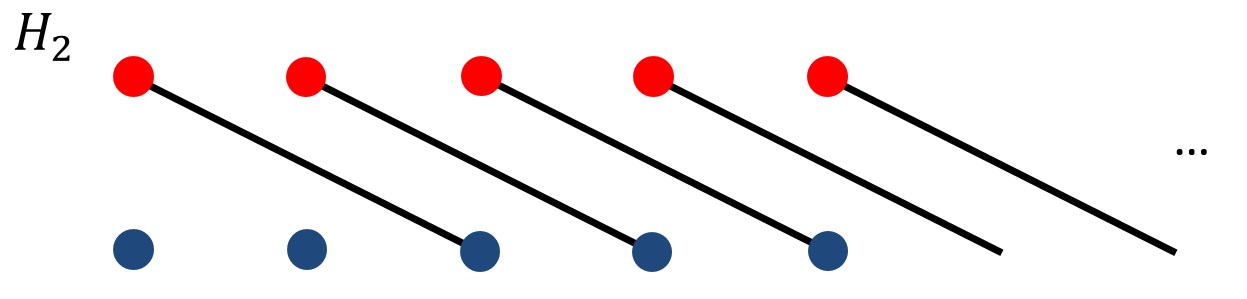}}\hskip 0.5cm
\subfloat[]{\includegraphics[width=0.32\textwidth]{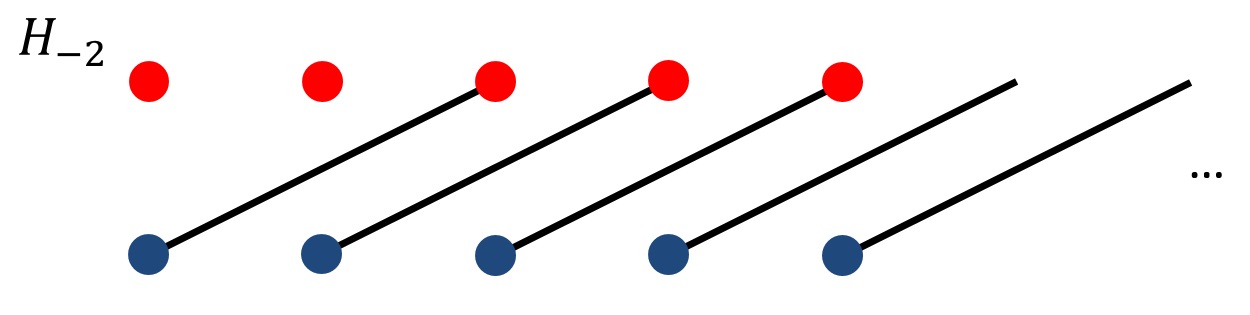}}\\
\caption{Schematic diagrams of $H_{\alpha}$ with $\alpha=-2, -1, 0, 1, 2$ in the right semi-infinite limits. The shadow part represents the unit cell.}
\label{figure of Ha}
\end{figure}
\begin{figure}[t!]
\centering
\subfloat[]{\includegraphics[width=0.32\textwidth]{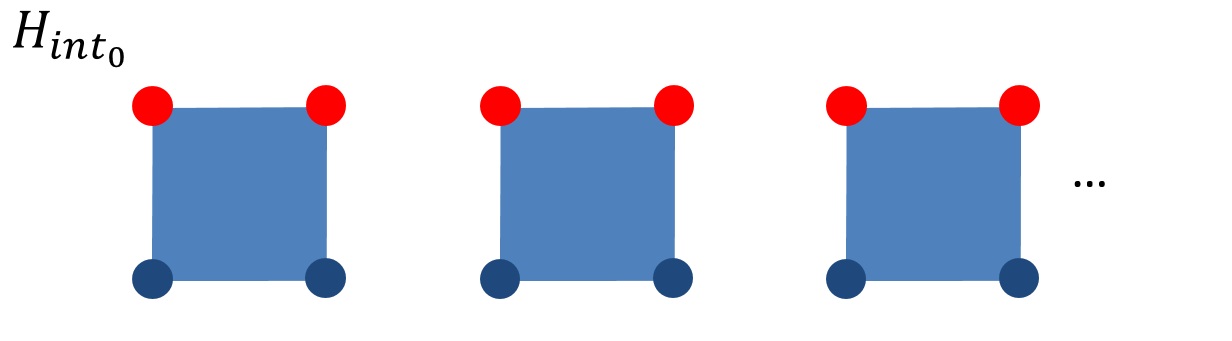}}\hskip 0.1cm
\subfloat[]{\includegraphics[width=0.32\textwidth]{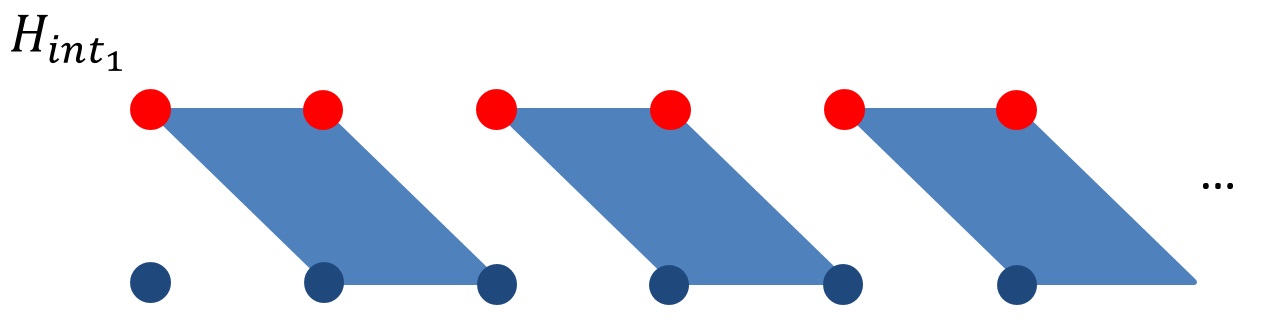}}\hskip 0.1cm
\subfloat[]{\includegraphics[width=0.32\textwidth]{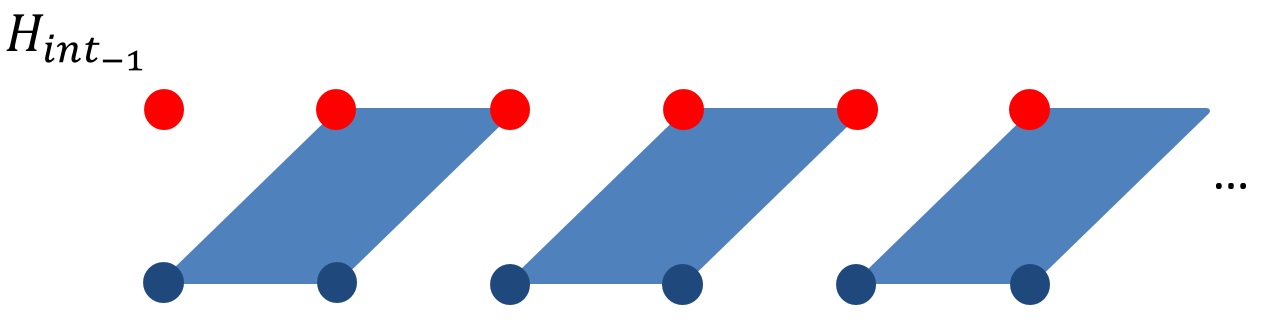}}\\
\subfloat[]{\includegraphics[width=0.32\textwidth]{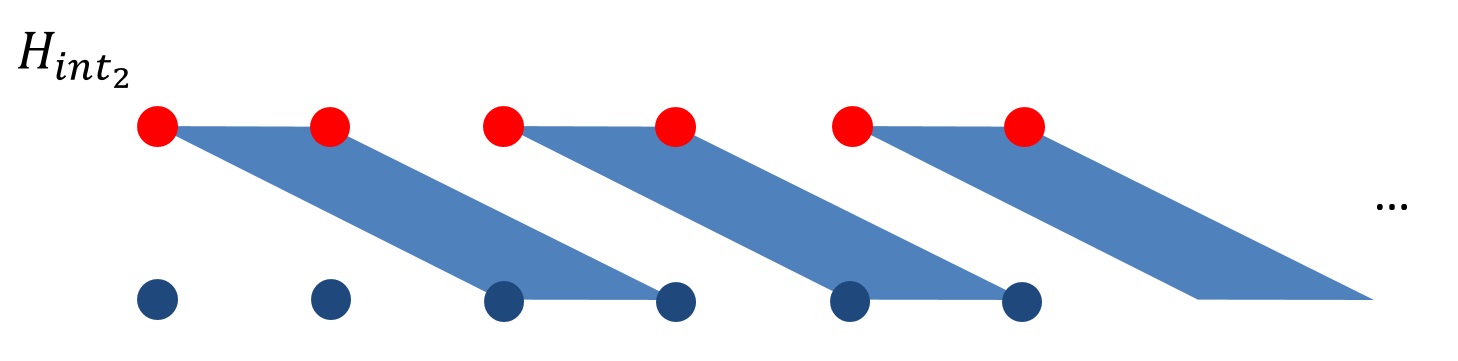}}\hskip 0.5cm
\subfloat[]{\includegraphics[width=0.32\textwidth]{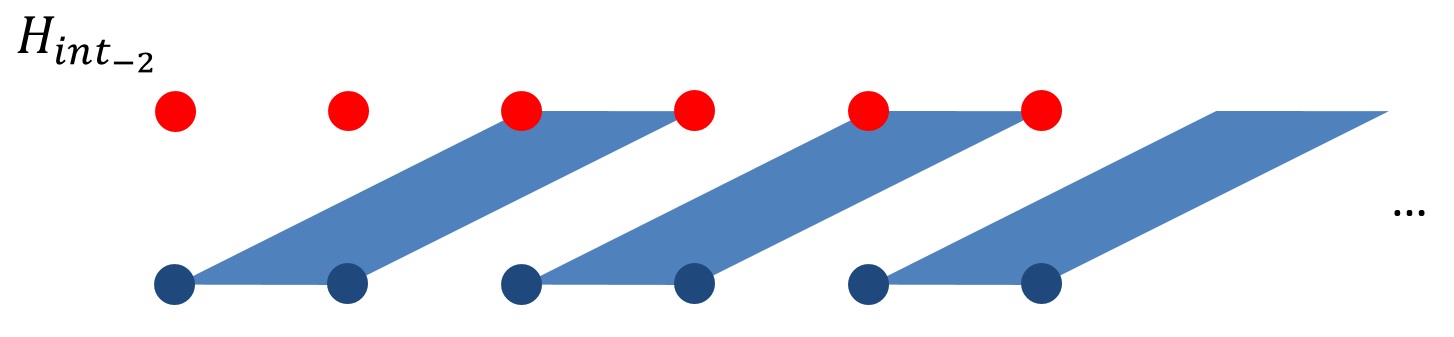}}\\
\caption{Schematic diagrams of $H_{int_\alpha}$ with $\alpha=-2, -1, 0, 1, 2$ in the right semi-infinite limits. The blue parallelograms represent the four-site fermions interacting through quartic interactions.}
\label{figure of Int}
\end{figure}

To study the interaction effects on the many-body topology of 1$d$ fermionic systems in the presence of chiral or reflection symmetry, we consider certain extensions of the Su–Schrieffer–Heeger (SSH) model without and with interactions:
\bea\label{free_systems}
H_{\alpha}=
\left\{ 
\begin{aligned}
    &\sum_{j=1}(c_{2j-1}^{\dag}c_{2j+2|\alpha|}+h.c.),\;\text{for}\; \alpha\leq0\\
    &\sum_{j=1}(c_{2j}^{\dag}c_{2j+2\alpha-1}+h.c.),\;\text{for}\; \alpha>0. 
\end{aligned}
\right.
\eea
\bea\label{interacting_systems}
H_{int_\alpha}=
\left\{ 
\begin{aligned}
&\sum_{j=1}(c_{4j-3}^{\dag}c^{\dag}_{4j-1}c_{4j+2|\alpha|-2}c_{4j+2|\alpha|}+h.c.),\;\text{for}\; \alpha\leq-1\\
&\sum_{j=1}(c_{4j-3}^{\dag}c_{4j-2}c^{\dag}_{4j-1}c_{4j}+h.c.),\;\text{for}\; \alpha=0\\
&\sum_{j=1}(c_{4j-2}^{\dag}c_{4j-1}c^{\dag}_{4j}c_{4j+1}+h.c.),\;\text{for}\; \alpha=1\\
&\sum_{j=1}(c_{4j-2}^{\dag}c^{\dag}_{4j}c_{4j+2\alpha-3}c_{4j+2\alpha-1}+h.c.),\;\text{for}\; \alpha\geq2.
\end{aligned}
\right.
\eea
Here $\alpha$ is an integer and the quartic interactions in $H_{int_\alpha}$ are constructed by coupling two nearest hopping terms of $H_{\alpha}$. Schematic diagrams of $H_{\alpha}$ and $H_{int_\alpha}$ are presented in Figure.~\ref{figure of Ha} and Figure~\ref{figure of Int}, respectively
\footnote{The non-interacting $H_{\alpha}$ are fermionic models analogous to the $\alpha$-chains \cite{Verresen:2017aa} with U(1) charge conservation.}.
The original SSH model is the linear superposition of $H_0$ and $H_1$. 

Both $H_{\alpha}$ and $H_{int_\alpha}$ respect the charge $U(1)$ symmetry $Uc_jU^{-1}=e^{i\alpha}c_j$, chiral symmetry $Sc_jS^{-1}=(-1)^jc_j^{\dag}$, and reflection symmetry $Rc_{j}^{\dag}R^{-1}=c_{N_{\text{tot}}-j+1}^{\dag}$, where $N_{\text{tot}}$ is the total number of sites. Additionally, the two types of lattice models are \textit{decomposable}, in the sense that the Hamiltonians can be decomposed into local and decoupled subsystems, and the ground states of the systems are just product states $\ket{GS}=\Pi_a\ket{\psi_a}$ with $\ket{\psi_a}$ being the ground state of each subsystem. 
Notice that the ground state(s) of a decomposable system defined here may in general not be unique. Nevertheless, either $H_{\alpha}$ or $H_{int_\alpha}$ at half-filling possesses a unique ground state; the uniqueness of the ground states is an essential ingredient for SPT phases.


\section{Many-body topological invariantss}\label{Many-body topological invariants}
In the operator formalism, SPT phases are typically characterized by non-local order parameters. To construct a many-body topological invariant for an SPT phase with a global symmetry $G$, one can encode such an order parameter in a Euclidean path integral (partition function)
\bea\label{partition function}
Z(X,\eta,A)=\int \prod_i D\phi_i \,\text{exp}[-S(X,\eta,A,\phi_i)],
\eea
which represents integrating over all matter degrees of freedom $\phi_i$ of the system in a background $(X,\eta,A)$. 
Here, $X$ denotes the Euclidean spacetime, which can be non-orientable if $G$ contains an orientation-reversing element (spatial reflection or time-reversal), $A$ is a background gauge field coupled to the matter fields associated with the internal or on-site sub-symmetry of $G$, and $\eta$ includes additional data such as the (s)pin structure if one considers a fermionic system.

For an invertible gapped phase, including an SPT phase, where the ground state is unique on all spatial manifolds, its partition function is expected to have a topological part which is represented by a $U(1)$ phase and independent of local data. This topological part of $Z(X,\eta,A)$ defines an invertible TQFT and can be classified by a (generalized) cohomology theory, such as the cobordism theory, which provides the classification of manifolds with certain structures. For pure-torsion topological classes, the $U(1)$ phase $\text{Arg}[Z(X,\eta,A)]$ is a cobordism invariant and its value on a generating manifold---a generator of the cobordism group $\Omega^\text{str}_d(BG)$ associated with the symmetry $G$---can serve as the many-body topological invariant \cite{PRLMTI,OnsiteMTI,crystalMTI} for the system. The classification of SPT phases is essentially given by classifying the phase factors $\text{Arg}[Z(X,\eta,A)]$.

\subsection{With chiral symmetry}\label{Z^S sec}
As a concrete example, we consider $1d$ charge-conserved fermionic systems with chiral symmetry, namely fermionic SPT phases with $G=U(1)\times \mathbb{Z}_2^{S}$, where the chiral symmetry $S=CT$ is the combination of charge-conjugation (particle-hole) and time-reversal symmetries. The classification of such SPT phases is given by Hom$(\Omega^{\text{pin}^c}_2(pt), U(1))=\mathbb{Z}_4$, where $\Omega^{\text{pin}^c}_2(pt)=\mathbb{Z}_4$ represents the cobordism group of 2d pin$^c$ manifolds and is generated by the real projective plane $\mathbb{R}P^2$ with a certain pin$^c$ structure.
The many-body topological invariant corresponds to the U(1) phase of the system's partition function~\eqref{partition function} on $\mathbb{R}P^2$ with a nontrivial pin$^c$ connection, which can be constructed using the partial transpose and chiral transformation in the operator formalism \cite{OnsiteMTI}:
\bea\label{chrial-respecting topological invariant}
Z^S=\text{Tr}_{I}[\rho_I U_S^{I_1} \rho_I^{T_1} [U_S^{I_1}]^{\dag}],
\eea
where $\rho_I$ is the reduced density matrix, $\rho_I^{T_1}$ is the partial transpose of $\rho_I$, and $U_S^{I_1}$ is the unitary part of the chiral symmetry. 

To proceed,  we will elaborate on how to evaluate $Z^S$. First, consider a chiral symmetric system with a unique ground state in the Hilbert space defined on the space manifold $S^1$. Introduce an interval $I=I_1 \cup I_2$ on $S^1$, where $I_1=\{c_i,...,c_j\}$ and $I_2=\{c_{j+1},...,c_n\}$ are two adjacent intervals, such as
\begin{equation*}
\ldots c_{i-1}\;\underbrace{c_i\;\ldots\;c_j}_{I_1}\; \underbrace{c_{j+1}\;\ldots\;c_{n}}_{I_2}\; c_{n+1}\ldots
\end{equation*}
For a given unique ground state $\ket{GS}$, $\rho_I$ can be obtained by tracing out the degrees of freedom outside $I$: $\rho_I=\text{Tr}_{S^1\backslash I}[\ket{GS}\bra{GS}]$. Given that $\rho_I$ can be expanded by the occupation number basis, $\rho_I=\sum_{n,\overline{n}} a_{n,\overline{n}}\ket{n}\bra{\overline{n}}=\sum_{n,\overline{n}} a_{n,\overline{n}}\ket{\{n_j\}_{j\in I_1},\{n_j\}_{j\in I_2}}\bra{\{\overline{n}_j\}_{j\in I_1},\{\overline{n}_j\}_{j\in I_2}}$, where $a_{n,\overline{n}}$ is a complex coefficient, we can study the fermionic partial transpose combined with $U_S^{I_1}$ in the occupation number basis, which is given by
\bea\label{chiral-respecting topo regarding occupied states}
\begin{aligned}
&U_S^{I_1} (\ket{\{n_j\}_{j\in I_1},\{n_j\}_{j\in I_2}}\bra{\{\overline{n}_j\}_{j\in I_1},\{\overline{n}_j\}_{j\in I_2}})^{T_1} [U_S^{I_1}]^{\dag}\\
&=i^{[\tau_1+\overline{\tau}_1]}(-1)i^{(\tau_1+\overline{\tau}_1)(\tau_2+\overline{\tau}_2)}U_S^{I_1}C_f^{I_1} (\ket{\{\overline{n}_j\}_{j\in I_1},\{n_j\}_{j\in I_2}}\bra{\{n_j\}_{j\in I_1},\{\overline{n}_j\}_{j\in I_2}})^{T_1}[C_f^{I_1}]^{\dag} [U_S^{I_1}]^{\dag},\\
\end{aligned}
\eea
where
\bea
\tau_i=\sum_{j\in I_i} n_j, \quad \overline{\tau}_i=\sum_{j\in I_i} \overline{n}_j, \quad [\tau]=\left\{ 
\begin{aligned}
0\quad& (\tau:\text{even})\\
1\quad& (\tau:\text{odd})
\end{aligned} 
\right.
, \quad C_f^{I_1}=\prod_{j\in I_1} (c_j^{\dag}+c_j).
\eea
In Appendix~\ref{analytic ZS of H1}, we present how to calculate $Z^S$ of $H_1=\sum_{j=1}(c_{2j}^{\dag}c_{2j+1}+h.c.)$ in the occupation number basis.

\subsection{With reflection symmetry}
If we replace the chiral symmetry $S$ in the above discussion by reflection symmetry $R$, we obtain the same topological classification; that is, fermionic SPT phase with $G=U(1)\times \mathbb{Z}_2^{R}$ are also classified by Hom$(\Omega^{\text{pin}^c}_2(pt), U(1))=\mathbb{Z}_4$. This is obvious from the perspective of Euclidean TQFT, as $S=CT$ and $R$ become the same symmetry in Euclidean signature.
However, the form of the partition function in the operator formalism differs significantly between the two cases. For reflection symmetry, the partition function being the many-body topological invariant of the system is evaluated using the partial reflection\cite{crystalMTI}:
\bea\label{reflection-respecting topological invariant}
Z^R=\bra{GS}[U_{\alpha}R]_I\ket{GS},
\eea
where $\ket{GS}$ is the unique ground state, and $[U_{\alpha}R]_I$ is the partial reflection. Specifically, we consider again a system with a unique ground state in the Hilbert space defined on the space manifold $S^1$. If it respects the reflection symmetry $Rc_{j}^{\dag}R^{-1}=c_{N_{\text{tot}}-j+1}^{\dag}$ and the $U(1)$ symmetry, we can pick up an interval $I=\{c_{a},\ldots,c_{b}\}\subset S^1$ ($a<b<N_{\text{tot}}\in \mathbb{Z}$), where the center of the interval $I$ is the same as the reflection center, to define the partial reflection $[U_{\alpha}R]_I$:
\bea\label{partial reflection}
[U_{\alpha}R]_I c_{j}^{\dag} ([U_{\alpha}R]_I)^{-1}=e^{-i\alpha}c_{b-j+a}^{\dag} \quad [U_{\alpha}R]_{I}\ket{0}=\ket{0}, \quad j\in\{a,\ldots,b\},
\eea
To achieve the $\mathbb{Z}_4$ classification, the appropriate $U(1)$ phase is $\alpha=\pm\pi/2$. For all the cases we study later, we will take $\alpha=-\pi/2$. Appendix~\ref{analytic ZR of H1} provides an analytical computation of  $Z^R$ of $H_1$ in the occupation number basis.

\section{Chiral-symmetric SPT phases}\label{Chiral-symmetric SPT phases}
Most interacting SPT phases can be understood by examining their free fermion counterparts in the presence of interactions. A renowned example was proposed by Fidkowski and Kitaev~\cite{edgedegeneracy1}, demonstrating that by introducing a quartic interaction, one can adiabatically deform a system of eight stacked Kitaev chains with periodic boundary conditions (PBC) in the topological phase into the trivial phase without closing the bulk gap. This highlights the connection between the interacting SPT phases of $1d$ superconductors in the BDI class, classified as $\mathbb{Z}_8$, and their free fermion analogs, which are classified as $\mathbb{Z}$. 
Tang and Wen's work \cite{edgedegeneracy2} also indicated the relation between interacting and free fermion SPT phases for $1d$ systems with on-site symmetries, illustrated by studying systems of stacked chains with open boundary conditions (OBCs). Inspired by these studies, we propose a method to understand the fermionic SPT phases of $1d$ charge-conserved systems with chiral symmetry from their free fermion counterparts.

Bulk-boundary correspondence is one of the notable properties of topological insulators \cite{TI,BBC1,BBC2,BBC3,BBC4,BBC5,BBC6,BBC7,BBC8,BBC9}. For $1d$ free fermion systems, this correspondence relates the bulk topological invariant to the presence of nontrivial edge states. For chiral symmetric systems in the semi-infinite limits, the number of robust zero-energy edge states is characterized by the winding number \cite{BBCthroK,Greensfunctions,free_chiral} defined on the bulk Bloch Hamiltonian. While this correspondence is limited to free fermion systems, it provides insights into understanding the interacting fermionic SPT phases, where the classification reduces from $\mathbb{Z}$ to $\mathbb{Z}_4$. In particular, for a decomposable system, the robust zero-energy states correspond to unbound fermions, and we can count the number of unbound fermions even in a many-body system. According to this idea, for $1d$ charge-conserved fermionic systems with the chiral symmetry $Sc_jS^{-1}=(-1)^jc_j^{\dag}$, we relabel fermions as
\begin{equation*}
\begin{aligned}
&c_{2j-1}=a_j,\quad &&Sa_jS^{-1}=-a^{\dag}_j;\\
&c_{2j}=b_j,\quad &&Sb_jS^{-1}=b^{\dag}_j,\\ 
\end{aligned}
\end{equation*}
and then propose the following bulk-boundary correspondence
\bea\label{ZS and Ne}
2\text{Arg}[Z^S(H_d)]/\pi=N_b-N_a\,\,\text{mod}\,\,4
\eea
for a decomposable system $H_d$. Here, $Z^S(H_d)$ represents the partial-transpose partition function \eqref{chrial-respecting topological invariant} of $H_d$ with PBC, which gives rise to a chiral-respecting many-body topological invariant $\text{Arg}[Z^S(H_d)]$, and $N_a$ ($N_b$) is the number of unbinding $a_j$ ($b_j$) of $H_d$ in the right semi-infinite limits. 
Note that this bulk-boundary correspondence holds only for the chiral symmetry $Sc_jS^{-1}=(-1)^jc_j^{\dag}$,  because the value of $Z^S$ depends on the unitary part $U_S^{I_1}$ of the chiral symmetry. 

Although this correspondence is only valid for decomposable systems, it can still be useful when discussing non-decomposable systems. For a gapped system which is a linear superposition of decomposable systems, assigning $N_b-N_a$ to its decomposable components allows us to determine whether a (topological) phase transition will occur when deforming the parameters of the combined system.

\subsection{Phase structures and many-body spectra of deformed systems}
\begin{table}[htb!]
\centering
\renewcommand{\arraystretch}{1.5}
\begin{tabular}{c|cccccc}
          & $\pm H_{2}$ & $\pm H_{1}$ & $\pm H_{0}$ & $\pm H_{-1}$ & $\pm 
 H_{-2}$ \\ \hline 
$N_a-N_b$ & $-2$ & $-1$ & $0$  & $1$ & $2$ \\ \hline
$Z^S$ & $-1/64$ & $-i/8$ & $1$ & $i/8$ & $-1/64$  \\
\end{tabular}
\end{table}
\begin{table}[htb!]
\centering
\renewcommand{\arraystretch}{1.5}
\begin{tabular}{c|cccccc}
          & $\pm H_{int_2}$ & $\pm H_{int_1}$ & $\pm H_{int_0}$ & $\pm H_{int_{-1}}$ & $\pm H_{int_{-2}}$ \\ \hline 
$N_a-N_b$ & $-2$ & $-1$ & $0$  & $1$ & $2$ \\ \hline
$Z^S$ & $-1/8$ & $-i/8$ & $1$ & $i/8$ & $-1/8$  \\
\end{tabular}
\caption{Numbers of unbound fermions $N_a-N_b$ and partial-transpose partition functions $Z^S$ for $H_{\alpha}$ and $H_{int_\alpha}$ at half-filling. $Z^S$ is evaluated for systems with $N_{\text{tot}}=12$ and we choose $I_1=\{c_1,...,c_4\}$ and $I_2=\{c_{5},...,c_8\}$.}
\label{Table for chiral H}
\end{table}
We now examine the bulk-boundary correspondence \eqref{ZS and Ne} by studying the models \eqref{free_systems} and \eqref{interacting_systems}. Since $H_{\alpha}$ and $H_{int_\alpha}$ are decomposable and their ground states at half-filling are product states, one can assign the quantum number $N_b-N_a$ to them in the right semi-infinite limits. This quantum number can be easily determined by referring to Fig.~\ref{figure of Ha} and Fig.~\ref{figure of Int}, where $N_a$ and $N_b$ are unbinding fermions colored blue and red in these figures, respectively. The partial-transpose partition function $Z^S$ and $N_b-N_a$ of $H_\alpha$ and $H_{int_\alpha}$ with $\alpha=-2, -1, 0, 1, 2$ are presented in Table~\ref{Table for chiral H}, showing that \eqref{ZS and Ne} holds. It is worth noting that Table~\ref{Table for chiral H} indicates our proposed bulk-boundary correspondence is valid even if the systems cannot be described in the single-particle basis.

\begin{figure}[hbt!]  
\centering
\subfloat[]{\includegraphics[width=0.31\textwidth]{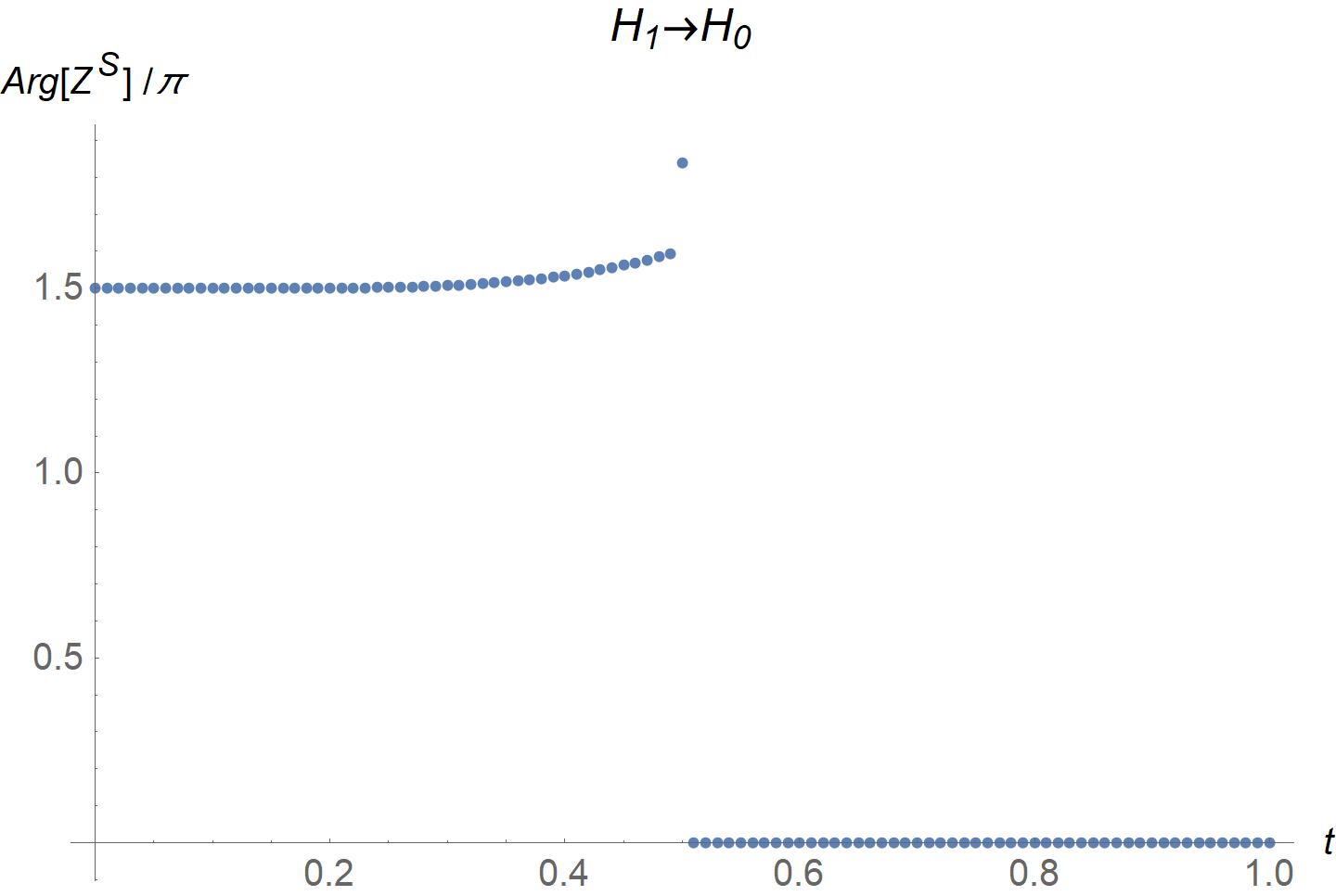}}\hskip 0.5cm
\subfloat[]{\includegraphics[width=0.31\textwidth]{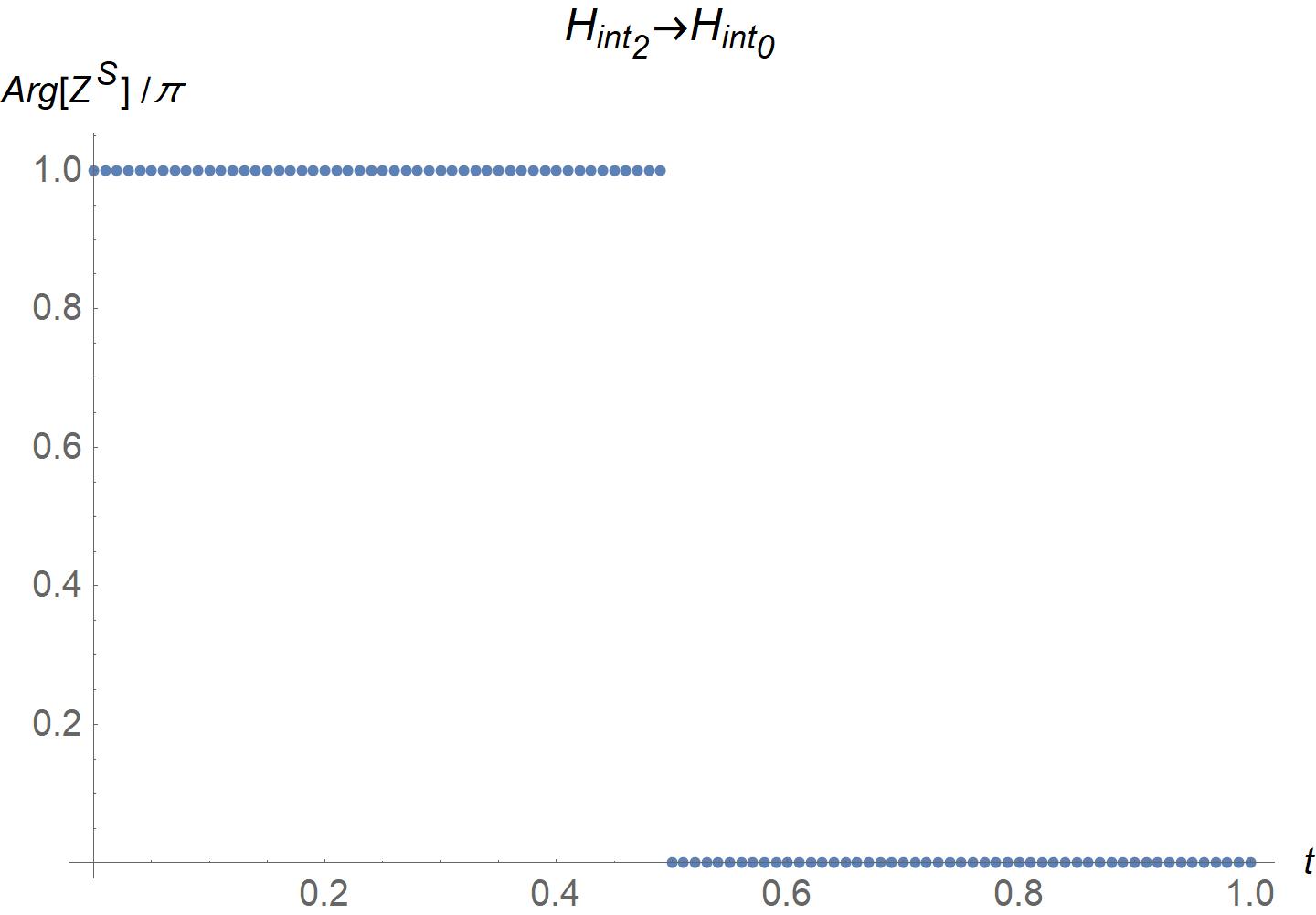}}\hskip 0.5cm
\subfloat[]{\includegraphics[width=0.31\textwidth]{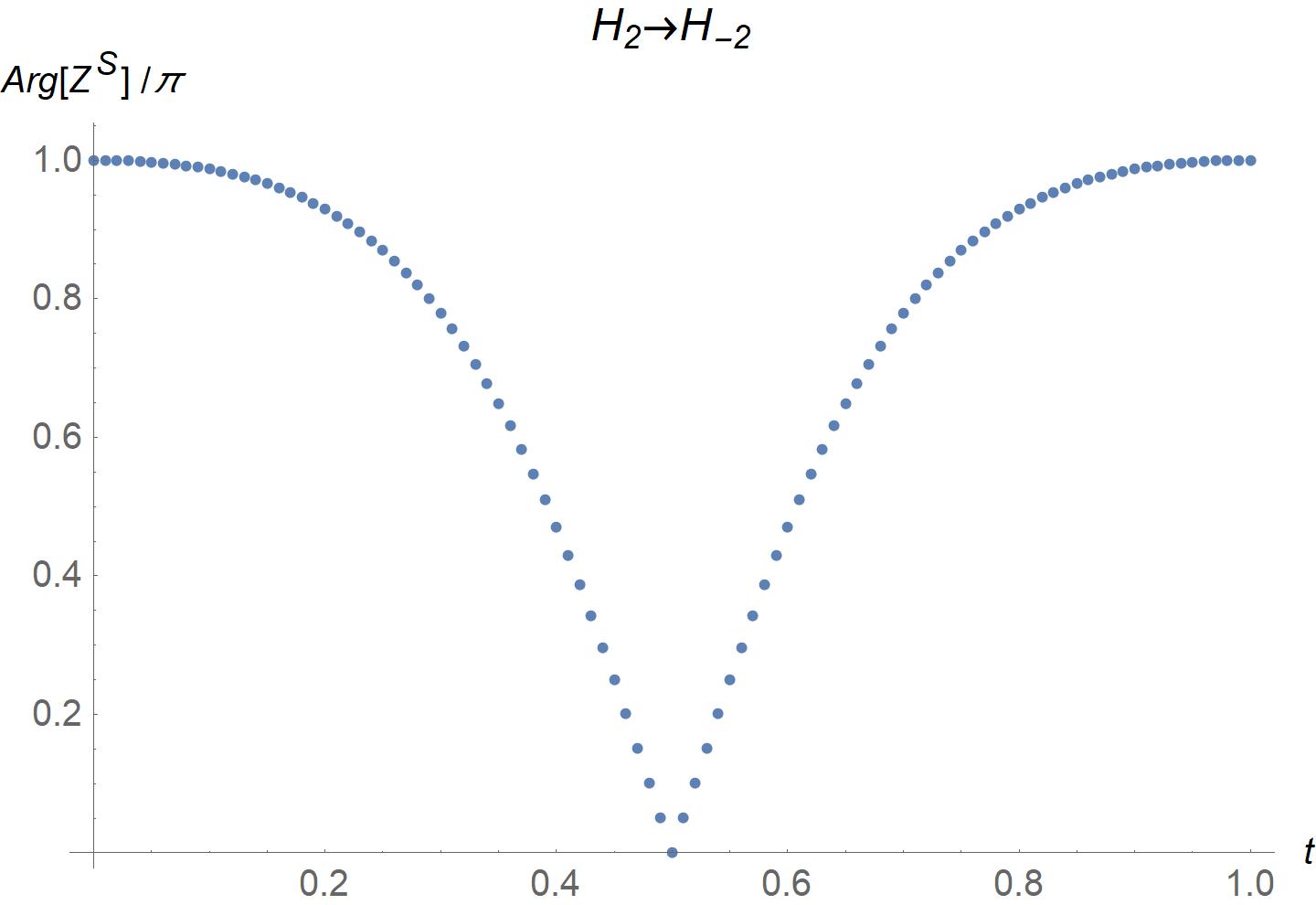}}\\
\subfloat[]{\includegraphics[width=0.31\textwidth]{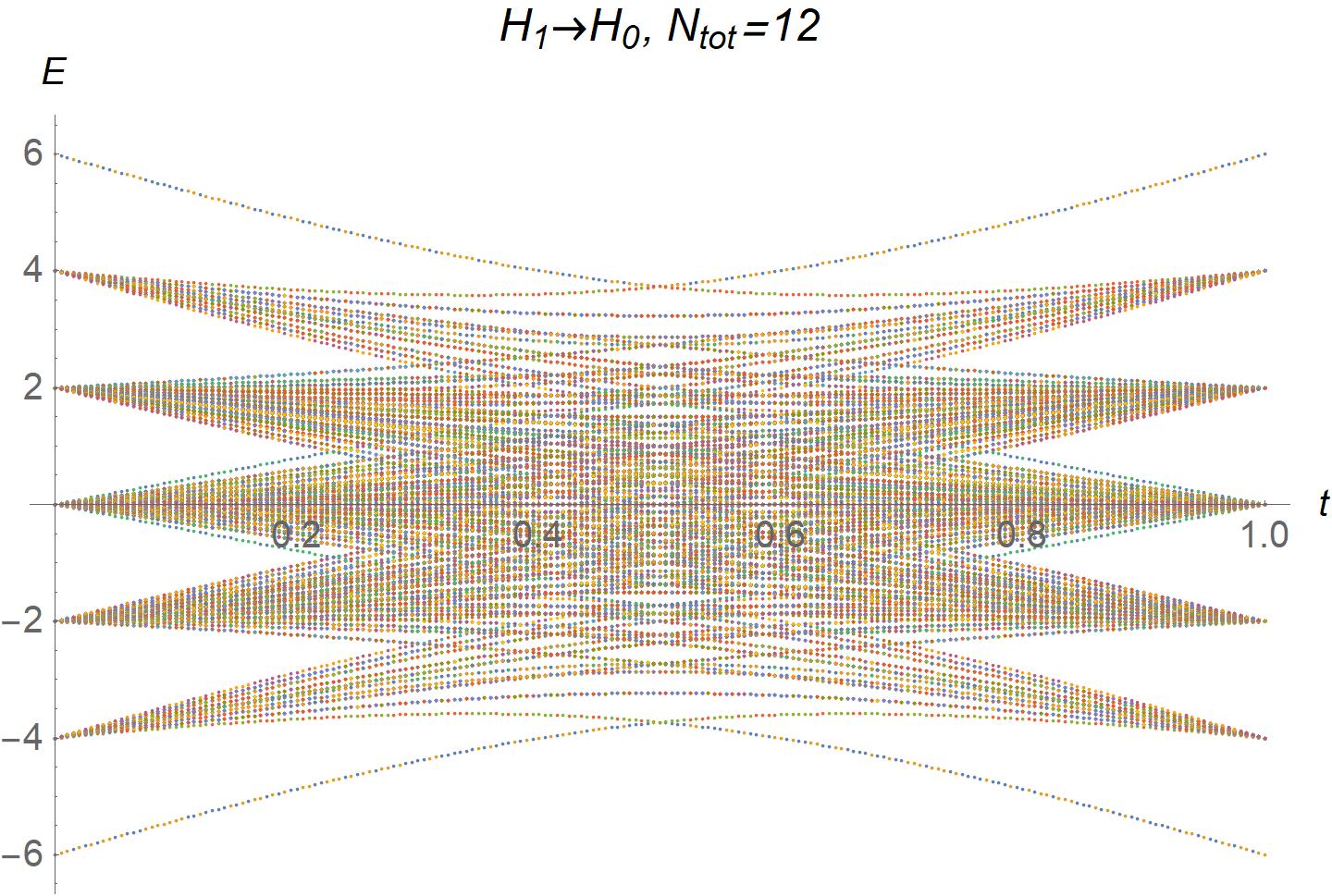}}\hskip 0.5cm
\subfloat[]{\includegraphics[width=0.31\textwidth]{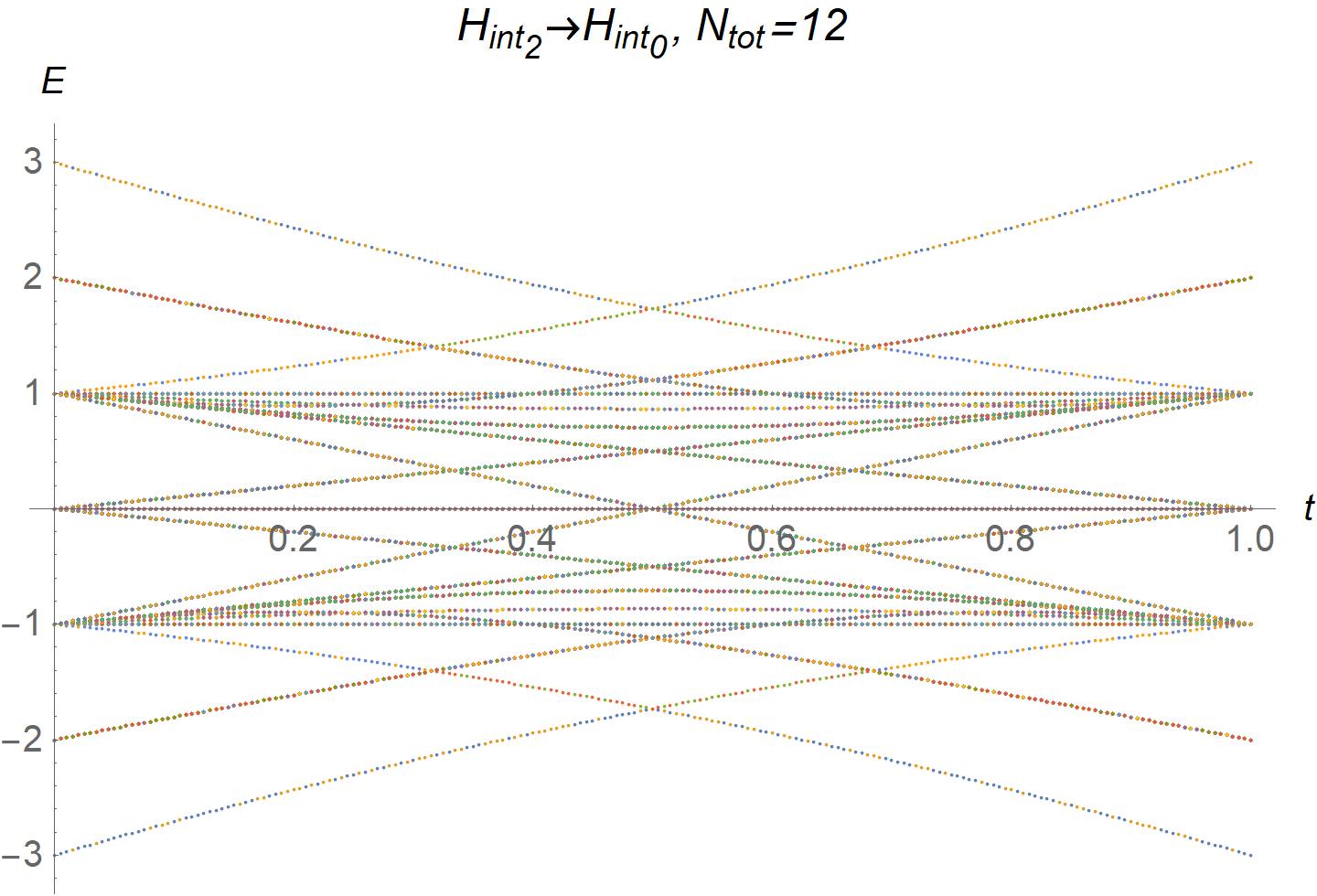}}\hskip 0.5cm
\subfloat[]{\includegraphics[width=0.31\textwidth]{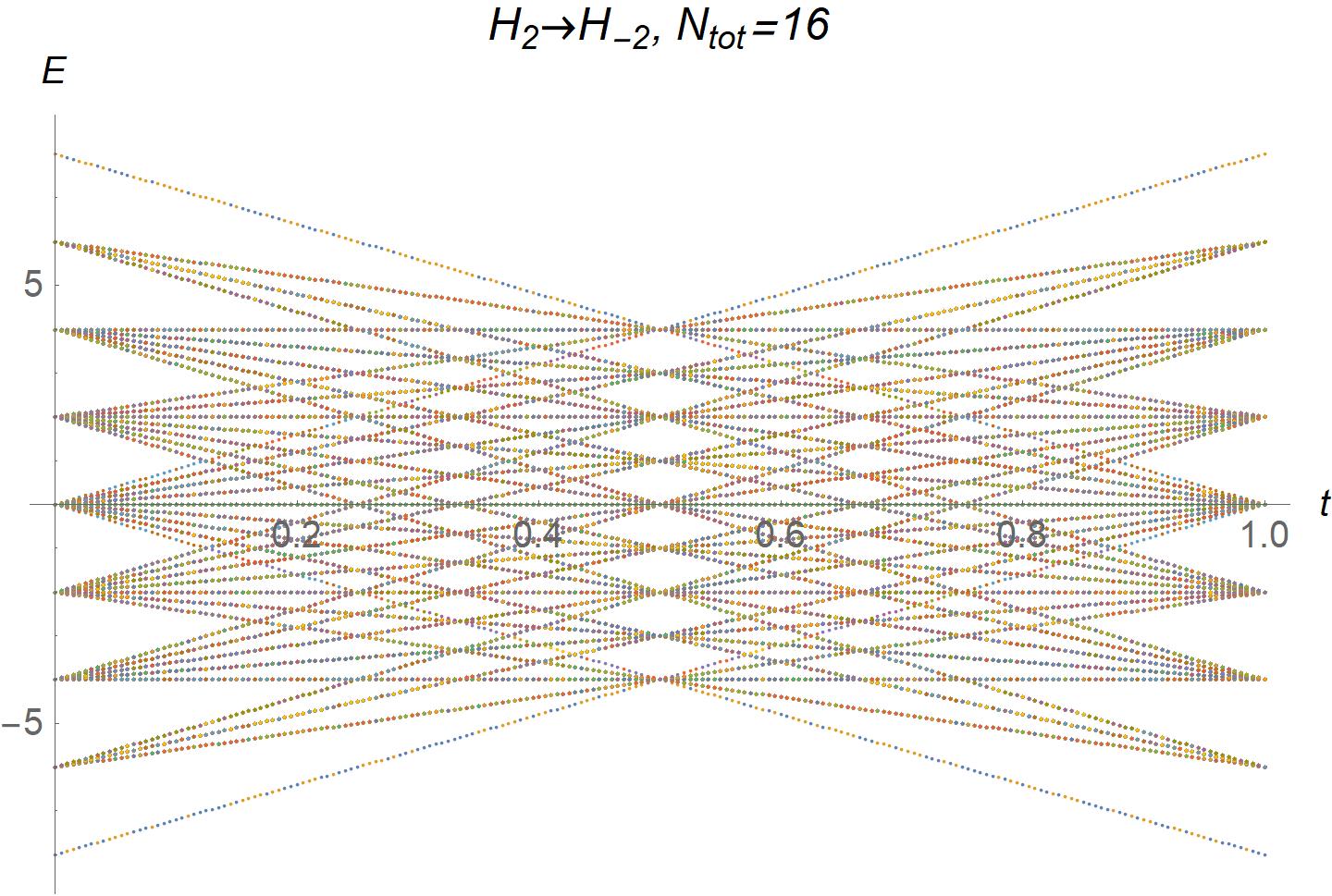}}\\
\caption{(a)(b)(c) $\text{Arg}[Z^S]$ and (d)(e)(f) spectra of the deformations $H_{1}\rightarrow H_{0}$, $H_{2}\rightarrow H_{-2}$, and $H_{int_{2}}\rightarrow H_{int_{0}}$ with PBCs.}
\label{ZS EN for w1tow0}
\end{figure}

\begin{figure}[htb!]
\centering
\subfloat[]{\includegraphics[width=0.31\textwidth]{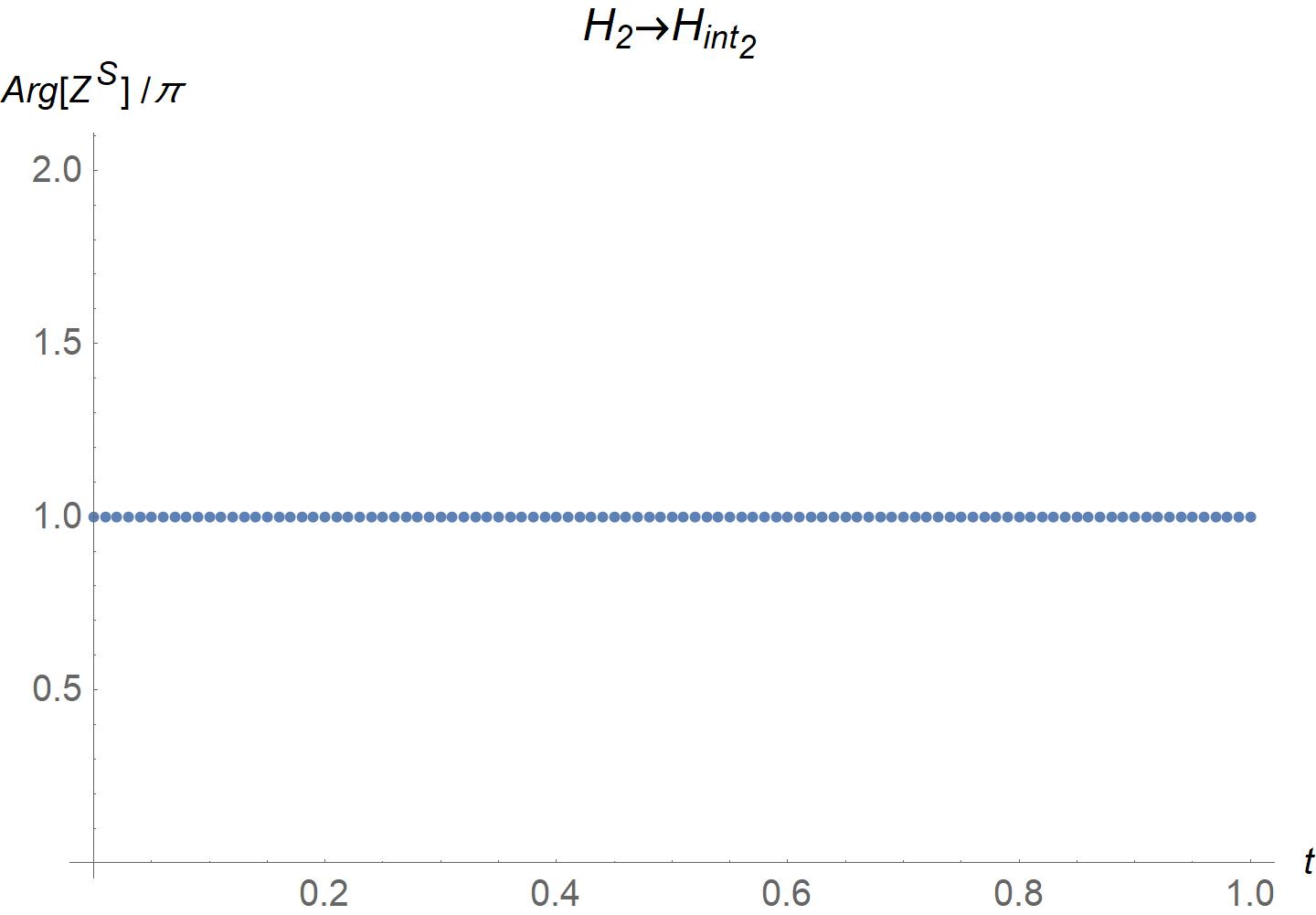}}\hskip 0.5cm
\subfloat[]{\includegraphics[width=0.31\textwidth]{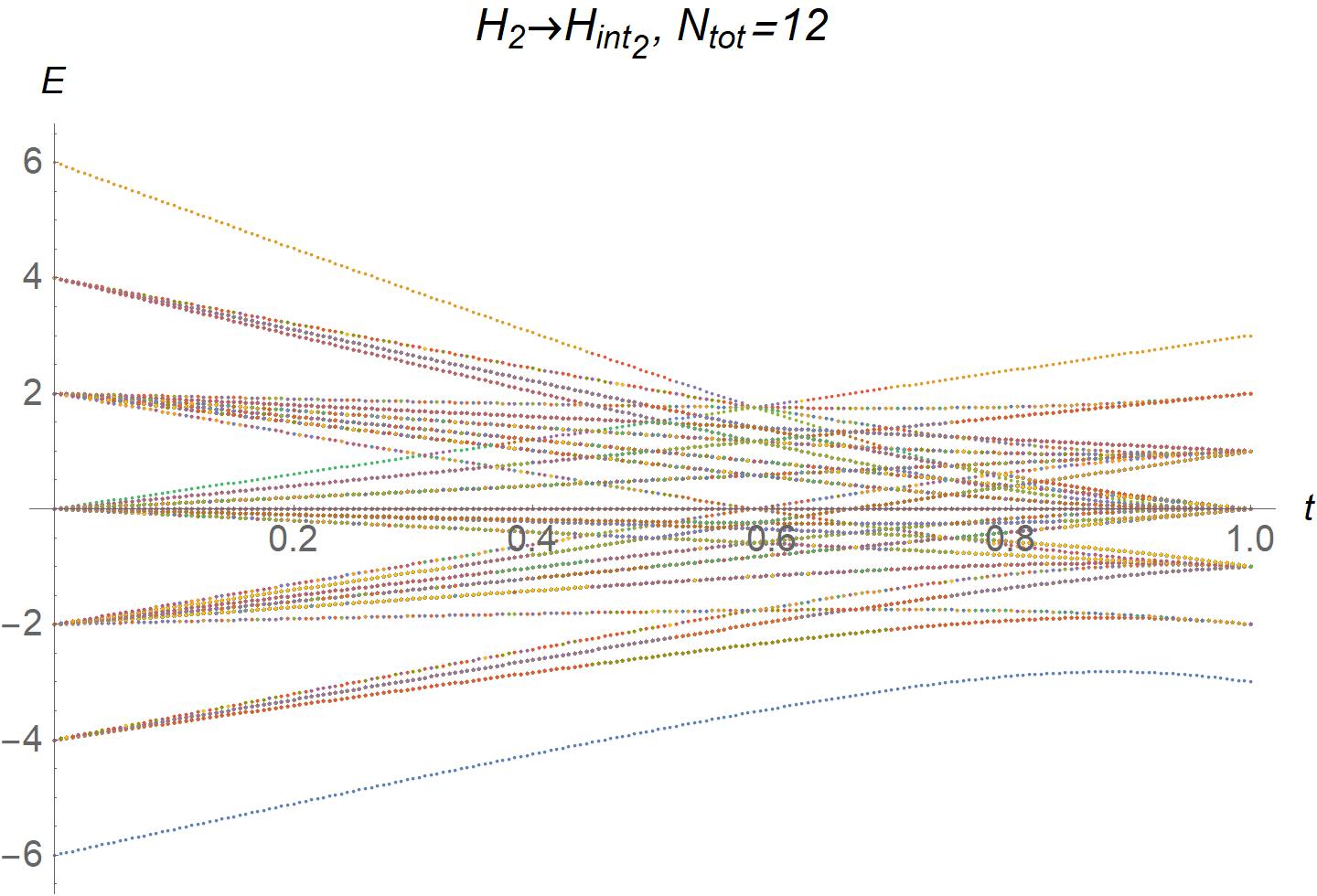}}\hskip 0.5cm
\subfloat[]{\includegraphics[width=0.31\textwidth]{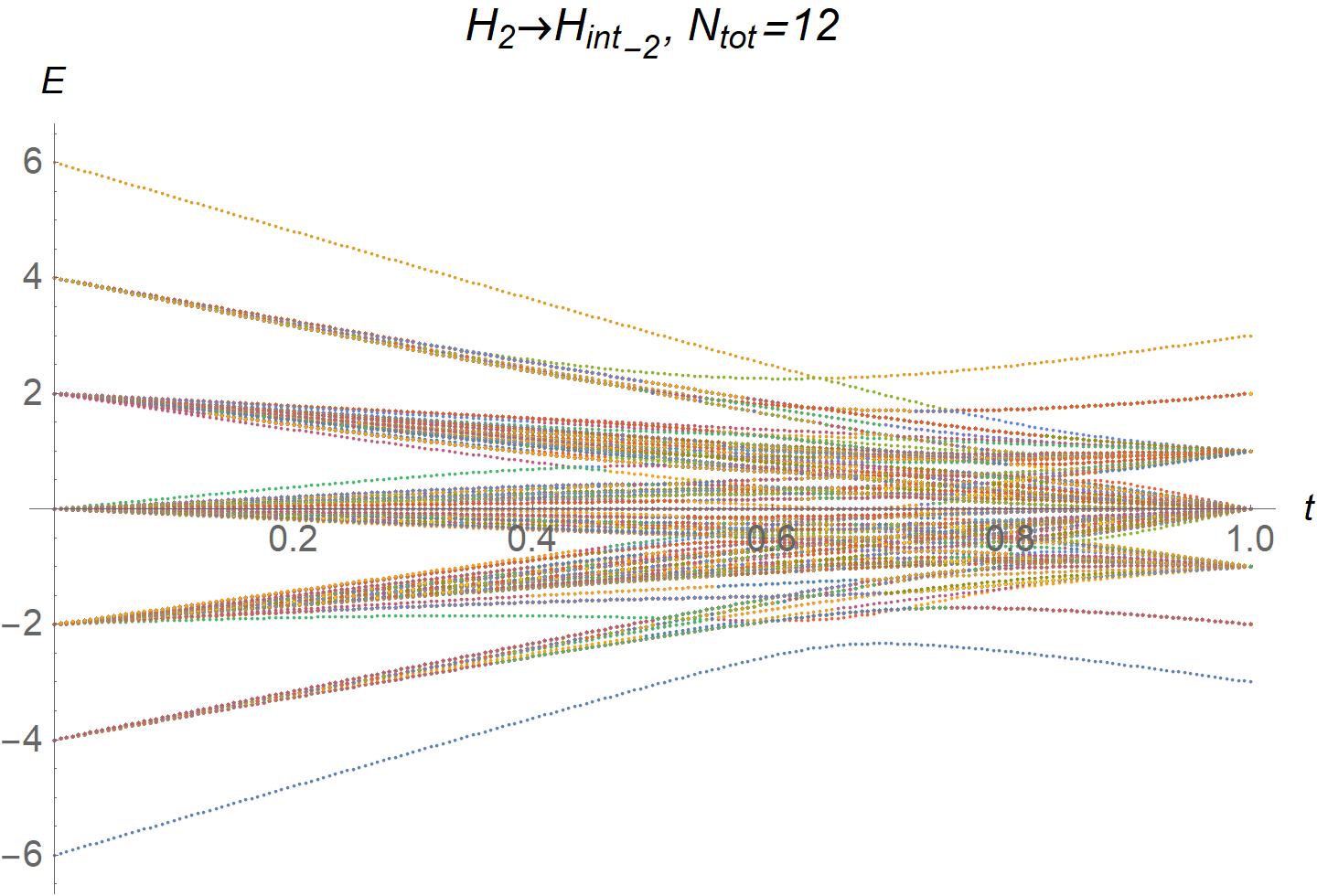}}\\
\subfloat[]{\includegraphics[width=0.31\textwidth]{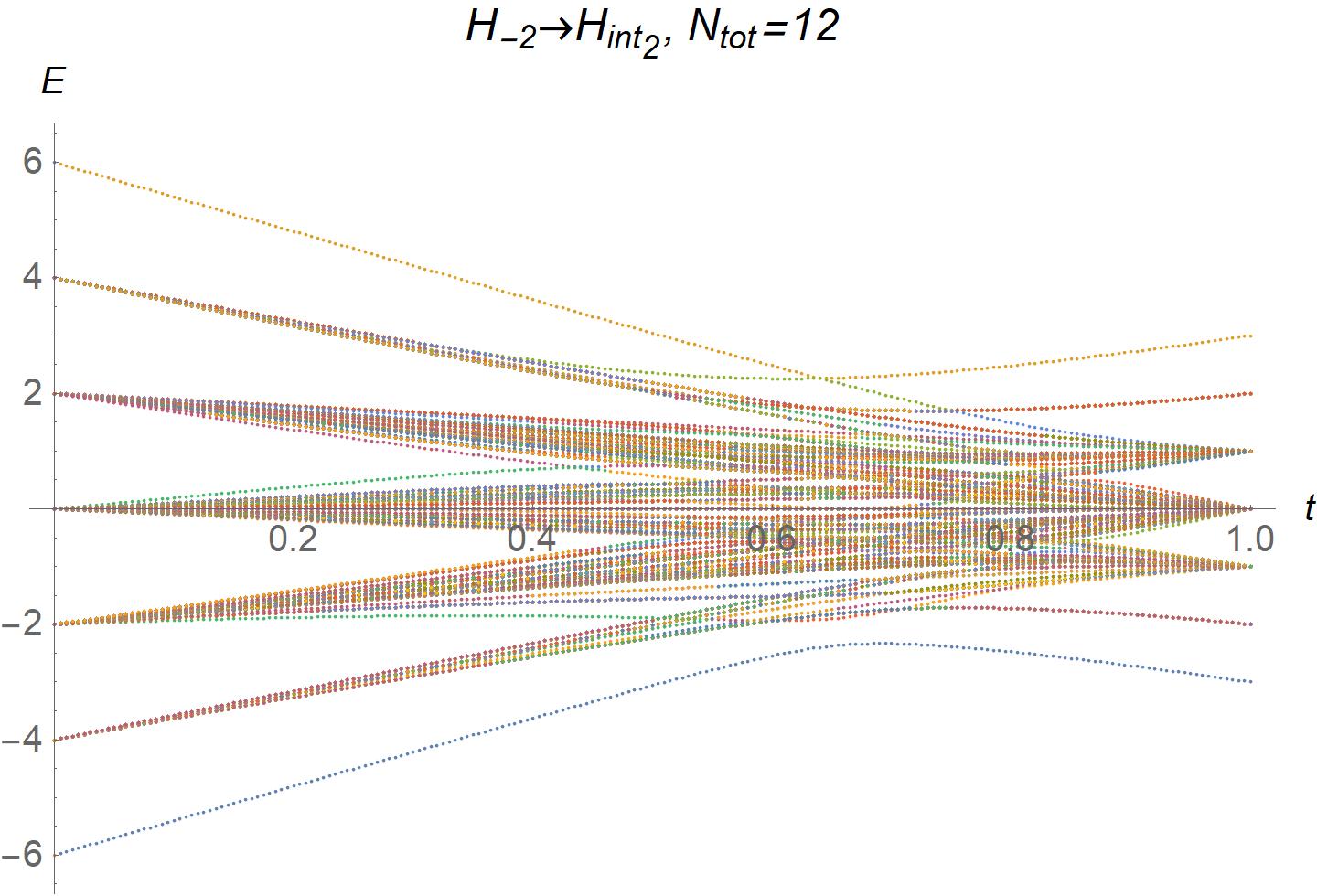}}\hskip 0.5cm
\subfloat[]{\includegraphics[width=0.31\textwidth]{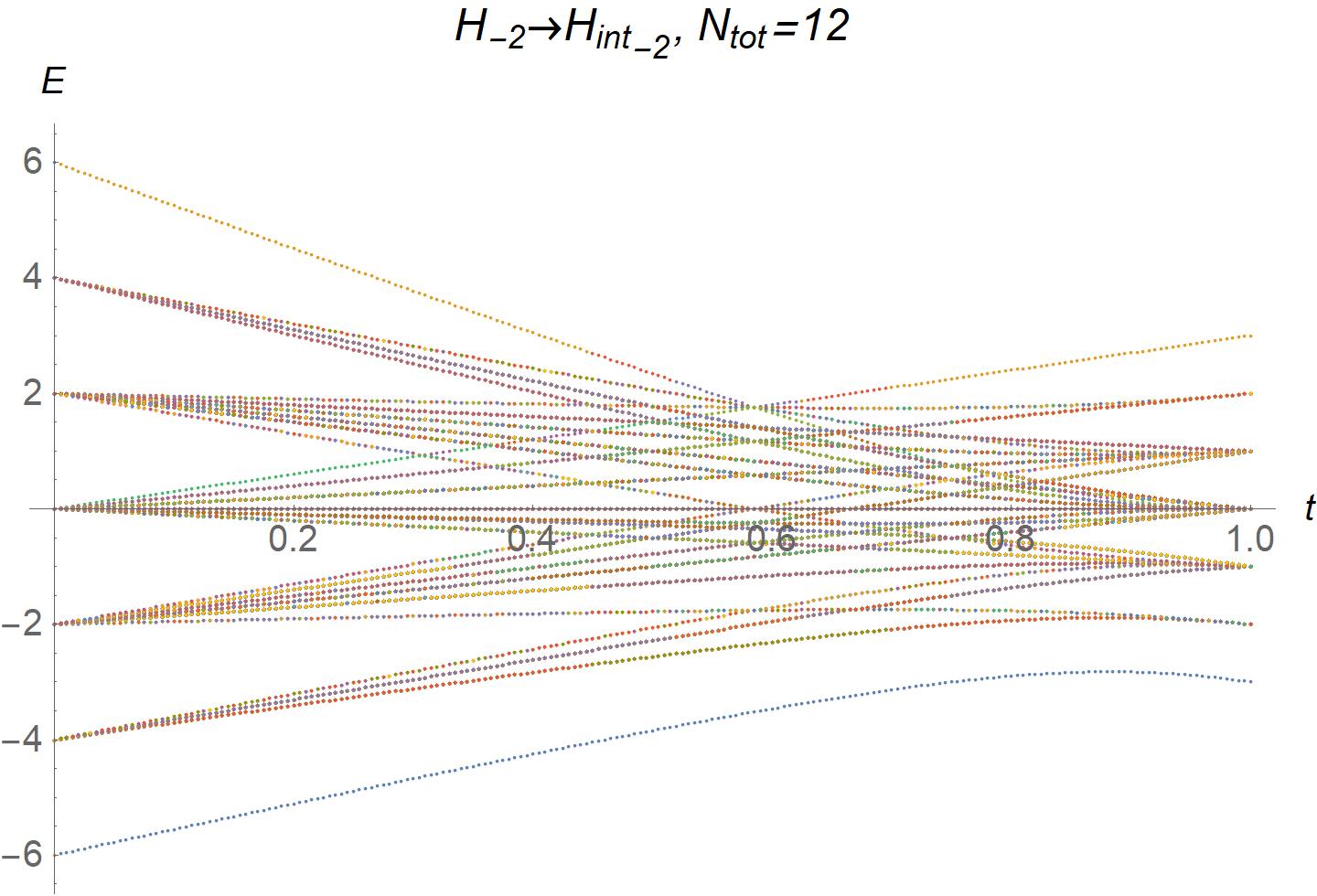}}\\
\caption{(a) $\text{Arg}[Z^S]$ of $H_{2}\rightarrow H_{int_{2}}$ (the same for $H_{2}\rightarrow H_{int_{-2}}$, $H_{-2}\rightarrow H_{int_{2}}$, and $H_{-2}\rightarrow H_{int_{-2}}$). (b)(c)(d)(e) Spectra of $H_{\pm2}\rightarrow H_{int_{\pm2}}$ with PBCs.}
\label{EN for H2_to_Intpm2 and Hm2_to_Intpm2}
\end{figure}

As a concrete example, let us consider the following deformed Hamiltonian:
\bea
(1-t)H_{a}+tH_{b}, \quad t\in\left[0,1\right].
\eea
Tuning the parameter $t$ from $0$ to $1$ in the above equation represents the adiabatic deformation from $H_{a}$ to $H_{b}$, which we will denote as $H_{a}\rightarrow H_{b}$ in the subsequent discussion. 
The energy spectra of such systems are solved by exact diagonalization. 
Here, we take the deformations $H_{1}\rightarrow H_{0}$, $H_{int_{2}}\rightarrow H_{int_{0}}$, $H_{2}\rightarrow H_{-2}$, and $H_{2}\rightarrow H_{int_{\pm2}}\rightarrow H_{-2}$ as examples to test our analytical results in Table~\ref{Table for chiral H} and their implications to phase transitions.
According to Table~\ref{Table for chiral H}, a phase transition is expected to occur in either $H_{1}\rightarrow H_{0}$ or $H_{int_{2}}\rightarrow H_{int_{0}}$, and the numerical results in Figure~\ref{ZS EN for w1tow0} align with this expectation, showing a phase transition at $t=0.5$ in both cases \footnote{The cobordism invariant partition functions, such as $Z^S$ and $Z^R$, may not be well quantized if the \textbf{subregion} we choose is not sufficiently larger than the correlation length \cite{crystalMTI}. Furthermore, they are ill-defined if systems have degenerate ground states (e.g. the midpoint of the deformation $H_{1}\rightarrow H_{0}$).}.
A more nontrivial and interesting case is deforming $H_{2}$ to $H_{-2}$. Since $H_{2}$ and $H_{-2}$ are characterized by different single-particle topological invariants (winding numbers), there must be a phase transition in the non-interacting deformation $H_{2}\rightarrow H_{-2}$, as shown in Figure~\ref{ZS EN for w1tow0}. On the other hand, $H_{2}$ and $H_{-2}$ share the same many-body topological invariant $\text{Arg}[Z^S]=\pi$ , meaning that there exists a chiral symmetric deformations with interactions enabling us to adiabatically deform $H_{2}$ to $H_{-2}$ without closing the many-body energy gap. As shown in Figure~\ref{EN for H2_to_Intpm2 and Hm2_to_Intpm2}, both the deformations $H_{2}\rightarrow H_{int_{\pm2}}\rightarrow H_{-2}$ do the job. Although our numerical calculations were performed on finite systems, the conclusion here should remain valid in the thermodynamic limit; see Appendix~\ref{Finite-size effect and many-body spectra} for further discussions on the finite-size effect.


\section{Reflection-symmetric SPT phases}\label{Reflection-symmetric SPT phases}
Now, let us consider charge-conserved systems with reflection symmetry, i.e., $G=U(1)\times \mathbb{Z}_2^{R}$. For topological systems with reflection symmetry, there is no bulk-edge correspondence in real space
\footnote{However, one still can use the mid-gap states in the entanglement spectrum to distinguish topologically trivial and non-trivial phase\cite{Turner:2010aa, Hughes:2011aa}.},
so the approach used for systems with chiral symmetry does not apply to such SPT phases. Nevertheless, using the Atiyah-Hirzebruch spectral sequence (AHSS) in generalized homology, we can still establish a relation between the partial reflection $Z^R$ defined in \eqref{reflection-respecting topological invariant} and certain quantum numbers assigned from a decomposable system. Generalized homology provides a classification of SPT phases associated with crystalline symmetries, and the AHSS is a common method for computing generalized homology \cite{ghomo2}. With the help of the AHSS, we will show that the topological invariant $Z^R$ actually corresponds to a pair of quantum numbers $(N_C,R_C)\in\mathbb{Z}\times\mathbb{Z}_2$, defined by an equivalence relation. Here, $N_C$ is the number of charges at the reflection center, and $R_C$ is defined as:
\bea\label{R2 def}
R_C=\left\{ 
\begin{aligned}
0,\quad&\text{if} \ R\psi_{c}R^{-1}=\psi_{c}&\\
1,\quad&\text{if} \ R\psi_{c}R^{-1}=-\psi_{c}&
\end{aligned}
,
\right.
\eea
where $\psi_{c}$ represents the charges localized on the reflection center. Since $(N_C,R_C)$ essentially represents the quantum numbers for the reflection center, which is a (0+1)$d$ system, the relation between $Z^R$ and $(N_C,R_C)$ can be regarded as a type of bulk-boundary correspondence, which we may call \textit{bulk-center correspondence}. 
This relation, similar to the bulk-edge correspondence for chiral symmetric systems, can also be extended to systems with short-range entangled ground states beyond decomposable systems, provided a more careful definition of the quantum numbers localized at the reflection center.

\subsection{Bulk-center correspondence from the AHSS}\label{classification and (N_C,R_C)}
\begin{figure}[htb!]
\centering
\includegraphics[width=0.45\textwidth]{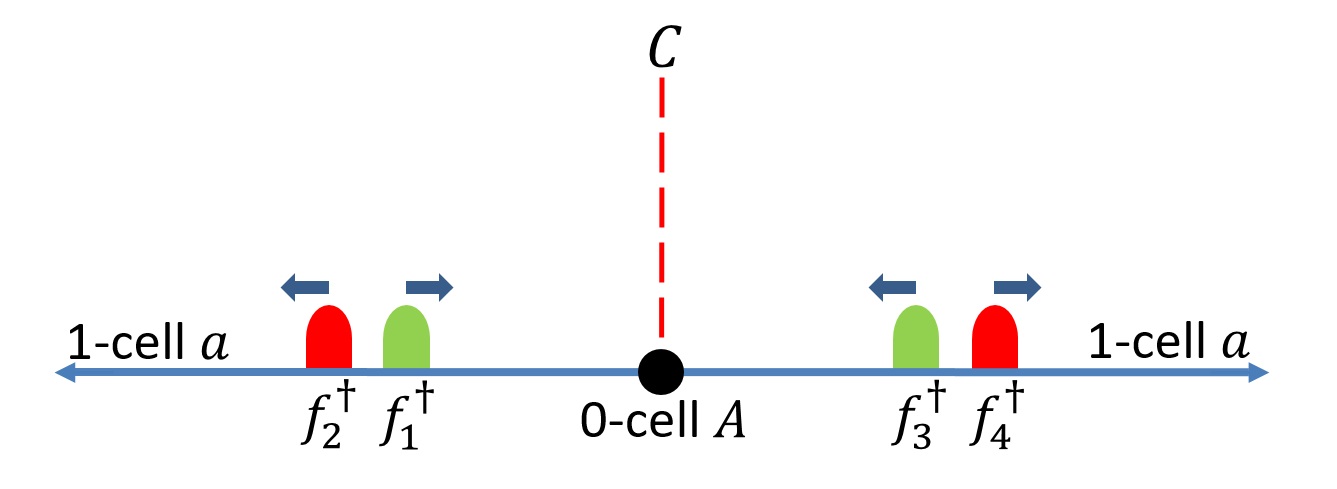}
\caption{Schematic of the first differential $d^{1}_{1,0}$. The red dashed line denotes the reflection center, while the green and red wave packets represent complex fermions with charges $+e$ and $-e$, respectively.}
\label{d1 picture}
\end{figure}
To see the relation between $Z^R$ and $(N_C, R_C)$, we need to delve into the details of generalized homology, which can be used to formulate SPT phases with crystalline symmetries \cite{ghomo2}. In Appendix \ref{The AHSS in generalized homology}, we provide a brief review of how to compute generalized homology using the AHSS. For our case, considering $1d$ systems with $G = U(1) \times \mathbb{Z}_2^{R}$ symmetry, where the crystalline symmetry is $\mathbb{Z}_2^{R}$, the classification is given by $h_{0}^{\mathbb{Z}_2}(\mathbb{R}, \partial \mathbb{R})$, which describes the SPT phases over $\mathbb{R}$ where anomalies may localize on $\partial \mathbb{R}$.

As the first step of the AHSS, we perform the $\mathbb{Z}_2$-symmetric cell decomposition, leading to one 0-cell ${A}$ that respects $G = U(1) \times \mathbb{Z}_2^{R}$ symmetry and one 1-cell ${a}$ with $U(1)$ symmetry, yielding the following $E^1$-page:
\bea\label{E-1 page for our case}
\centering
\renewcommand{\arraystretch}{1.5}
\begin{tabular}{c|cc}
$q=0$        &  $\mathbb{Z}\times \mathbb{Z}_2$ & $\mathbb{Z}$ \\
$q=1$        & $0$                              & $0$ \\ \hline
$E^1_{p,-q}$ & $p=0$                          & $p=1$   \\
\end{tabular}.
\eea
The $E^2$-pages are then given by
\bea\label{E2-page for our case}
\begin{aligned}
&E^{2}_{0,0}=E^{1}_{0,0}/\text{Im}(d^1_{1,0}),\\
&E^{2}_{1,-1}=0.\\
\end{aligned}
\eea
As the $E^2$-pages are the limiting pages for $1d$ systems, the generalized homology $h_{0}^{\mathbb{Z}_2}(\mathbb{R}, \partial \mathbb{R})$ fits into the short exact sequence:
\bea\label{short exact sequences for our case}
0\rightarrow E^{2}_{0,0}\rightarrow h_{0}^{\mathbb{Z}_2}(\mathbb{R},\partial \mathbb{R})\rightarrow E^{2}_{1,-1}\rightarrow 0.
\eea
Since $E^{2}_{1,-1} = 0$, the classification here is given by $h_{0}^{\mathbb{Z}_2}(\mathbb{R}, \partial \mathbb{R}) \cong E^{2}_{0,0} = E^{1}_{0,0} / \text{Im}(d^1_{1,0})$. Without loss of generality, we can assign quantum numbers $(N_C, R_C)$ to $E^{1}_{0,0}$, which characterize the degree-0 SPT phases at the reflection center, giving a $\mathbb{Z} \times \mathbb{Z}_2$ classification. Here, $N_C$ is the number of charges located at the reflection center, and $R_C$ is defined by eq.~\eqref{R2 def}. With the quantum numbers $(N_C, R_C)$, the quotient $E^{1}_{0,0} / \text{Im}(d^1_{1,0})$, which essentially represents a process where the SPT states on the 0-cell are trivialized by pair creation of SPT states on the adjacent 1-cell \cite{d^1PhysRevX.7.011020}, can be readily understood. Let's first focus on $\text{Im}(d^1_{1,0})$. As shown in Fig.~\ref{d1 picture}, $d^1_{1,0}$ represents the adiabatic pump as follows. First, we create a pair of complex fermions with charges $+e$ and $-e$, denoted by $f_1^{\dag} f_2^{\dag}$ (and their reflection counterparts $f_3^{\dag} f_4^{\dag}$) at the 1-cell $\{a\}$. Then, we move $f_1^{\dag}$ and $f_3^{\dag}$ to the reflection center and move $f_2^{\dag}$ and $f_4^{\dag}$ to infinity while preserving reflection symmetry. This adiabatic pump results in $f_1^{\dag} f_3^{\dag}$ being located at the reflection center (0-cell ${A}$), so we can assign $(N_C, R_C)$ to $f_1^{\dag} f_3^{\dag}$, which is given by $(2,1)$ because $R f_1^{\dag} f_3^{\dag} R^{-1} = f_3^{\dag} f_1^{\dag} = -f_1^{\dag} f_3^{\dag}$. Given that this procedure can be repeated many times and that we can exchange the roles of $f_1^{\dag}$ ($f_3^{\dag}$) and $f_2^{\dag}$ ($f_4^{\dag}$), we have $\text{Im}(d^1_{1,0}) = \mathbb{Z}(2,1)$, leading to $E^{1}_{0,0} / \text{Im}(d^1_{1,0}) \cong (\mathbb{Z} \times \mathbb{Z}_2) / \mathbb{Z}(2,1)$. More specifically, the equivalence relation of the quotient $E^{1}_{0,0} / \text{Im}(d^1_{1,0})$ can be written as: 
\bea\label{eqivalence relation}
(N_C,R_C)+n(2,1)\sim (N_C,R_C), \quad n \in \mathbb{Z},
\eea
and the corresponding equivalence classes can be defined as $[(0,0)], [(1,0)], [(2,0)],$ and $[(3,0)]$, yielding the $\mathbb{Z}_4$ classification for $1d$ systems with $U(1) \times \mathbb{Z}_2^{R}$ symmetry.

The above discussion implies that if we can define $\psi_C$ in lattice models, the SPT phases here can be physically realized by assigning the quantum numbers $(N_C, R_C)$ to the lattice models and considering the relation \eqref{eqivalence relation}. In the rest of this section, we will demonstrate that this idea is applicable to decomposable systems, and we can establish the bulk-center correspondence as
\bea\label{classes and ZR}
(N_C,R_C)\sim (2\,\text{Arg}[Z^R(H_d)]/\pi,0), \quad \text{for} \ \alpha=-\pi/2,
\eea
where $H_d$ is a decomposable system, and $Z^R(H_d)$ is the reflection-respecting topological invariant of $H_d$. (The above relation may change if $\alpha \neq -\pi/2$.)

\subsection{Phase structures and many-body spectra of deformed systems}\label{Phase structures and many-body spectra of deformed systems}
\begin{figure}[htb!]
\centering
\includegraphics[width=0.4\textwidth]{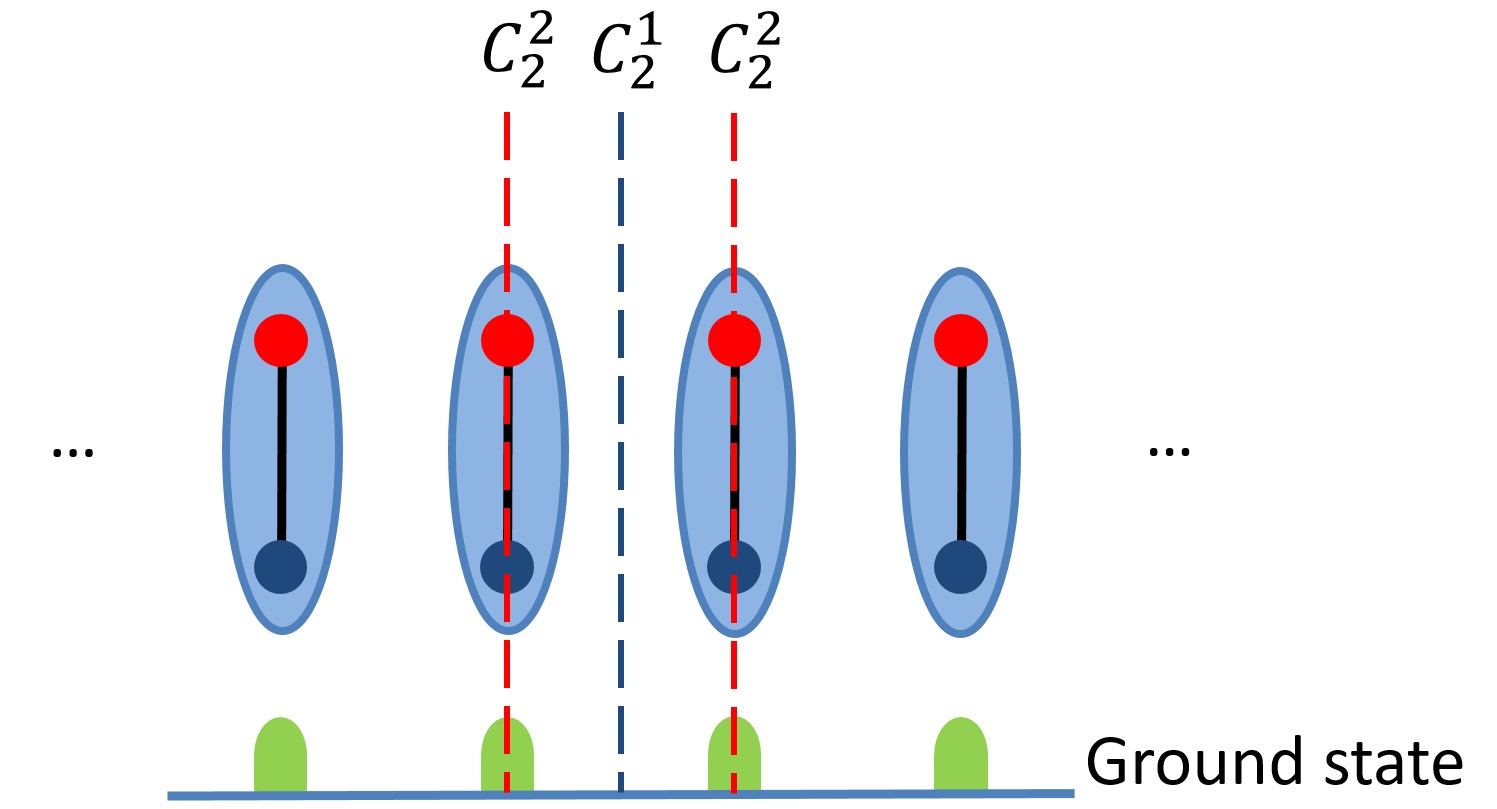}
\caption{For $\pm H_0$, the green wave packets represent the charges $\psi_j=(c^{\dag}_{2j-1}\mp c^{\dag}_{2j})/\sqrt{2}$. There are two types of reflection symmetry that $\pm H_0$ (or generally $\pm H_{\alpha}$) can respect, where the reflection centers are $C_2^1$ and $C_2^2$. We highlight them with blue and red dashed lines, respectively.}
\label{reflection_H0}
\end{figure}
\begin{table}[htb!]
\centering
\renewcommand{\arraystretch}{1.5}
\begin{tabular}{c|cccccc}
          & $H_{2}$ & $H_{1}$ & $H_{0}$ & $H_{-1}$ & $H_{-2}$ \\ \hline 
$(N_{C_2^1},R_{C_2^1})$ & $(0,0)$ & $(1,1)$ & $(0,0)$  & $(1,1)$ & $(0,0)$ \\ \hline
$Z^R_{C_2^1}$ & $1/4$ & $-i/2$ & $1$ & $-i/2$ & $1/4$  \\ \hline
$(N_{C_2^2},R_{C_2^2})$ & $(1,1)$ & $(0,0)$ & $(1,1)$  & $(0,0)$ & $(1,1)$ \\ \hline
$Z^R_{C_2^2}$ & $-i/4$ & $1/2$ & $-i$ & $1/2$ & $-i/4$  \\
\end{tabular}
\caption{The quantum number $(N_C,R_C)$ and $Z^R_C$ of $H_{\alpha}$ with $\alpha=-2\sim 2$ at half-filling concerning two types of reflection symmetry. $Z^R_C$ denotes the reflection-respecting topological invariant related to the reflection symmetry with reflection center $C$. To evaluate $Z^R_{C_2^1}$ and $Z^R_{C_2^2}$, we consider the systems with PBCs and the total number of sites is $N_{\text{tot}}=12$. For $C_2^1$, we make the reflection center located between the fermions $c_2$ and $c_3$ and pick the interval $I=\{c_1,...c_4\}$.  For $C_2^2$, we choose the interval $I=\{c_1,...c_6\}$, where the reflection center is situated between the fermions $c_3$ and $c_4$.}
\label{Table for reflection Ha}
\end{table}
\begin{table}[htb!]
\centering
\renewcommand{\arraystretch}{1.5}
\begin{tabular}{c|cccccc}
          & $-H_{2}$ & $-H_{1}$ & $-H_{0}$ & $-H_{-1}$ & $- H_{-2}$ \\ \hline 
$(N_{C_2^1},R_{C_2^1})$ & $(0,0)$ & $(1,0)$ & $(0,0)$  & $(1,0)$ & $(0,0)$ \\ \hline
$Z^R_{C_2^1}$ & $1/4$ & $i/2$ & $1$ & $i/2$ & $1/4$  \\ \hline
$(N_{C_2^2},R_{C_2^2})$ & $(1,0)$ & $(0,0)$ & $(1,0)$  & $(0,0)$ & $(1,0)$ \\ \hline
$Z^R_{C_2^2}$ & $i/4$ & $1/2$ & $i$ & $1/2$ & $i/4$  \\
\end{tabular}
\caption{The quantum number $(N_C,R_C)$ and $Z^R_C$ of $-H_{\alpha}$ with $\alpha=-2\sim 2$ at half-filling concerning two types of reflection symmetry. The way to calculate $Z^R_{C_2^1}$ and $Z^R_{C_2^1}$ is the same as that when discussing $H_{\alpha}$ in Table~\ref{Table for reflection Ha}.}
\label{Table for reflection mHa}
\end{table}
We now show how to assign $(N_C,R_C)$ to decomposable systems. As a simple example, let's start with $\pm H_0$. The unique ground state of $\pm H_0$ with PBC can be written as
\bea
\ket{GS(\pm H_0)}=\prod_{j=1} \left[\frac{1}{\sqrt{2}}(c^{\dag}_{2j-1}\mp c^{\dag}_{2j})\right]\ket{0}=\prod_{j=1} \psi_j \ket{0}.
\eea
Here we regard the configuration $(c^{\dag}_{2j-1}\mp c^{\dag}_{2j})/\sqrt{2}$ as the charge $\psi_j$, where $\psi_j$ locates between $c^{\dag}_{2j-1}$ and $c^{\dag}_{2j}$. With this idea, the lattice model $\pm H_0$ can be pictorially represented, as shown in Fig.~\ref{reflection_H0}. Depending on the number of cells, $\pm H_0$ can respect two types of reflection symmetry, where the reflection center is situated between two adjacent unit cells or at the midpoint of a unit cell. We label them as $C_2^1$ and $C_2^2$, respectively. If we consider $\pm H_0$ with PBC, implying the existence of translation symmetry, they will respect the reflection symmetry with $C_2^1$ and $C_2^2$ simultaneously. Obviously, there is no charge located at $C_2^1$, so the charge located at the center $C_2^1$ is $\psi_{C_2^1}(\pm H_0)=0$. For the reflection center $C_2^2$, we can define $\psi_{C_2^2} (\pm H_0)= (c^{\dag}_{2L-1}\mp c^{\dag}_{2L})/\sqrt{2}$ where the reflection operator $R_{C_2^2}$ acts as $R_{C_2^2}c^{\dag}_{2L-1-x}R_{C_2^2}^{-1}=c^{\dag}_{2L+x}$.

Now, we can assign the quantum number $(N_C,R_C)$ to $\pm H_0$. Here, $N_C$ can be determined by
\bea
\bra{\psi_C}\hat{N}\ket{\psi_C}, \;\; \text{with}\; \ket{\psi_C}=\psi^\dag_C\ket{0},
\eea
and $R_C$ is defined as eq.~\eqref{R2 def}. By these definitions, for the reflection symmetry with $C_2^1$, the quantum number of $\pm H_0$ is given by $(0,0)$. For the reflection symmetry with $C_2^2$, the quantum numbers of $H_0$ and $-H_0$ are $(1,1)$ and $(1,0)$, respectively. One can evaluate the reflection-respecting topological invariant $Z^R$ of them, as shown in Tables~\ref{Table for reflection Ha} and~\ref{Table for reflection mHa}, and will see the bulk-center correspondence~\eqref{classes and ZR} established. We also provide the values of $(N_C,R_C)$ and $Z^R$ for $\pm H_{\alpha}$ with $\alpha=-2, -1, 0, 1, 2$ in these two tables. As the above method, one can
define the charges of $\pm H_{\alpha}$ by considering their ground states as the product of charges, and then determine the quantum number $(N_C,R_C)$ by studying these charges. Additionally, depending on the total number of cells, all $H_{\alpha}$ can respect the reflection symmetry with $C_2^1$ or $C_2^2$.

The bulk-center correspondence~\eqref{classes and ZR} clearly illustrates the mechanism of these SPT phases. If one directly evaluates the $Z^R$ of $\pm H_0$ concerning two different types of reflection symmetry, it would be confusing why $H_0$ and $-H_0$ are topologically identical for the reflection center $C_2^1$ but fall into different phases for $C_2^2$. However, after assigning $(N_C,R_C)$ to them, this question can be physically answered.

One may notice that $H_0$ and $H_2$ belong to the same SPT phase with either $C_2^1$ or $C_2^2$ centered reflection symmetry, according to Tables~\ref{Table for reflection Ha}. This contrasts with the fact that $H_0$ and $H_2$ are in different SPT phases with chiral symmetry, as shown in Table~\ref{Table for chiral H}. As a result, a direct deformation $H_0 \rightarrow H_2$ that respects all these symmetries would exhibit a phase transition, while there exists a chiral-symmetry-breaking deformation that connects $H_0$ and $H_2$ without a phase transition. In Appendix~\ref{Deformation between $H_0$ and $H_2$ in the free fermion scope}, we provide one such deformation, which can be performed even without interactions.

\begin{figure}[H]
\centering
\subfloat[]{\includegraphics[width=0.32\textwidth]{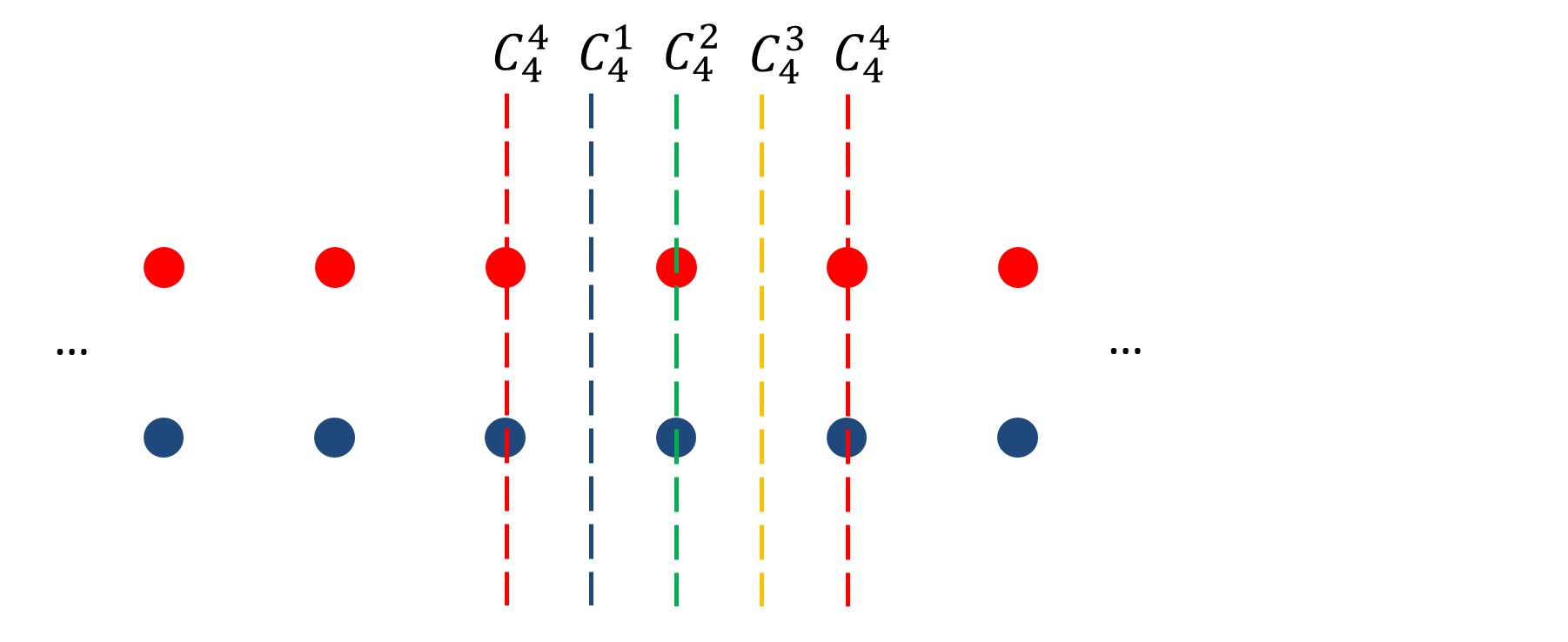}}\hskip 0.1cm
\subfloat[]{\includegraphics[width=0.32\textwidth]{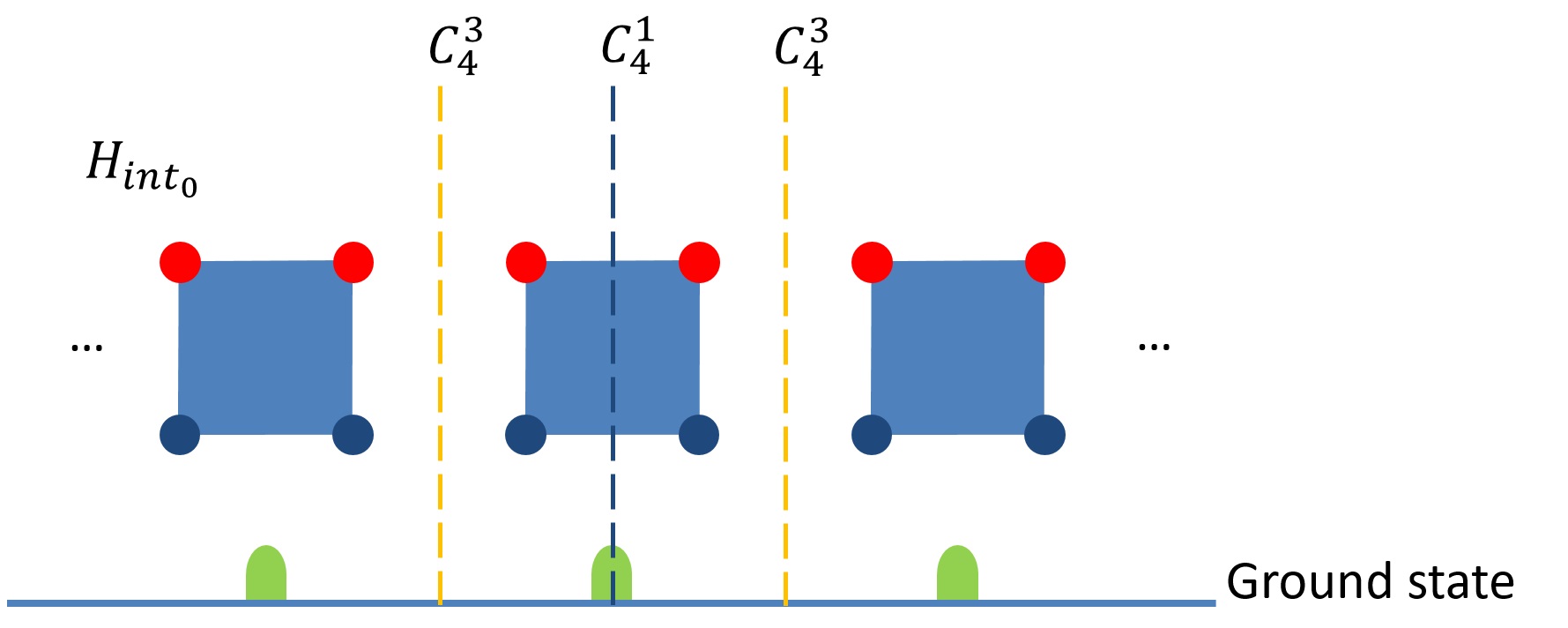}}\hskip 0.1cm
\subfloat[]{\includegraphics[width=0.32\textwidth]{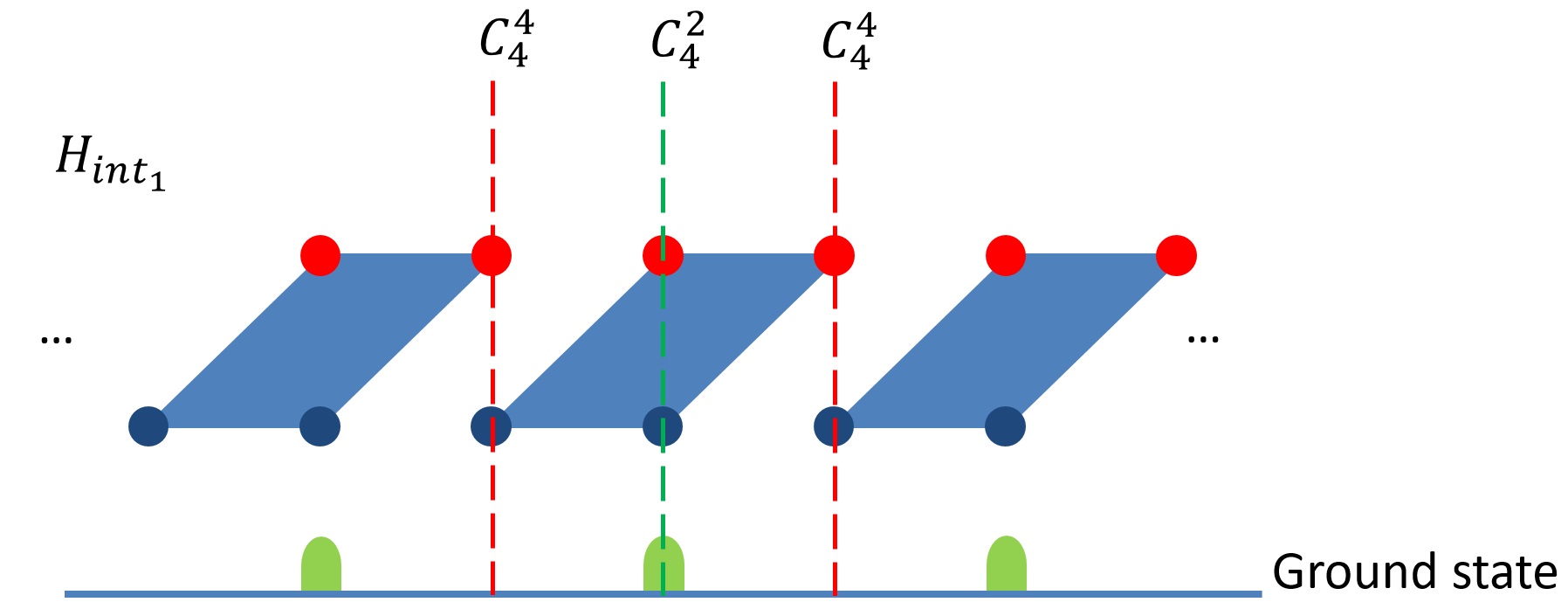}}\\
\caption{For $\pm H_{int_{\alpha}}$ with even $\alpha$, two possible types of reflection centers are labeled by $C_4^1$ and $C_4^3$ and highlighted with dashed blue and yellow lines. For odd $\alpha$, we employ red and green lines to characterize the possible types of reflection centers with the labels $C_4^2$ and $C_4^4$.}
\label{Reflection_Inta}
\end{figure}


We can assign the quantum number $(N_C,R_C)$ to $\pm H_{int_{\alpha}}$ as well. To achieve it, as an analogy of unit cells, here we introduce the concept of a "subsystem" that contains four fermions forming a single quartic interaction. Then, we can define the charge $\psi_j$ as the configurations of each subsystem's unique ground state, where the locations of $\psi_j$ are at the midpoint of each subsystem. For example, each subsystem of $\pm H_{int_0}$ is composed of the fermion $c_{4j-3}, c_{4j-2}, c_{4j-1},$ and $c_{4j}$, where $j$ is the number of subsystems. By considering the explicit form of the ground state of $\pm H_{int_0}$ with PBC
\bea
\ket{GS(\pm H_{int_0})}=\prod_{j=1} \left[\frac{1}{\sqrt{2}}(c^{\dag}_{4j-3}c^{\dag}_{4j-2}\mp c^{\dag}_{4j-1}c^{\dag}_{4j})\right]\ket{0}=\prod_{j=1} \psi_j \ket{0},
\eea
the charges here can be read as $\psi_j =(c^{\dag}_{4j-3}c^{\dag}_{4j-1}\pm c^{\dag}_{4j-2}c^{\dag}_{4j})/\sqrt{2}$. With these $\psi_j$, we can determine the $(N_C,R_C)$ of $\pm H_{int_0}$, which gives a more physical and intuitive topological classification of the systems.
\begin{table}[htb!]
\centering
\renewcommand{\arraystretch}{1.5}
\begin{tabular}{c|cccccc}
          & $H_{int_1}$ & $-H_{int_1}$ & $H_{int_{-1}}$ & $-H_{int_{-1}}$  \\ \hline 
$(N_{C_4^2},R_{C_4^2})$ & $(2,0)$ & $(2,1)$ & $(2,1)$  & $(2,0)$  \\ \hline
$Z^R_{C_4^2}$ & $-1/2$ & $1/2$ & $1/2$ & $-1/2$  \\ \hline
$(N_{C_4^4},R_{C_4^4})$ & $(0,0)$ & $(0,0)$ & $(0,0)$  & $(0,0)$ \\ \hline
$Z^R_{C_4^4}$ & $1/2$ & $1/2$ & $1/2$ & $1/2$   \\
\end{tabular}
\caption{The quantum number $(N_C,R_C)$ and $Z^R_C$ of $\pm H_{int_{\alpha}}$ with $\alpha=-1$ and $1$ at half-filling concerning two types of reflection symmetry. To evaluate $Z^R_{C_4^2}$ and $Z^R_{C_4^4}$, we consider the systems with PBCs and the total number of sites is $N_{\text{tot}}=12$. For $C_4^2$, we make the reflection center located between the fermions $c_3$ and $c_4$ and pick the interval $I=\{c_1,...c_6\}$.  For $C_4^4$, we choose the interval $I=\{c_3,...c_{8}\}$, where the reflection center is situated between the fermions $c_5$ and $c_6$.}
\label{Table for reflection oddInta}
\end{table}

\begin{table}[htb!]
\centering
\renewcommand{\arraystretch}{1.5}
\begin{tabular}{c|cccccc}
          & $H_{int_2}$ & $-H_{int_2}$ & $H_{int_0}$ & $-H_{int_{0}}$ & $H_{int_{-2}}$ & $-H_{int_{2}}$ \\ \hline 
$(N_{C_4^1},R_{C_4^1})$ & $(0,0)$ & $(0,0)$ & $(2,0)$  & $(2,1)$ & $(0,0)$ & $(0,0)$ \\ \hline
$Z^R_{C_4^1}$ & $1/2$ & $1/2$ & $-1$ & $1$ & $1/2$ & $1/2$  \\ \hline
$(N_{C_4^3},R_{C_4^3})$ & $(2,1)$ & $(2,0)$ & $(0,0)$  & $(0,0)$ & $(2,1)$ & $(2,0)$ \\ \hline
$Z^R_{C_4^3}$ & $1/2$ & $-1/2$ & $1$ & $1$ & $1/2$ & $-1/2$  \\
\end{tabular}
\caption{The quantum number $(N_C,R_C)$ and $Z^R_C$ of $\pm H_{int_{\alpha}}$ with $\alpha=-2, 0,$ and $2$ at half-filling concerning two types of reflection symmetry. To evaluate $Z^R_{C_4^1}$ and $Z^R_{C_4^3}$, we consider the systems with PBCs and the total number of sites is $N_{\text{tot}}=12$. For $C_4^1$, we make the reflection center located between the fermions $c_2$ and $c_3$ and pick the interval $I=\{c_1,...c_4\}$.  For $C_4^3$, we choose the interval $I=\{c_1,...c_8\}$, where the reflection center is situated between the fermions $c_4$ and $c_5$.}
\label{Table for reflection evenInta}
\end{table}

Like the lattice models $\pm H_{\alpha}$, all $\pm H_{int_{\alpha}}$ respect two types of reflection symmetry, where the reflection centers are located between two adjacent subsystems or at the midpoint of a subsystem. But the reflection centers of $\pm H_{int_{\alpha}}$ with $\alpha\in \text{even}$ and $\alpha\in \text{odd}$ are different, as shown in Fig.~\ref{Reflection_Inta}. For $H_{int_{\alpha}}$ with even $\alpha$, we denote the possible types of reflection centers as $C_4^1$ and $C_4^3$, and for $H_{int_{\alpha}}$ with odd $\alpha$, the possible types of reflection centers are labeled by $C_4^2$ and $C_4^4$. In Tables~\ref{Table for reflection evenInta} and~\ref{Table for reflection oddInta}, we present the $(N_C,R_C)$ and $Z^R$ of $\pm H_{int_{\alpha}}$ with $\alpha=-2, -1, 0, 1, 2$ concerning different reflection symmetries, indicating that the bulk-center correspondence~\eqref{classes and ZR} holds.
Note that for $\pm H_{\alpha}$, the reflection symmetry with the center $C_4^1$ or $C_4^3$ ($C_4^2$ or $C_4^4$) is the same as that with $C_2^1$ ($C_2^2$). 

\begin{figure}[htb!]
\centering
\subfloat[]{\includegraphics[width=0.31\textwidth]{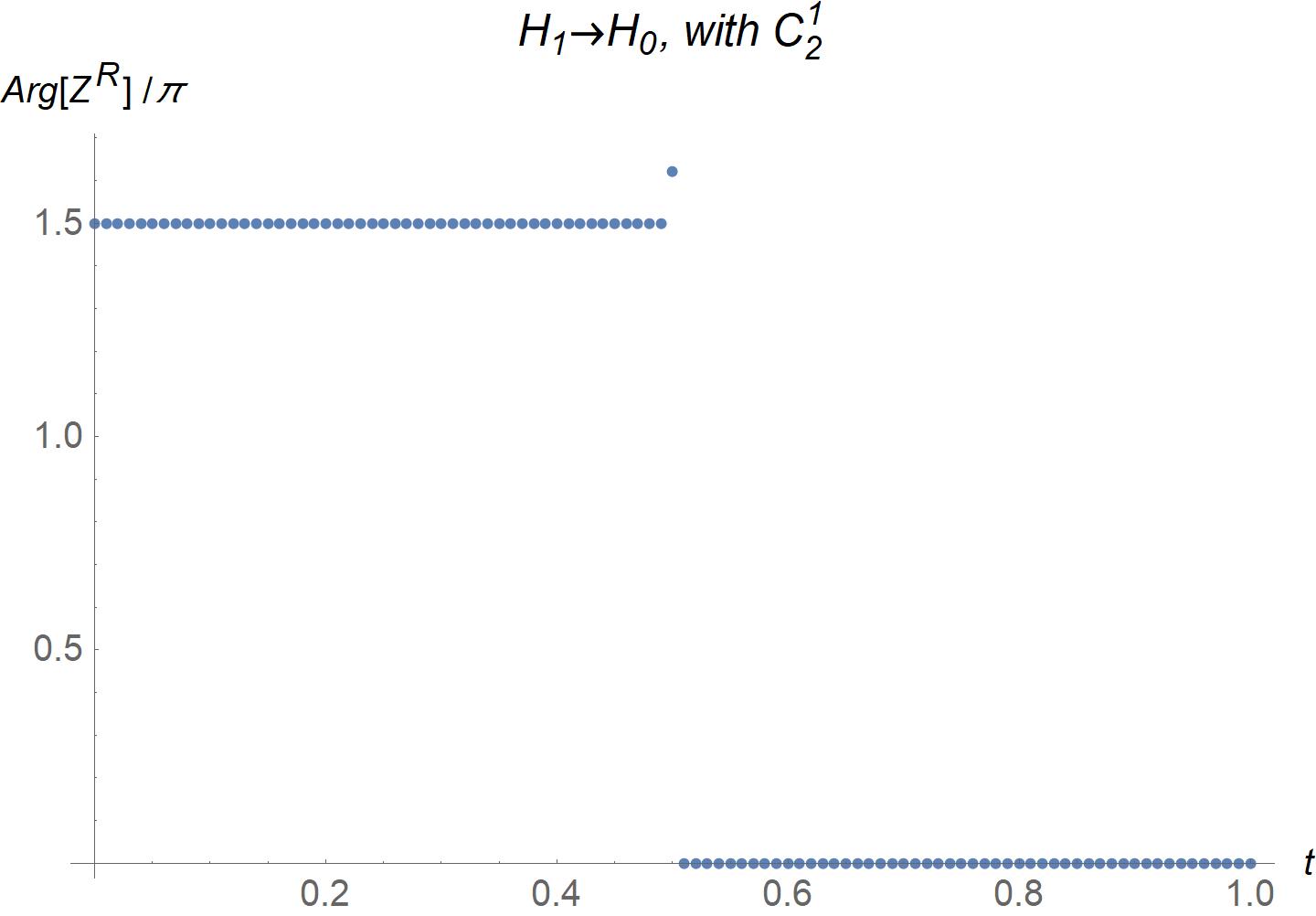}}\hskip 0.5cm
\subfloat[]{\includegraphics[width=0.31\textwidth]{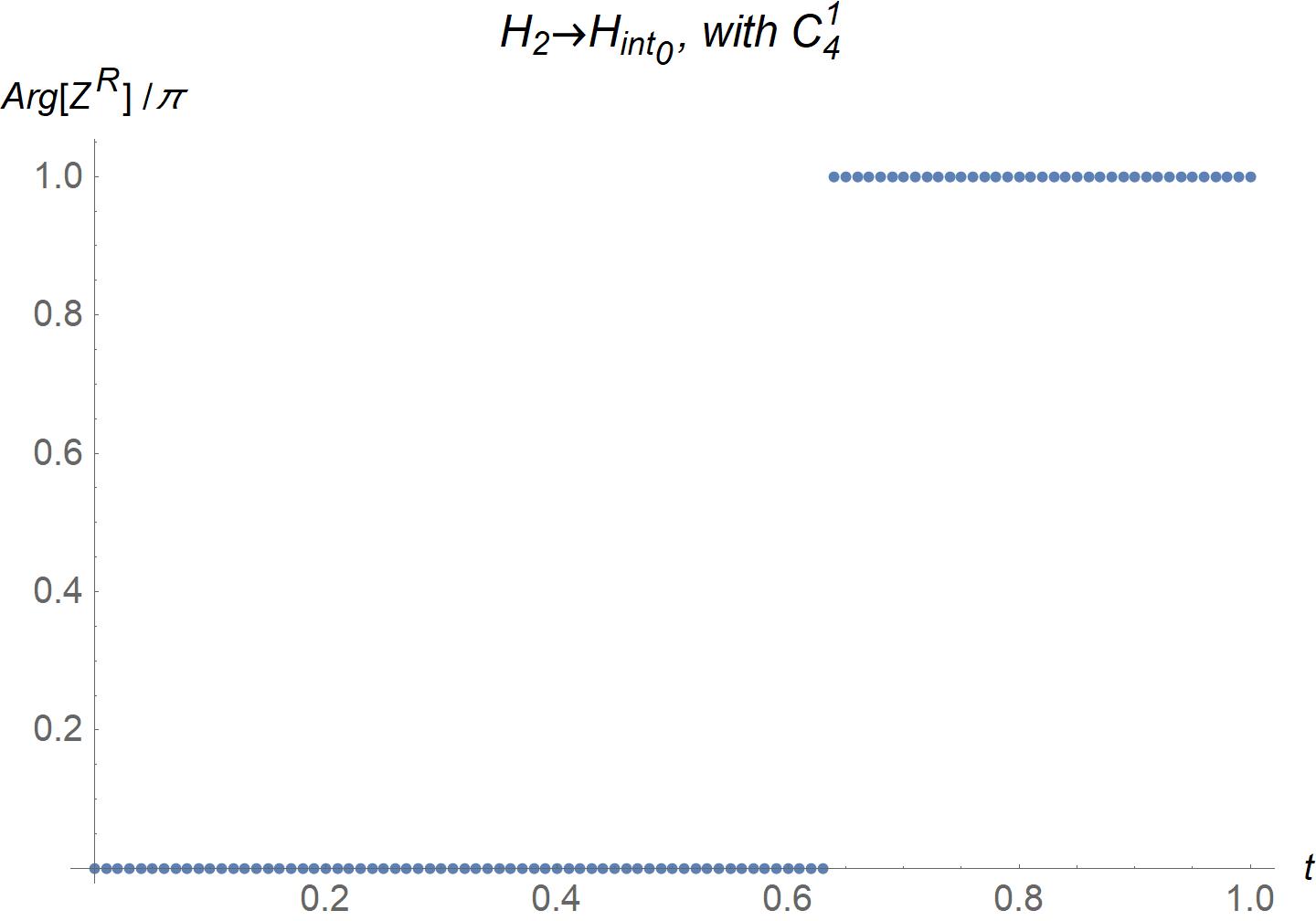}}\hskip 0.5cm
\subfloat[]{\includegraphics[width=0.31\textwidth]{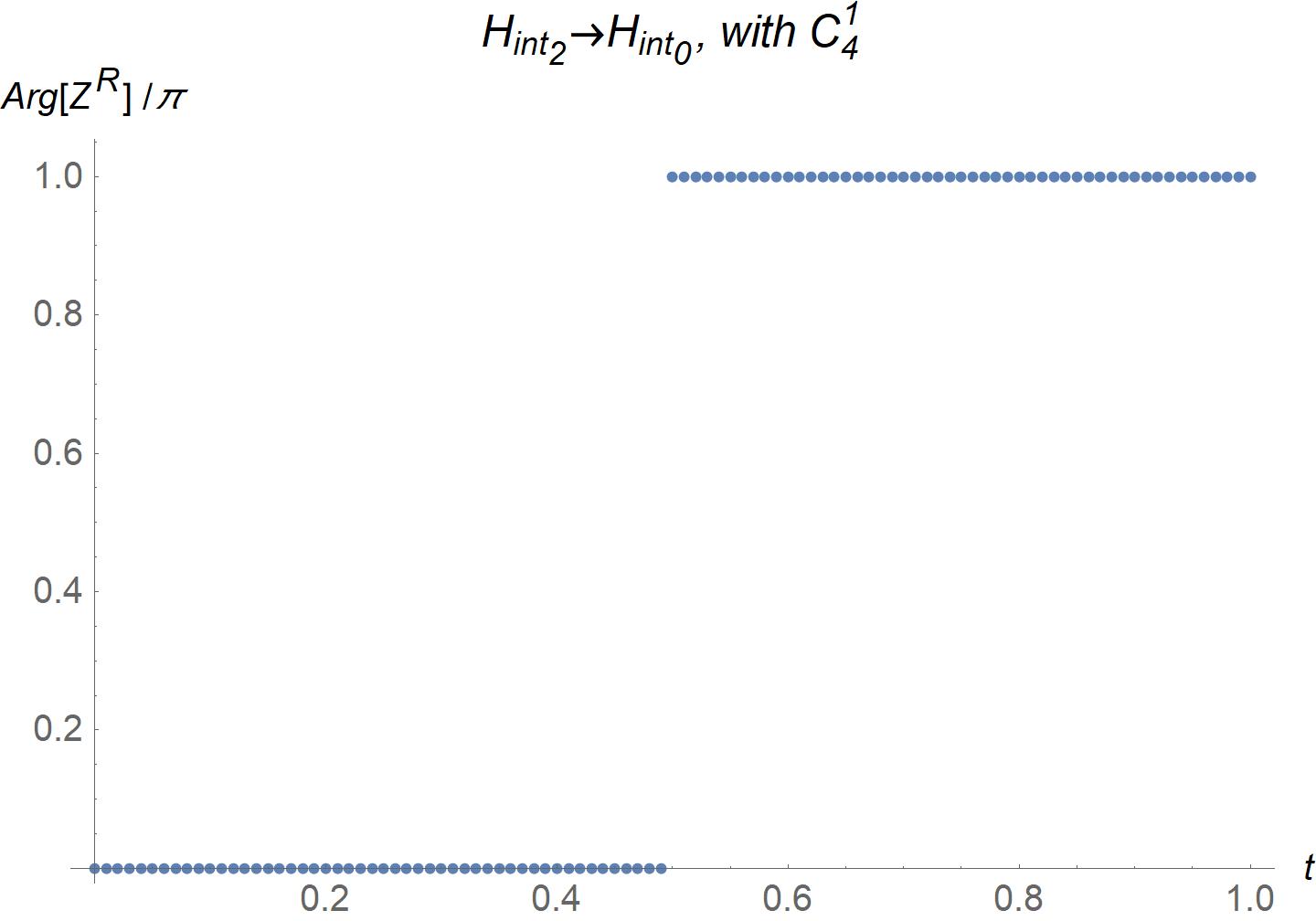}}\\
\subfloat[]{\includegraphics[width=0.31\textwidth]{EN_w1tow0_12.jpg}}\hskip 0.5cm
\subfloat[]{\includegraphics[width=0.31\textwidth]{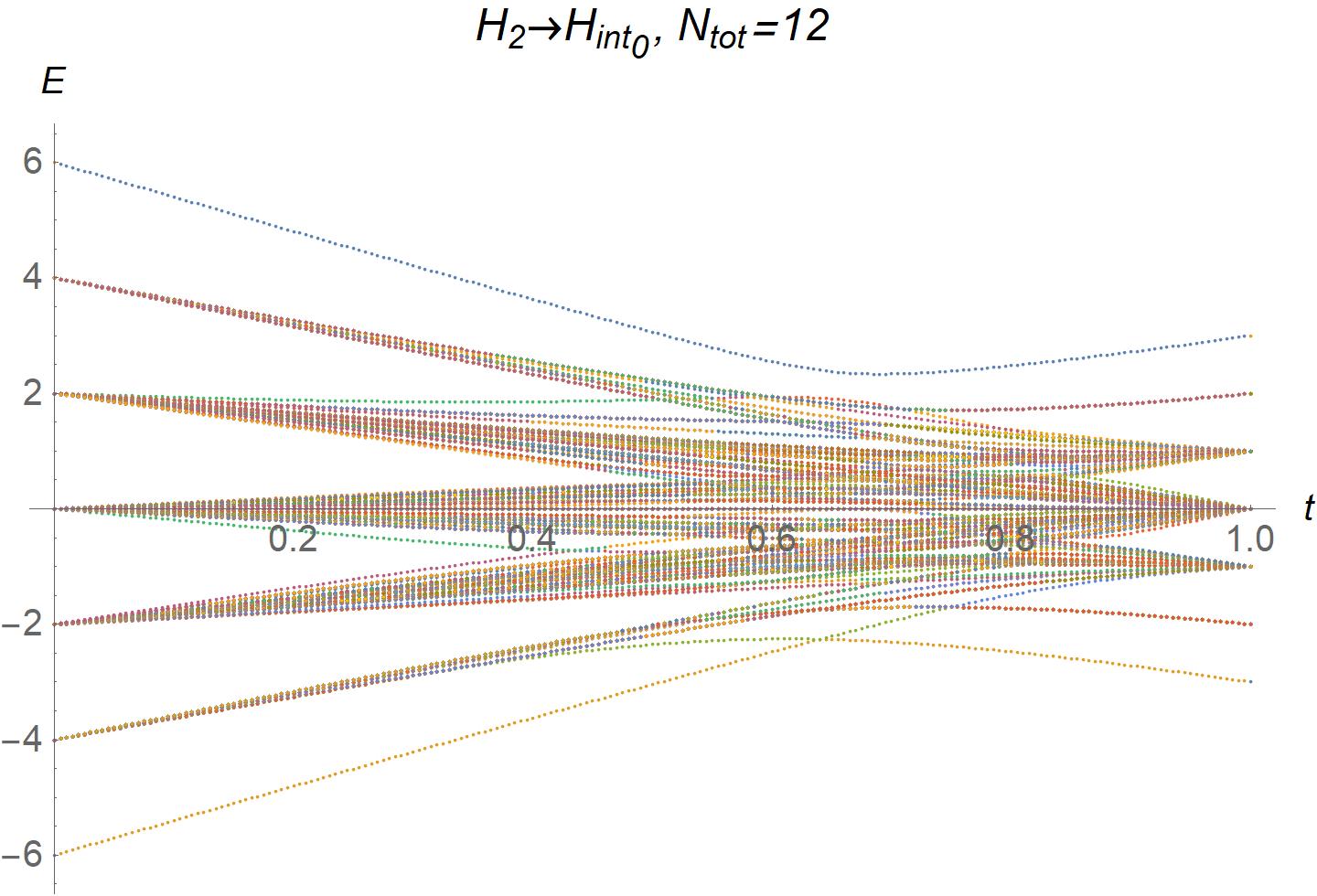}}\hskip 0.5cm
\subfloat[]{\includegraphics[width=0.31\textwidth]{EN_Int2toInt0_12.jpg}}\\
\caption{We consider the systems with $N_{\text{tot}}=12$. (a) $\text{Arg}[Z^R]$ of $H_{1}\rightarrow H_{0}$ with the reflection center $C_2^1$. The interval $I=\{c_2,...,c_7\}$ with the reflection center located between $c_4$ and $c_5$ is assigned here. (b)(c) $\text{Arg}[Z^R]$ of $H_{2}\rightarrow H_{int_0}$ and $H_{int_2}\rightarrow H_{int_0}$ with the reflection center $C_4^1$. We make the reflection center situated between $c_2$ and $c_3$ and choose the interval $I=\{c_1,...,c_4\}$ (d)(e)(f) spectra of the deformations $H_{1}\rightarrow H_{0}$, $H_{2}\rightarrow H_{int_0}$, and $H_{int_{2}}\rightarrow H_{int_{0}}$ with PBCs..}
\label{ZR_deformation}
\end{figure}

\begin{figure}[htb!]
\centering
\subfloat[]{\includegraphics[width=0.35\textwidth]{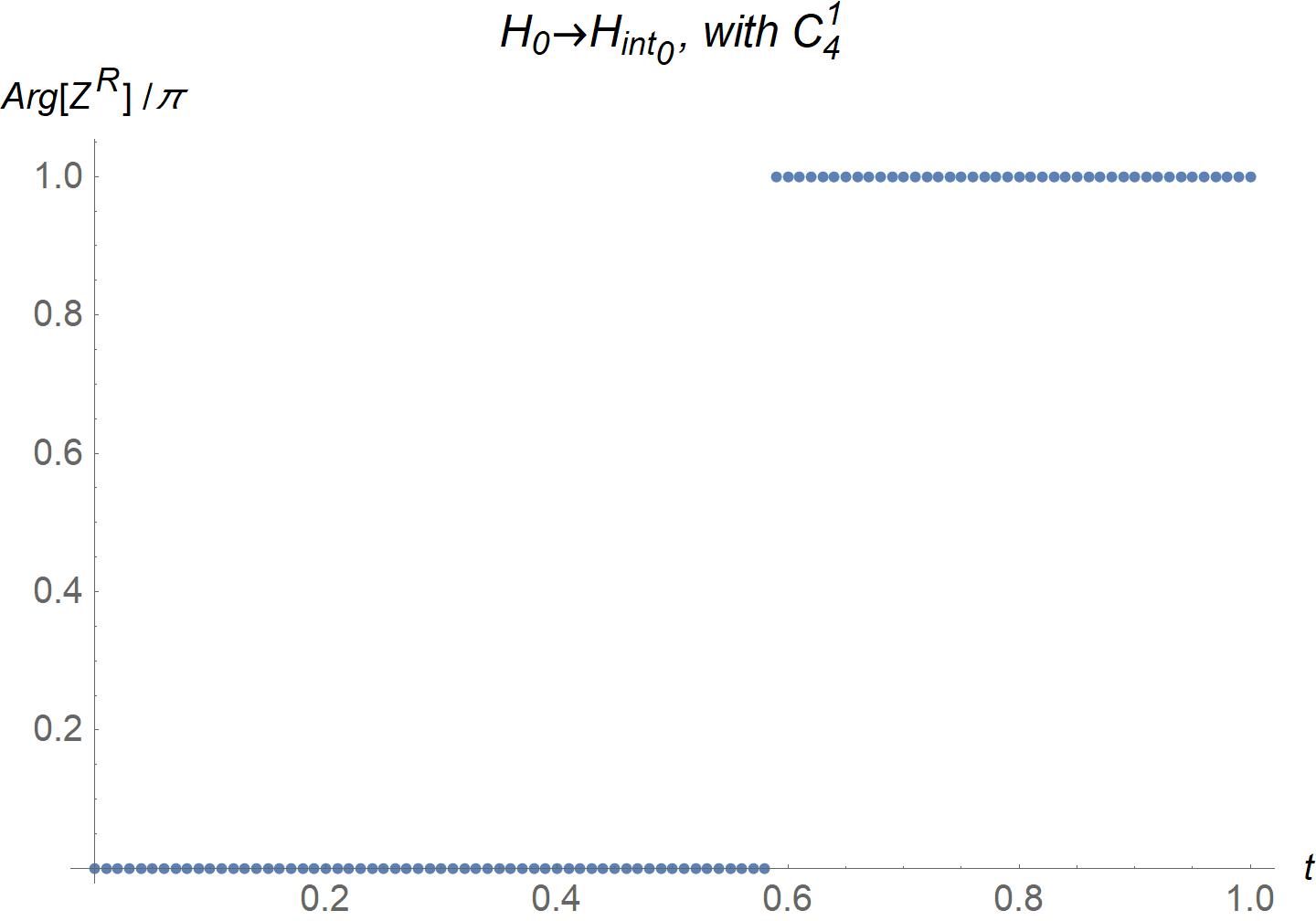}}\hskip 0.5cm
\subfloat[]{\includegraphics[width=0.35\textwidth]{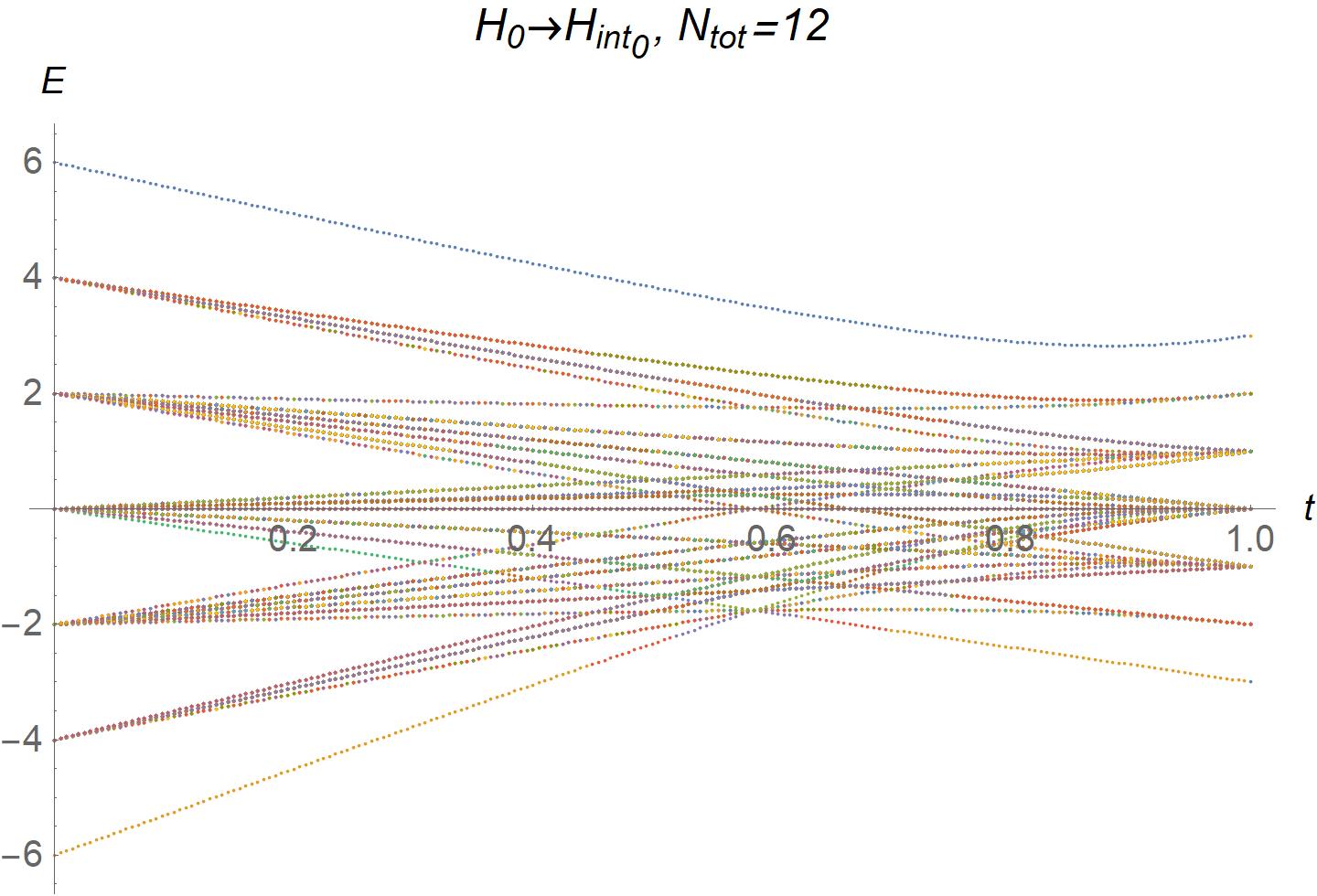}}\\
\caption{We consider the system with $N_{\text{tot}}=12$. (a) $\text{Arg}[Z^R]$ of $H_{0}\rightarrow H_{int_0}$ with the reflection center $C_4^1$. We pick up the interval $I=\{c_1,...,c_4\}$ with the reflection center situated between $c_2$ and $c_3$. (b) The energy spectrum of $H_{0}\rightarrow H_{int_0}$ with PBC.}
\label{ZR and EN for H0_to_Int0}
\end{figure}
\begin{figure}[htb!]
\centering
\subfloat[]{\includegraphics[width=0.4\textwidth]{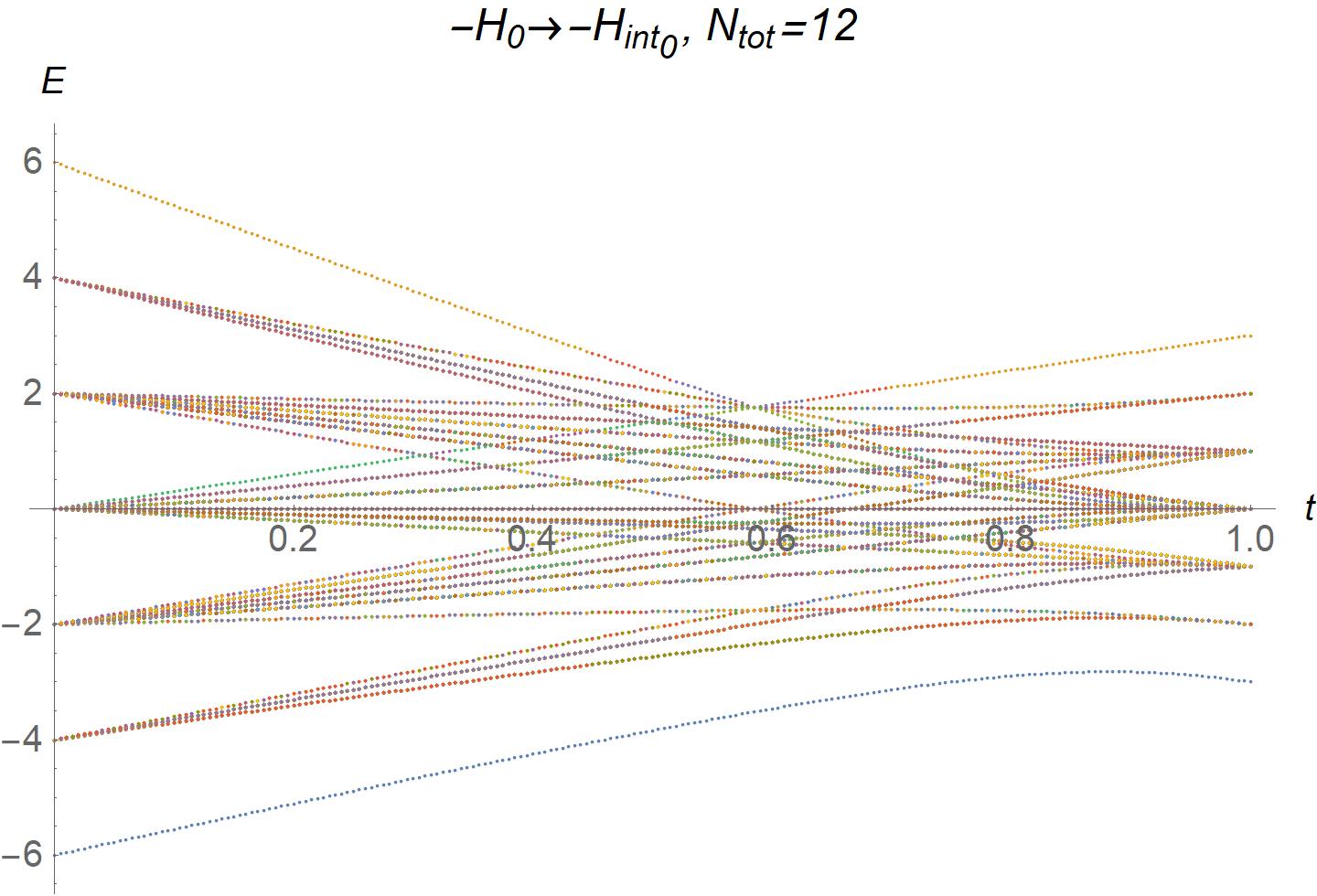}}\\
\caption{The energy spectrum of $-H_{0}\rightarrow -H_{int_0}$ with PBC.}
\label{EN for negH0_to_Int0}
\end{figure}

As $\pm H_{\alpha}$ and $\pm H_{\text{int}_{\alpha}}$ respect multiple symmetries simultaneously, including chiral symmetry and various reflection symmetries discussed here, they are characterized by multiple SPT phases. If the topological invariant for at least one of these SPT phases changes during a deformation, a phase transition will occur.
We numerically evaluate $Z^R$ for different types of reflection centers during the deformations $H_{1} \rightarrow H_{0}$, $H_{2} \rightarrow H_{\text{int}0}$, and $H_{\text{int}2} \rightarrow H_{\text{int}_0}$, as shown in Fig.~\ref{ZR_deformation}. In all cases, the phase transition points are identical to the gap-closing points.
A more nontrivial example is the deformation $H_0 \rightarrow H_{\text{int}_0}$, where the interaction couples the two nearest-neighbor subsystems (unit cells) in $H_0$. One might expect no phase transition to occur when this interaction is turned on. However, during this deformation, although the topological invariants associated with chiral symmetry and $C_4^3$ reflection symmetry remain unchanged, the one associated with $C_4^1$ changes, indicating a phase transition. This is confirmed in Fig.~\ref{ZR and EN for H0_to_Int0}, where the transition point coincides with the gap-closing point.
On the other hand, for the deformation $-H_0 \rightarrow -H_{\text{int}_0}$, the topological invariants for both chiral symmetry and $C_4^{1/3}$ ($C_2^1$) reflection symmetry remain unchanged, consistent with the absence of gap closing in the spectrum for this deformation, as shown in Fig.~\ref{EN for negH0_to_Int0}.

\section{Role of translation symmetry}\label{Role of translation symmetry}
In this section, we mainly discuss, when the translation symmetry gets involved, how the SPT phases of the $1d$ charge-conserved systems with chiral symmetry and with reflection symmetry will change. First, roughly speaking, because the translation symmetry is spatial symmetry and the chiral symmetry is on-site symmetry, there is no interplay between them. Therefore, there is no new classification of the SPT phases regarding the systems with chiral symmetry and translation symmetry. 
On the other hand, since both translation symmetry and reflection symmetry are spatial symmetry, new SPT phases related to the combination of these two symmetries may emerge. In the latter discussion, we will show that these new SPT phases are classified by $\mathbb{Z}\times \mathbb{Z}_2\times \mathbb{Z}_4$ .
\begin{figure}[htb!]
\centering
\includegraphics[width=0.45\textwidth]{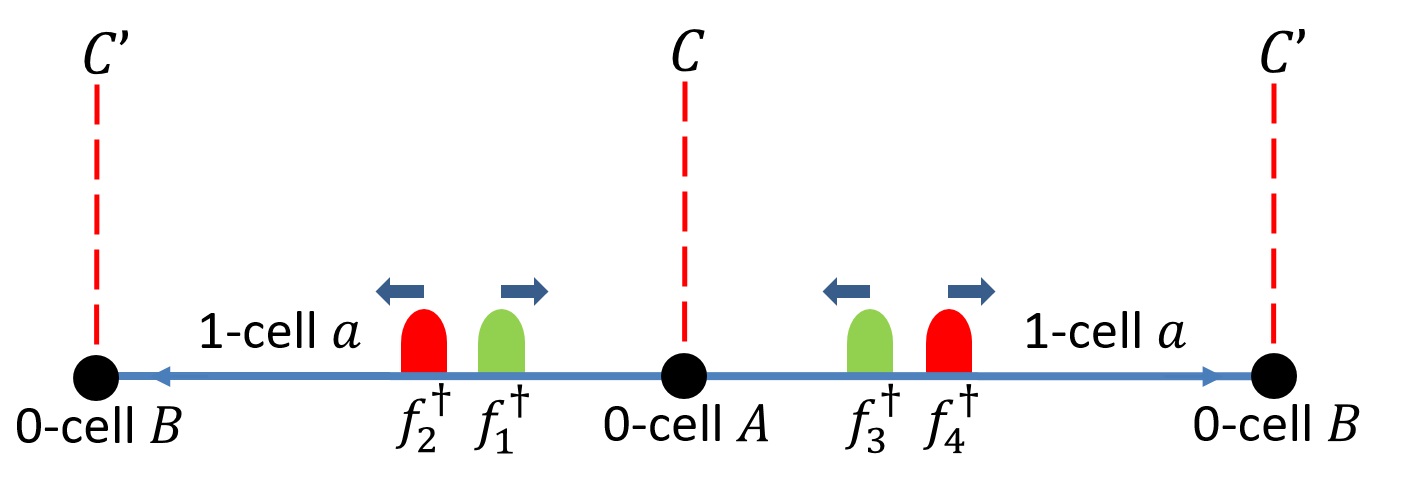}
\caption{The cell decomposition and the schematic of the first differential $d^{1}_{1,0}$ for $1d$ fermionic systems with $G=U(1)\times \mathbb{Z}_2^{R}$ and translation symmetry. This adiabatic pump is almost the same as that of the systems without translation symmetry. The only difference is $f^{\dag}_3$ and $f^{\dag}_4$ will meet on $c'$ instead of going to infinity, so $\text{Im}(d^1_{1,0})=\mathbb{Z}(2,1,-2,1)$ here.}
\label{translation d1 picture}
\end{figure}
In addition to $G=U(1)\times \mathbb{Z}_2^{R}$ symmetry, now we involve the translation symmetry $\mathbb{Z} \ni n:j\rightarrow j+n$. The corresponding SPT phases can be classified by the generalized homology $h_{0}^{\mathbb{Z}\rtimes \mathbb{Z}_2}(\mathbb{R},\partial \mathbb{R})$, noting that the total symmetry $G$ is $\mathbb{Z}\rtimes \mathbb{Z}_2$ because the reflection here $r\in\mathbb{Z}_2$ acts on $\mathbb{Z}$ as reflection. As studying the cases with only $G=U(1)\times \mathbb{Z}_2^{R}$ symmetry, one can use the AHSS to compute this generalized homology as well. After doing cell decomposition shown in Fig~\ref{translation d1 picture}, we have two $0$-cells $\{A,B\}$, which correspond to the reflection centers $\{C,C'\}$ and respect $G=U(1)\times \mathbb{Z}_2^{R}$ symmetry, and a $1$-cell $\{a\}$ with $U(1)$ symmetry, leading to the following $E^1$-page 

\bea\label{E-1 page for translation}
\centering
\renewcommand{\arraystretch}{1.5}
\begin{tabular}{c|cc}
$q=0$        &  $(\mathbb{Z}\times \mathbb{Z}_2)\times(\mathbb{Z}\times \mathbb{Z}_2)$ & $\mathbb{Z}$ \\
$q=1$        & $0$                              & $0$ \\ \hline
$E^1_{p,-q}$ & $p=0$                          & $p=1$   \\
\end{tabular}
\eea
Given that the limiting page here is $E^2$-page, the generalized homology $h_{0}^{\mathbb{Z}\rtimes \mathbb{Z}_2}(\mathbb{R},\partial \mathbb{R})$ fits into the short exact sequence
\bea\label{short exact sequences for translation}
0\rightarrow E^{2}_{0,0}\rightarrow h_{0}^{\mathbb{Z}\rtimes \mathbb{Z}_2}(\mathbb{R},\partial \mathbb{R})\rightarrow E^{2}_{1,-1}\rightarrow 0.
\eea
By using eq.~\eqref{r-differential and Er+1-page}, these $E^2$-pages are given by
\bea\label{E2-page for translation}
\begin{aligned}
&E^{2}_{0,0}=E^{1}_{0,0}/\text{Im}(d^1_{1,0}),\\
&E^{2}_{1,-1}=0,\\
\end{aligned}
\eea
which leads to $h_{0}^{\mathbb{Z}\rtimes \mathbb{Z}_2}(\mathbb{R},\partial \mathbb{R})\cong E^{1}_{0,0}/\text{Im}(d^1_{1,0})$. As discussed in Sec.~\ref{classification and (N_C,R_C)}, we can assign the quantum number $(N_C,R_C,N_{C'},R_{C'})$ to $E^{1}_{0,0}$ here, and $\text{Im}(d^1_{1,0})$ represents the adiabatic pump shown in Fig~\ref{translation d1 picture}, so we have
\bea\label{G homo for translation}
\begin{aligned}
h_{0}^{\mathbb{Z}\rtimes \mathbb{Z}_2}(\mathbb{R},\partial \mathbb{R})&\cong E^{1}_{0,0}/\text{Im}(d^1_{1,0})\cong (\mathbb{Z}\times \mathbb{Z}_2\times\mathbb{Z}\times \mathbb{Z}_2)/\mathbb{Z}(2,1,-2,1)\\
&\cong \mathbb{Z}\times \mathbb{Z}_2\times \mathbb{Z}_4.
\end{aligned}
\eea
Therefore, the classification of these SPT phases is $\mathbb{Z}\times \mathbb{Z}_2\times \mathbb{Z}_4$.

One can assign the quantum number $(N_C,R_C,N_{C'},R_{C'})$ to decomposable systems as well, but it should be careful. Recalling that for the decomposable systems without translation symmetry, we use $(N_C,R_C)$ to describe their SPT phases, and $(N_C,R_C)$ is not unique because we can move the charges in 1-cell to the reflection center while preserving reflection symmetry. Nevertheless, because of eq.~\eqref{eqivalence relation}, it will not change the topology of the system, although $(N_C,R_C)$ will be different. However, for the cases with translation symmetry, we should impose a restriction to $(N_C,R_C,N_{C'},R_{C'})$. Due to the common sense about translation symmetry where $\mathbb{Z}$ is related to the $filling$ $Z_f$ in each subsystem, we have
\bea\label{charges restriction}
Z_f=N_C+N_{C'},
\eea
The definition of $Z_f$ is
\bea\label{filling}
Z_f=\frac{\bra{GS} \hat{N} \ket{GS}}{L},
\eea
where $\hat{N}$ is the particle number operator and $L$ is an integer and is defined as $N_{\text{tot}}/n$ for the translation symmetry $\mathbb{Z} \ni n:j\rightarrow j+n$. If we consider a decomposable system, $L$ is the same as the number of unit cells (or subsystems). Note that $Z_f$ can be evaluated even if systems are not decomposable. The restriction \eqref{charges restriction} implies that in a subsystem, we have to make all the charges located at two reflection centers only. The reason is that we require $(N_C,R_C,N_{C'},R_{C'})$ to represent an SPT phase in $\mathbb{Z}\times \mathbb{Z}_2\times \mathbb{Z}_4$ classification, and if we don't consider the restriction \eqref{charges restriction}, $(N_C,R_C,N_{C'},R_{C'})$ will not give us the correct information about $\mathbb{Z}$. In Sec.~\ref{j to j+2} and~\ref{j to j+4}, we will demonstrate how to assign the quantum number $(N_C,R_C,N_{C'},R_{C'})$ to the decomposable systems with the translation symmetry $j\rightarrow j+2$ and $j\rightarrow j+4$, while showing that the restriction~\eqref{charges restriction} is necessary.

\subsection{Constructing the topological invariants}

Since translation symmetry is not an element of a point group, the cobordism group $\Omega^{\text{pin}^c}_2(pt)$ itself does not include translation symmetry, making the classification cannot be given by $\Omega^{\text{pin}^c}_2(pt)$ and the corresponding many-body topological invariant cannot be constructed by the method mentioned in Sec.~\ref{Many-body topological invariants}. However, with the help of the AHSS, we indicate that the topological invariants can be constructed in an alternative way, which is given by
\bea\label{translation topo inv}
(Z_f,2\,\text{Arg}[Z^R_{C}]/\pi,2\,\text{Arg}[Z^R_{C'}]/\pi).
\eea
Here we express the translation as the composition of two reflections with reflection centers $C$ and $C'$. Thus, $Z^R_{C}$ and $Z^R_{C'}$ denote the topological invariants~\eqref{reflection-respecting topological invariant} with respect to $C$ and $C'$.

The idea of employing eq.~\eqref{translation topo inv} to describe $\mathbb{Z}\times \mathbb{Z}_2\times \mathbb{Z}_4$ classification is based on two reasons. First, since the whole system respects $U(1)$ and translation symmetry, it's straightforward to think $\mathbb{Z}$ is related to the number of charges in each subsystem, which corresponds to $Z_f$. Secondly, recall that for the cases without translation symmetry, $Z^R$ responds to the quantum number $(N_C,R_C)$. We can extend this concept to systems with translation symmetry. Because the SPT phases here can be determined by the (0+1)$d$ SPT phases on two different reflection centers separately, which is described by $(N_C,R_C,N_{C'},R_{C'})$, we can naively conceive that $Z^R_{C}$ and $Z^R_{C'}$ may serve as a part of the topological invariants here. Taking these into account, we speculate $(Z_f,2\,\text{Arg}[Z^R_{C}]/\pi,2\,\text{Arg}[Z^R_{C'}]/\pi)$ can be assigned to this $\mathbb{Z}\times \mathbb{Z}_2\times \mathbb{Z}_4$ classification. Our strategy to prove this surmise is right is as follows. We first find the generators of $\mathbb{Z}$, $\mathbb{Z}_2$, and $\mathbb{Z}_4$ in terms of quantum numbers $(N_C,R_C,N_{C'},R_{C'})$, which are denoted as $g_{\mathbb{Z}}$, $g_{\mathbb{Z}_2}$, and $g_{\mathbb{Z}_4}$. Then we consider the following statements
\bea\label{topo statements}
\begin{aligned}
&(Z_f(H(g_{\mathbb{Z}})),2\,\text{Arg}[Z^R_{C}(H(g_{\mathbb{Z}}))]/\pi,2\,\text{Arg}[Z^R_{C'}(H(g_{\mathbb{Z}}))]/\pi)\in \mathbb{Z},\\
&(Z_f(H(g_{\mathbb{Z}_2})),2\,\text{Arg}[Z^R_{C}(H(g_{\mathbb{Z}_2}))]/\pi,2\,\text{Arg}[Z^R_{C'}(H(g_{\mathbb{Z}_2}))]/\pi)\in \mathbb{Z}_2,\\
&(Z_f(H(g_{\mathbb{Z}_4})),2\,\text{Arg}[Z^R_{C}(H(g_{\mathbb{Z}_4}))]/\pi,2\,\text{Arg}[Z^R_{C'}(H(g_{\mathbb{Z}_4}))]/\pi)\in \mathbb{Z}_4,\\
\end{aligned}
\eea
where $H(g_{x})$ is the corresponding decomposable system of $g_{x}$. If the above statements are satisfied, this classification can be described by the set of topological invariants \eqref{translation topo inv}. Here, we choose the following set of generators
\bea\label{generators}
\begin{aligned}
&g_{\mathbb{Z}}=(1,0,0,0),\\
&g_{\mathbb{Z}_2}=(0,1,0,0),\\
&g_{\mathbb{Z}_4}=(1,0,-1,0).\\
\end{aligned}
\eea
As shown in Appendix \hyperref[prove statement]{D}, for these three generators, we have
\bea
\begin{aligned}
&(Z_f(H(ng_{\mathbb{Z}})),2\,\text{Arg}[Z^R_{C}(H(ng_{\mathbb{Z}}))]/\pi,2\,\text{Arg}[Z^R_{C'}(H(ng_{\mathbb{Z}}))]/\pi)=\left(n,n\,\,\text{mod}\,\,4,0\right)\in \mathbb{Z},\\
&(Z_f(H(ng_{\mathbb{Z}_2})),2\,\text{Arg}[Z^R_{C}(H(ng_{\mathbb{Z}_2}))]/\pi,2\,\text{Arg}[Z^R_{C'}(H(ng_{\mathbb{Z}_2}))]/\pi)=\left(0,2n\,\,\text{mod}\,\,4,0\right)\in \mathbb{Z}_2,\\
&(Z_f(H(ng_{\mathbb{Z}_4})),2\,\text{Arg}[Z^R_{C}(H(ng_{\mathbb{Z}_4}))]/\pi,2\,\text{Arg}[Z^R_{C'}(H(ng_{\mathbb{Z}_4}))]/\pi)=\left(0,n\,\,\text{mod}\,\,4,3n\,\,\text{mod}\,\,4\right)\in \mathbb{Z}_4,\\
\end{aligned}
\eea
where $n$ is an integer, and $H(ng)=\bigoplus_{i=1}^{n}H(g)$. The above equation satisfies the statements in eq.~\eqref{topo statements}, so $(Z_f,2\,\text{Arg}[Z^R_{C}]/\pi,$ $2\,\text{Arg}[Z^R_{C'}]/\pi)$ is sufficient to describe the $\mathbb{Z}\times \mathbb{Z}_2\times \mathbb{Z}_4$ classification.

\subsection{Translation symmetry \texorpdfstring{$j\rightarrow j+2$}{\textmu}}\label{j to j+2}

For the systems that respect the translation symmetry $j\rightarrow j+2$ and reflection symmetry simultaneously, this translation symmetry can be composed of two reflection symmetries with the centers $C_2^1$ and $C_2^2$ defined in Sec.~\ref{Phase structures and many-body spectra of deformed systems}. Therefore, the set of topological invariants here is $(Z_f,2\,\text{Arg}[Z^R_{C_2^1}]/\pi,$ $2\,\text{Arg}[Z^R_{C_2^2}]/\pi)$.

The lattice models $\pm H_{\alpha}$ with PBCs respect this kind of symmetry. At half-filling, we have $Z_f=1$ and can assign the quantum number $(N_{C_2^1},R_{C_2^1},N_{C_2^2},R_{C_2^2})$ to describe their SPT phases. The values of their quantum numbers, $Z^R_{C_2^1}$, and $Z^R_{C_2^2}$ are the same as those provided in Tables~\ref{Table for reflection Ha} and~\ref{Table for reflection mHa}.

\subsection{Translation symmetry \texorpdfstring{$j\rightarrow j+4$}{\textmu}}\label{j to j+4}
There are two ways to compose the translation symmetry $j\rightarrow j+4$: choosing two reflection symmetries with the centers $C_4^1$ and $C_4^3$ or with the centers $C_4^2$ and $C_4^4$ defined in Sec.~\ref{Phase structures and many-body spectra of deformed systems}.

For the lattice models $\pm H_{int_{\alpha}}$ with even $\alpha$, if we consider PBCs, they respect the translation symmetry composed of reflection symmetries with the centers $C_4^1$ and $C_4^3$. Hence, at half-filling, the corresponding set of topological invariants is $(2,2\,\text{Arg}[Z^R_{C_4^1}]/\pi,$ $2\,\text{Arg}[Z^R_{C_4^3}]/\pi)$, and the quantum number is $(N_{C_4^1},R_{C_4^1},N_{C_4^3},R_{C_4^3})$. The values of their topological invariants and quantum numbers can be obtained in Table~\ref{Table for reflection evenInta}. On the other hand, for odd $\alpha$, the translation symmetry of $\pm H_{int_{\alpha}}$ with PBCs is composed of reflection symmetries with the centers $C_4^2$ and $C_4^4$, leading to the set of topological invariants $(2,2\,\text{Arg}[Z^R_{C_4^2}]/\pi,$ $2\,\text{Arg}[Z^R_{C_4^4}]/\pi)$ and the quantum number $(N_{C_4^2},R_{C_4^2},N_{C_4^4},R_{C_4^4})$ at half-filling. Table~\ref{Table for reflection evenInta} provides the values of their topological invariants and quantum numbers.

The lattice models $\pm H_{\alpha}$ with PBCs also can respect this translation symmetry, whether composed of two reflection symmetries with the centers $C_4^1$ and $C_4^3$ or with the centers $C_4^2$ and $C_4^4$. At half-filling, for the translation symmetry with $C_4^1$ and $C_4^3$, the set of topological invariants is $(1,2\,\text{Arg}[Z^R_{C_2^1}]/\pi,$ $2\,\text{Arg}[Z^R_{C_2^1}]/\pi)$, and for the translation symmetry with $C_4^2$ and $C_4^4$, their SPT phases are described by $(1,2\,\text{Arg}[Z^R_{C_2^2}]/\pi,$ $2\,\text{Arg}[Z^R_{C_2^2}]/\pi)$. The topological invariants $Z^R_{C_2^1}$ and $Z^R_{C_2^2}$ are provided in Tables~\ref{Table for reflection Ha} and~\ref{Table for reflection mHa}. However, the quantum numbers here cannot be determined in the same manner ad in Sec.~\ref{Phase structures and many-body spectra of deformed systems} because the way used there doesn't satisfy the restriction~\eqref{charges restriction}. Taking $-H_0$ with the translation symmetry composed of two reflection symmetries with the centers $C_4^1$ and $C_4^3$ as an example, the picture of its charges in a subsystem is
\begin{equation*}
\centering
\includegraphics[width=0.5\textwidth]{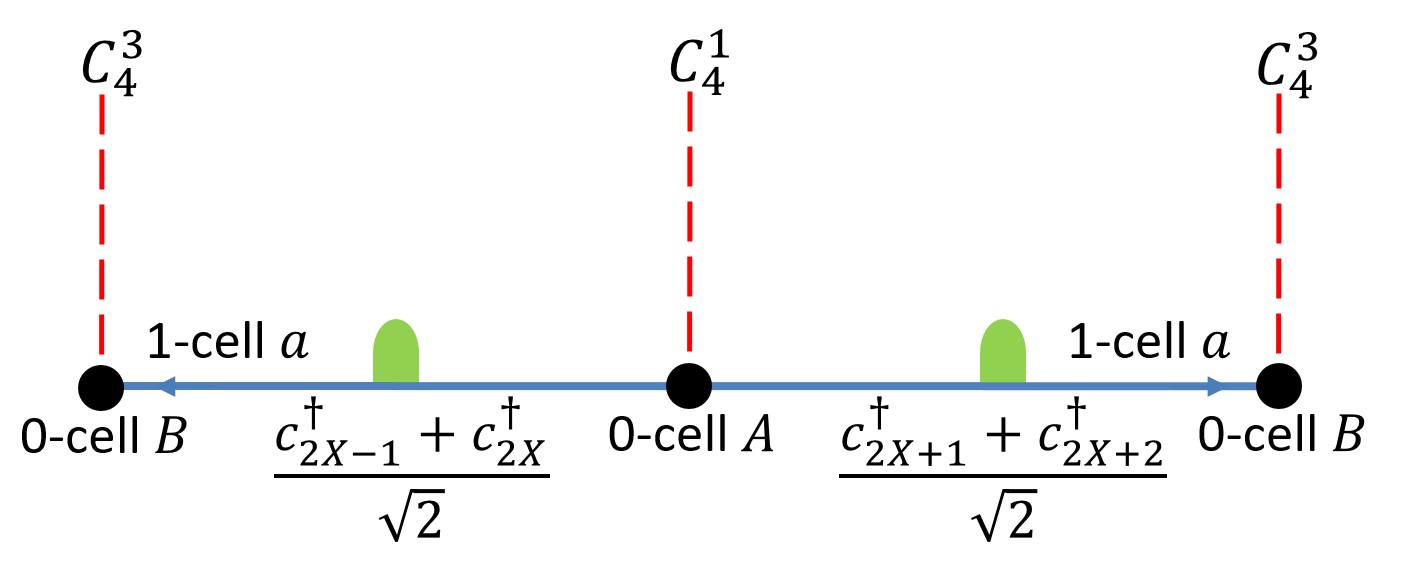} 
\end{equation*}
Here we make the reflection center $C_4^1$ located between $c_{2X}$ and $c_{2X+1}$. At first sight, one may think the corresponding quantum number is $(0,0,0,0)$. If we only want to know the information about $Z^R_{C^1_4}$ and $Z^R_{C^3_4}$, this quantum number is fine, but now we need all the information including $Z_f$, so it doesn't work, and we need a quantum number that satisfies the restriction \eqref{charges restriction}. The valid quantum number can be obtained by moving charges while preserving reflection and translation symmetry, such as
\begin{equation*}
\centering
\includegraphics[width=0.48\textwidth]{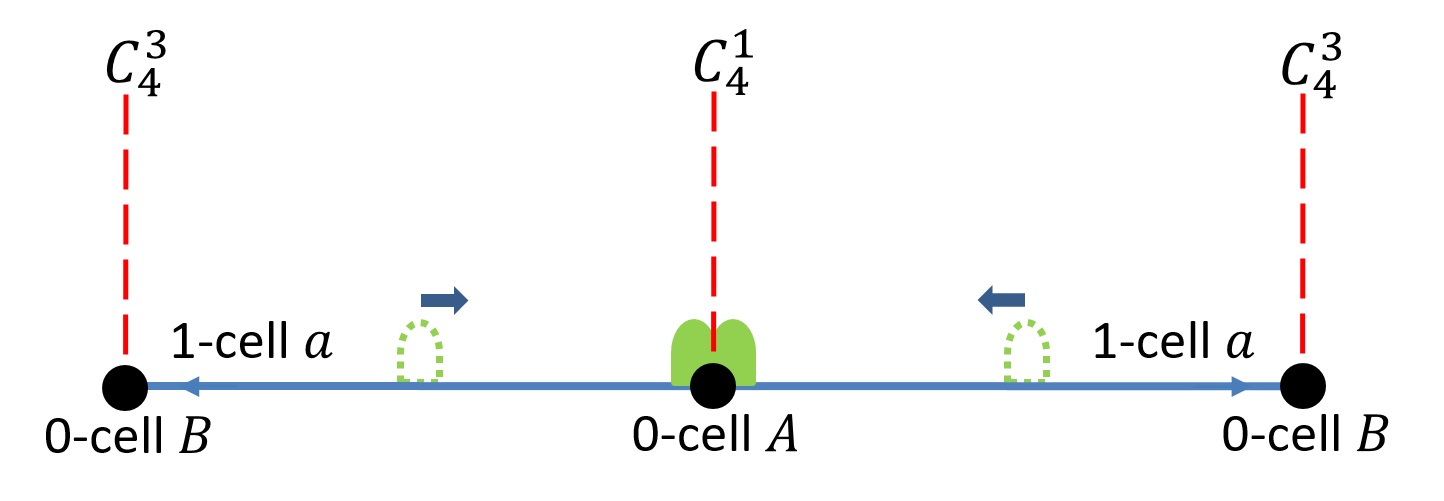} 
\end{equation*}
Because of $R(c_{2X-1}^{\dag}+c_{2X}^{\dag})(c_{2X+1}^{\dag}-c_{2X+2}^{\dag})R^{-1}=(c_{2X+2}^{\dag}+c_{2X+1}^{\dag})(c_{2X}^{\dag}+c_{2X-1}^{\dag})=-(c_{2X-1}^{\dag}+c_{2X}^{\dag})(c_{2X+1}^{\dag}-c_{2X+2}^{\dag})$, the quantum number here is $(2,1,0,0)$. We also can move charges to another reflection center $C_4^3$, and the quantum number becomes $(0,0,2,1)$. Note that $(2,1,0,0)$ and $(0,0,2,1)$ are topologically identical because $(N_C,R_C,N_{C'},R_{C'})+n(2,1,-2,1)\sim (N_C,R_C,N_{C'},R_{C'})$ where $n$ is an integer. As a result, for $-H_0$, the valid quantum number is $(0,0,2,1)$ or $(2,1,0,0)$ instead of $(0,0,0,0)$ because $(0,0,0,0)$ doesn't provide the right information about $\mathbb{Z}$ and therefore it cannot describe $\mathbb{Z}\times \mathbb{Z}_2\times \mathbb{Z}_4$ classification. The necessity of restriction~\eqref{charges restriction} is supported by the energy spectrum of $-H_0 \rightarrow -H_{int_0}$ shown in Fig~\ref{EN for negH0_to_Int0}, As stated in Sec.~\ref{Phase structures and many-body spectra of deformed systems}, this energy spectrum implies that $-H_0$ and $-H_{int_0}$ are topologically identical. Concerning the translation symmetry with $C_4^1$ and $C_4^3$, the quantum number of $-H_{int_0}$ is $(2,1,0,0)$, so $(0,0,0,0)$ cannot be the quantum number of $H_0$, otherwise, it will contradict the fact that $-H_0 \sim -H_{int_0}$.

\section{Conclusion}\label{Conclusion}

In this work, we studied $1d$ fermionic SPT phases with chiral and reflection symmetries. These two types of topological phases are related by the crystalline equivalence principle, which predicts the same classification for both free-fermion and many-body systems. However, their topological characteristics can be very different in non-relativistic condensed matter systems and are quite dependent on microscopic details. To investigate this, we consider specific models of fermions that respect both chiral and reflection symmetries and exam their topological properties in detail. In particular, we consider deformations among these models, where we observe the correspondence between transitions of the many-body topological invariants and gap-closing points (level crossings between the ground and first excited states) in the many-body energy spectra. As expected, gapped deformations occur when the topological invariants remain unchanged.

\begin{figure}[htb!]
\centering
\includegraphics[width=0.4\textwidth]{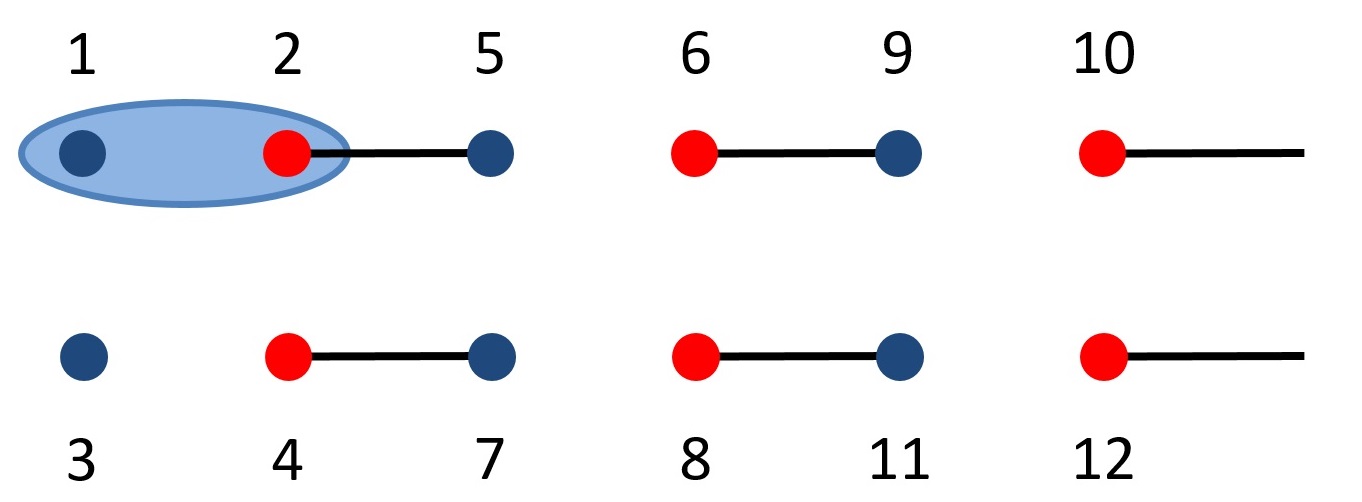}
\caption{After rearranging, $H_2$ is equivalent to stacking two dimerized SSH chains $H_1$.}
\label{Stacking two chains}
\end{figure}
\begin{figure}[htb!]
\centering
\subfloat[]{\includegraphics[width=0.35\textwidth]{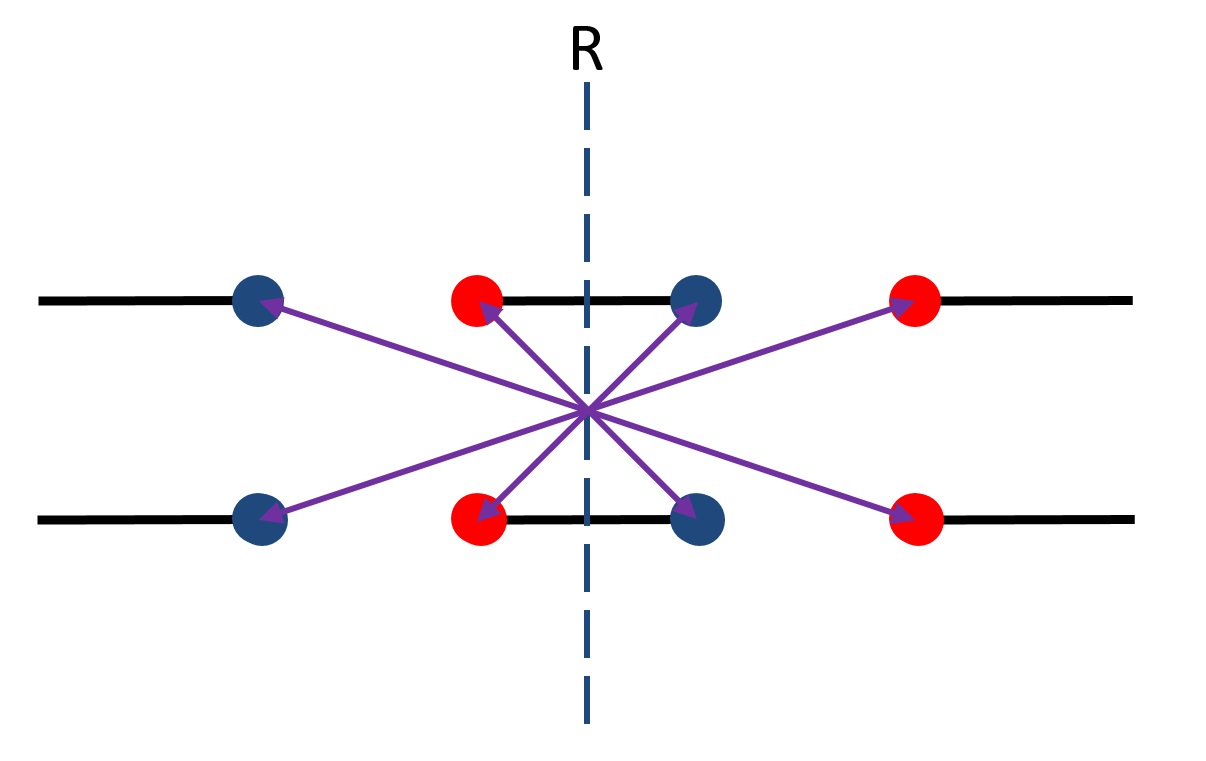}}\hskip 0.5cm
\subfloat[]{\includegraphics[width=0.35\textwidth]{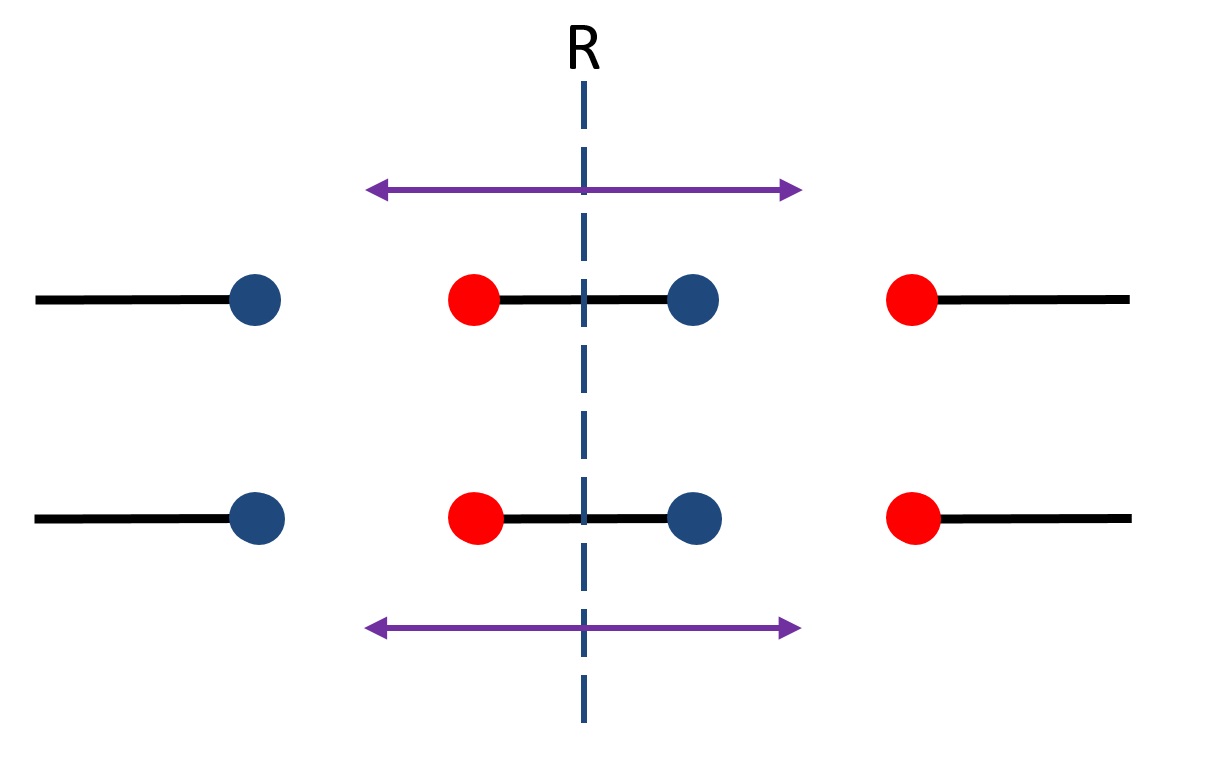}}\\
\caption{The purple arrows represent the directions of reflection symmetry. (a) The reflection symmetry of $H_2$. (b) The reflection symmetry of stacking two $H_1$.}
\label{R of H2 and H1}
\end{figure}

One intriguing distinction between the two SPT phases is that, for the non-interacting model $H_{\alpha}$ given in (\ref{free_systems}), the free-fermion (many-body) topological invariants associated with chiral symmetry directly correspond to the hopping range $\alpha$ ($\alpha$ mod 4). In contrast, those associated with reflection symmetry do not exhibit this correspondence. This difference arises because $H_{\alpha}$ can be regarded as a stack of $|\alpha|$ chains with $H_{\mathrm{sign}(\alpha)}$ by rearranging the site numbering (illustrated in Fig.~\ref{Stacking two chains} for $\alpha=2$), which does not affect the action of the on-site chiral symmetry but does change the action of the non-on-site reflection symmetry, as depicted in Fig.~\ref{R of H2 and H1}.

Furthermore, we provide more physical interpretations of the rather abstract many-body topological invariants for decomposable systems in terms of symmetry quantum numbers, thereby simplifying the computation of these invariants. Such interpretations naturally give the bulk-boundary correspondence for chiral-symmetric systems and the bulk-center correspondence for reflection-symmetric systems. The latter can be particularly justified using the Atiyah-Hirzebruch spectral sequence in generalized homology theory. We expect similar expressions to be applicable in higher-dimensional crystalline SPT phases.

\medskip
Acknowledgments---
C.-S. L. is supported by National Science and Technology Council Graduate Research Fellowship Pilot Program under Grant No. 113-2926-I-002-002-MY3.
K.S. is supported by JST CREST Grant No.~JPMJCR19T2 and JSPS KAKENHI Grant Nos.~22H05118 and 23H01097. 
C.-T. H. is supported by the Yushan (Young) Scholar Program under Grant No. NTU-111VV016 and by the National Science and Technology Council (NSTC) of Taiwan under Grant No. 112-2112-M-002-048-MY3.

\begin{appendices}
\section{Many-body topological invariants of \texorpdfstring{$H_{1}$}{\textmu}} \label{Many-body topological invariants of H1}
To illustrate how to evaluate the topological invariants $Z^S$ and $Z^R$ in section~\ref{Many-body topological invariants}, we analytically calculate the corresponding topological invariants of $H_1$ at half-filling in this section. 
\subsection{Chiral-respecting topological invariant of \texorpdfstring{$H_{1}$}{\textmu}}\label{analytic ZS of H1}
For simplicity, we consider the system with $N_{\text{tot}}=6$, where the ground state can be written as
\bea
\ket{GS(H_1)}=\frac{1}{2^{3/2}}(c_{2}^{\dag}-c_{3}^{\dag})(c_{4}^{\dag}-c_{5}^{\dag})(c_{6}^{\dag}-c_{1}^{\dag})\ket{0}.
\eea
Here we choose the adjacent intervals $I_1=\{c_1,c_2\}$ and $I_2=\{c_{3},c_4\}$, such as
\begin{equation*}
\underbrace{c_1\;c_2}_{I_1}\; \underbrace{c_{3}\;c_{4}}_{I_2}\; c_{5}\;c_{6}
\end{equation*}
With this choice of intervals, the reduced density matrix is given by
\bea\label{density matrix of H1}
\begin{aligned}
\rho_I&=\text{Tr}_{\{c_5,c_6\}}[\ket{GS}\bra{GS}]\\
&\begin{aligned}
=2^{-3}[&\ket{1101}\bra{1101}+\ket{1101}\bra{1011}+\ket{1100}\bra{1100}+\ket{1100}\bra{1010}\\
&\ket{1011}\bra{1101}+\ket{1011}\bra{1011}+\ket{1010}\bra{1100}+\ket{1010}\bra{1010}\\
&\ket{0101}\bra{0101}+\ket{0101}\bra{0011}+\ket{0100}\bra{0100}+\ket{0100}\bra{0010}\\
&\ket{0011}\bra{0101}+\ket{0011}\bra{0011}+\ket{0010}\bra{0100}+\ket{0010}\bra{0010}],\\
\end{aligned}\\
\end{aligned}
\eea
with the occupied states defined as
\begin{equation*}
\ket{n_1 n_2 n_3 n_4}=(c_{1}^{\dag})^{n_1}(c_{2}^{\dag})^{n_2}(c_{3}^{\dag})^{n_3}(c_{4}^{\dag})^{n_4}\ket{0}.
\end{equation*}
Since the system respects the chiral symmetry $Sc_jS^{-1}=(-1)^jc_j^{\dag}$, the $U_S^{I_1}$ acts on occupied states as
\begin{equation*}
U_S^{I_1}\ket{n_1 n_2 n_3 n_4}=(-1)^{n_1}(1)^{n_2} \ket{n_1 n_2 n_3 n_4}.
\end{equation*}
Combining the above equation with $C_f^{I_1}=(c_1^{\dag}+c_1)(c_2^{\dag}+c_2)$ leads to
\bea\label{chiral-related operators}
U_S^{I_1}C_f^{I_1}\ket{n_1 n_2 n_3 n_4}=-\ket{\widetilde{n}_1 \widetilde{n}_2 n_3 n_4}, \quad \text{with}\; \widetilde{n}_j=|n_j-1|.
\eea
By using eqs.~\eqref{chiral-related operators} and~\eqref{chiral-respecting topo regarding occupied states}, we have
\bea
\begin{aligned}
U_S^{I_1} \rho_I^{T_1} [U_S^{I_1}]^{\dag}=2^{-3}[&\ket{0001}\bra{0001}-i\ket{0101}\bra{0011}+\ket{0000}\bra{0000}-i\ket{0100}\bra{0001}\\
&-i\ket{0011}\bra{0101}+\ket{0111}\bra{0111}-i\ket{0010}\bra{0100}+\ket{0110}\bra{0110}\\
&\ket{1001}\bra{1001}-i\ket{1101}\bra{1011}+\ket{1000}\bra{1000}-i\ket{1100}\bra{1010}\\
&-i\ket{1011}\bra{1101}+\ket{1111}\bra{1111}-i\ket{1010}\bra{1100}+\ket{1110}\bra{1110}].\\
\end{aligned}
\eea
Finally, we obtain
\bea
Z^S(H_1)=\text{Tr}_{I}[\rho_I U_S^{I_1} \rho_I^{T_1} [U_S^{I_1}]^{\dag}]=-i/8
\eea
\subsection{Reflection-respecting topological invariant of \texorpdfstring{$H_{1}$}{\textmu}}\label{analytic ZR of H1}
Let's consider the system with $N_{\text{tot}}=2L+2$. Since it respects reflection symmetry $Rc_{j}^{\dag}R^{-1}=c_{2L+3-j}^{\dag}$, by the definition~\eqref{partial reflection}, we can pick up the interval $I=\{c_{2N+1},\ldots,c_{2M}\}$, where its center is the same as reflection center, and then define the following partial reflection $[U_{\alpha}R]_I$
\bea
[U_{\alpha}R]_{I}c_{j}^{\dag}([U_{\alpha}R]_{I})^{-1}=i c_{2(M+N)-j+1}^{\dag}\quad [U_{\alpha}R]_{I}\ket{0}=\ket{0}, \quad j\in\{2N+1,\ldots,2M\}.
\eea
Here we suppose $0\leq N<M<L$. The partial reflection acts on the ground state as
\bea
[U_{\alpha}R]_{I}\ket{GS(H_1)}=\frac{1}{2^{(L+1)/2}}[\ldots(c_{2N}^{\dag}-ic_{2M}^{\dag})A_{\text{part}}(ic_{2N+1}^{\dag}-c_{2M+1}^{\dag})\ldots]\ket{0},
\eea
where
\bea
\begin{aligned}
A_{\text{part}}&=i^{(N_{\text{p}}-1)}(c_{2M-1}^{\dag}-c_{2M-2}^{\dag})\ldots(c_{2N+3}^{\dag}-c_{2N+2}^{\dag})\\
&=i^{(N_{\text{p}}-1)}(-1)^{\sum_{j=1}^{N_{\text{p}}-1}(j-1)}(c_{2N+3}^{\dag}-c_{2N+2}^{\dag})\ldots(c_{2M-1}^{\dag}-c_{2M-2}^{\dag})\\
&=i^{(N_{\text{p}}-1)}(-1)^{(N_{\text{p}}-1)(N_{\text{p}}-2)/2}(-1)^{(N_{\text{p}}-1)}(c_{2N+2}^{\dag}-c_{2N+3}^{\dag})\ldots(c_{2M-2}^{\dag}-c_{2M-1}^{\dag}),\\
\end{aligned}
\eea
with $N_{\text{p}}$ defined as the number of unit cells in the interval $I$. It's straightforward to see that
\bea\label{H1 topological invariant}
\begin{aligned}
Z^R(H_1)&=\bra{GS(H_1)}[U_{\alpha}R]_{I}\ket{GS(H_1)}\\
&=\frac{2^{(L-1)/2}}{2^{(L+1)/2}}(-1)^{(N_{\text{p}}-1)(N_{\text{p}}-2)/2}(-1)^{(N_{\text{p}}-1)}i^{(N_{\text{p}}-1)}B_{\text{part}}\\
&=\frac{i^{(N_{\text{p}}-1)}}{4}(-1)^{(N_{\text{p}}-1)(N_{\text{p}}-2)/2}(-1)^{(N_{\text{p}}-1)}(1-i^2)\\
&=\frac{i^{(N_{\text{p}}-1)}}{2}(-1)^{(N_{\text{p}}-1)(N_{\text{p}}-2)/2}(-1)^{(N_{\text{p}}-1)},\\
\end{aligned}
\eea
with $B_{\text{part}}=\bra{0}(c_{2M}-c_{2M+1})(c_{2N}-c_{2N+1})(c_{2N}^{\dag}-ic_{2M}^{\dag})(ic_{2N+1}^{\dag}-c_{2M+1}^{\dag})\ket{0}$. The even and odd $N_{\text{p}}$ correspond to the reflection centers $C_2^1$ and $C_2^2$ defined in Sec.~\ref{Phase structures and many-body spectra of deformed systems}, respectively.

Because we consider the system with translation symmetry when evaluating the reflection-respecting topological invariant, it respects two reflection symmetries with $C_2^1$ and $C_2^2$ simultaneously. Therefore, it's worth noting that one can choose the interval used in Tables~\ref{Table for reflection Ha} and~\ref{Table for reflection mHa} to calculate the $Z^R$ concerning $C_2^1$ and $C_2^2$, alternatively. 

\section{Finite-size effect and many-body spectra} \label{Finite-size effect and many-body spectra}

\begin{figure}[hbt!]
\centering
\subfloat[]{\includegraphics[width=0.31\textwidth]{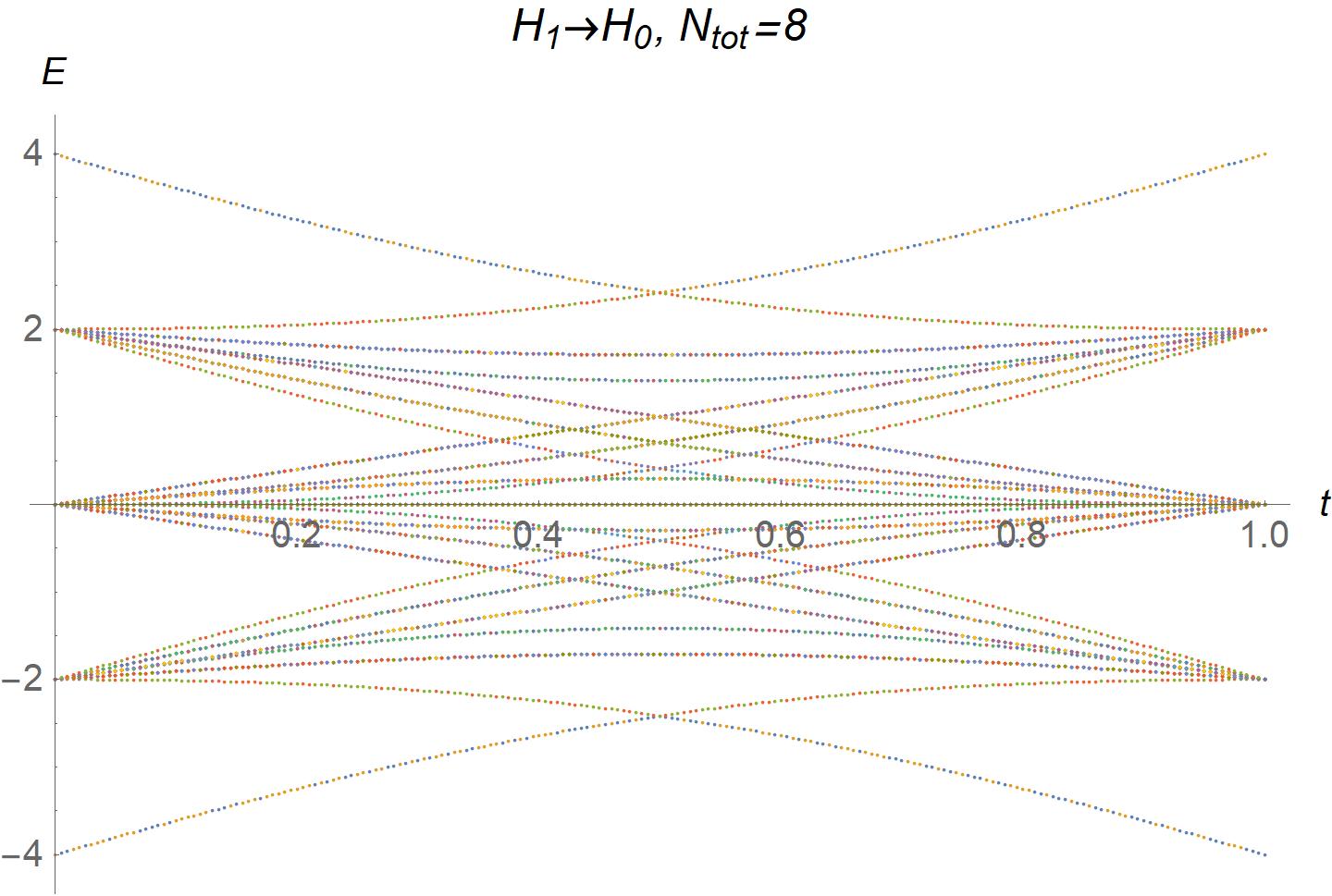}}\hskip 0.5cm
\subfloat[]{\includegraphics[width=0.31\textwidth]{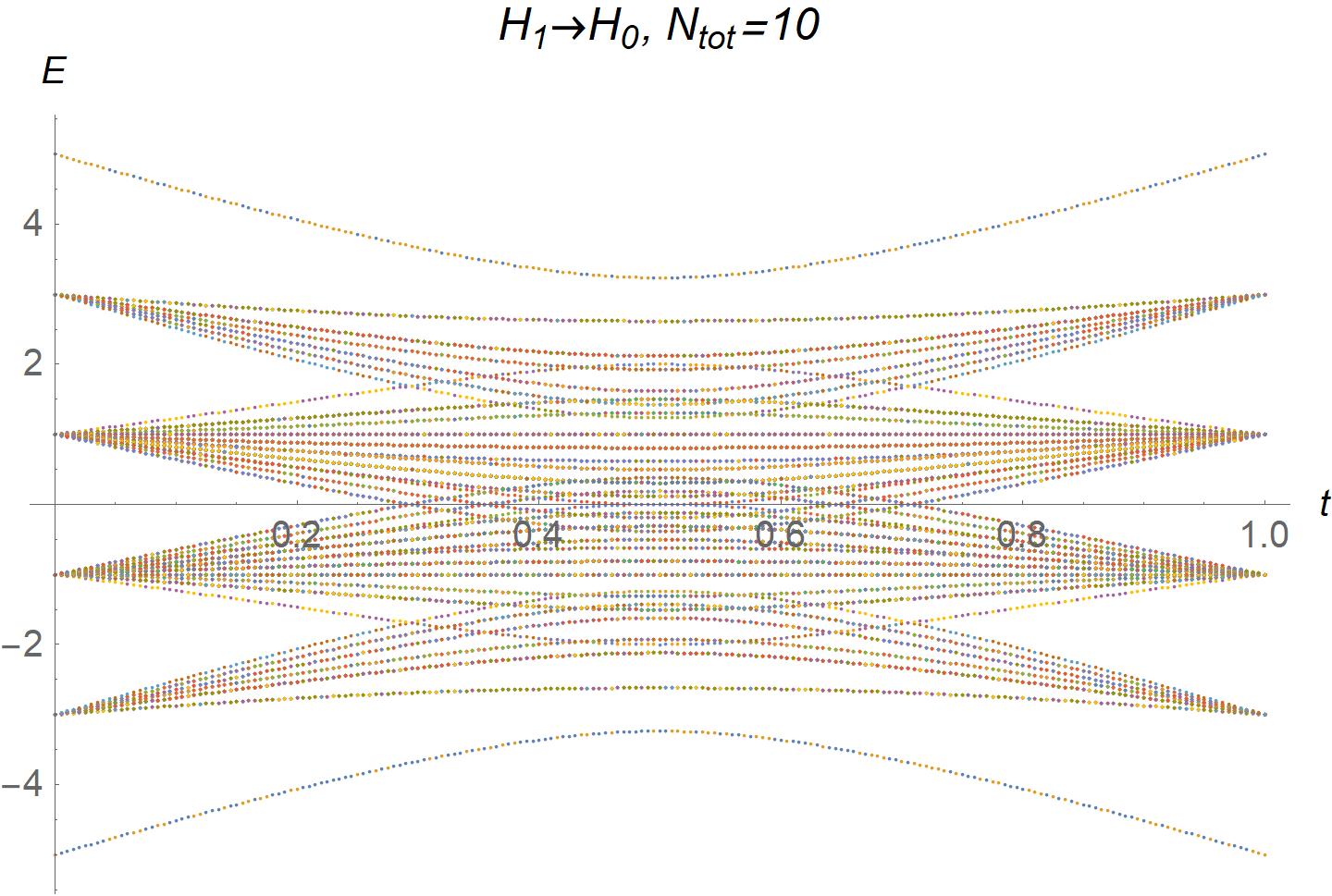}}\hskip 0.5cm
\subfloat[]{\includegraphics[width=0.31\textwidth]{EN_w1tow0_12.jpg}}\\
\caption{The energy spectra of $H_{1}\rightarrow H_{0}$ with different values of $N_{\text{tot}}$. A level crossing occurs for the lowest two states (at $t=0.5$) for $N_{\text{tot}}=8$ and  $N_{\text{tot}}=12$, indicating a phase transition for this deformation.}
\label{ZS EN for w1tow0 in appendix}
\end{figure}

\begin{figure}[htb!]
\centering
\subfloat[]{\includegraphics[width=0.31\textwidth]{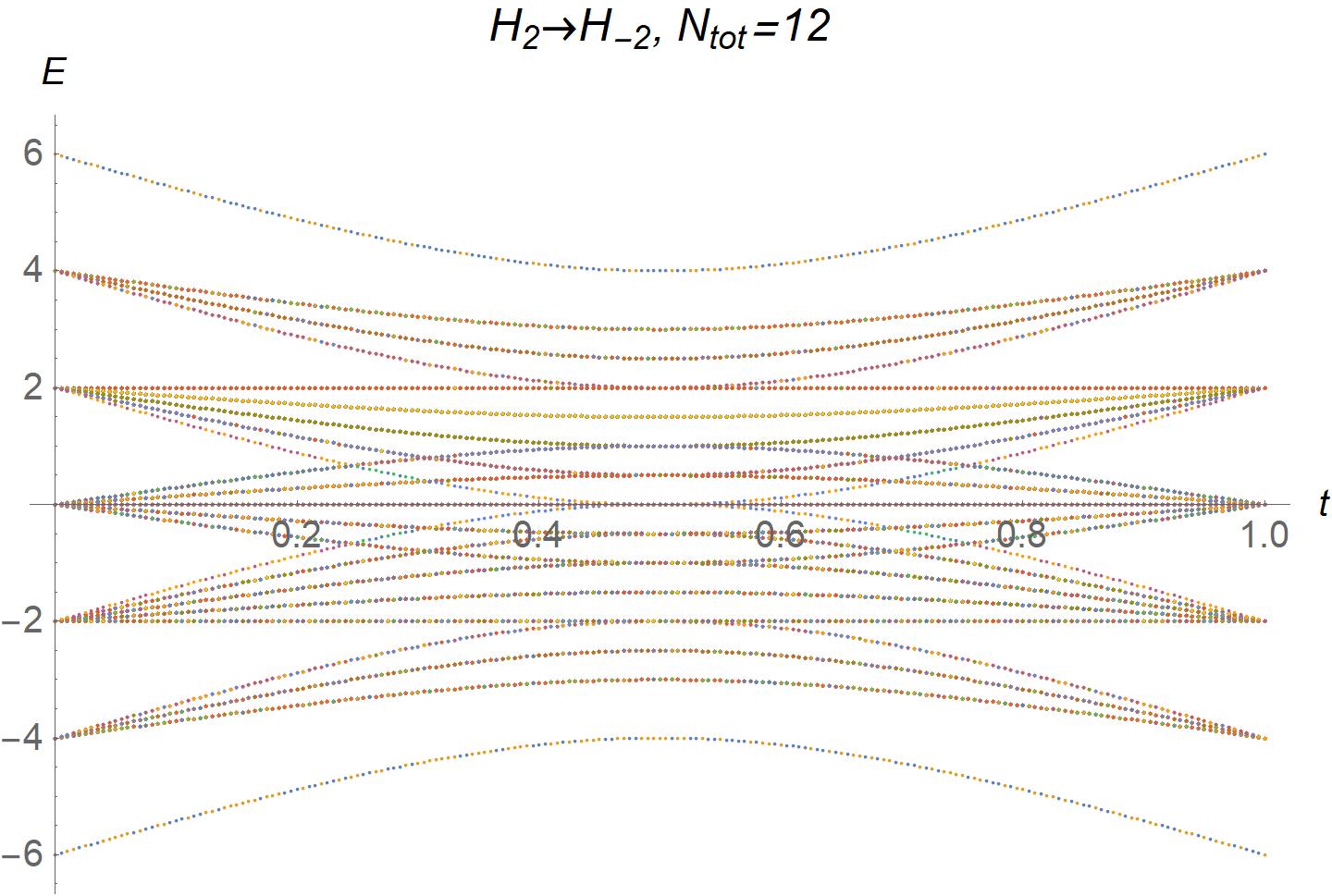}}\hskip 0.5cm
\subfloat[]{\includegraphics[width=0.31\textwidth]{EN_w2toInt0_12.jpg}}\hskip 0.5cm
\subfloat[]{\includegraphics[width=0.31\textwidth]{EN_Int2toInt0_12.jpg}}\\
\subfloat[]{\includegraphics[width=0.31\textwidth]{EN_w2towm2_16.jpg}}\hskip 0.5cm
\subfloat[]{\includegraphics[width=0.31\textwidth]{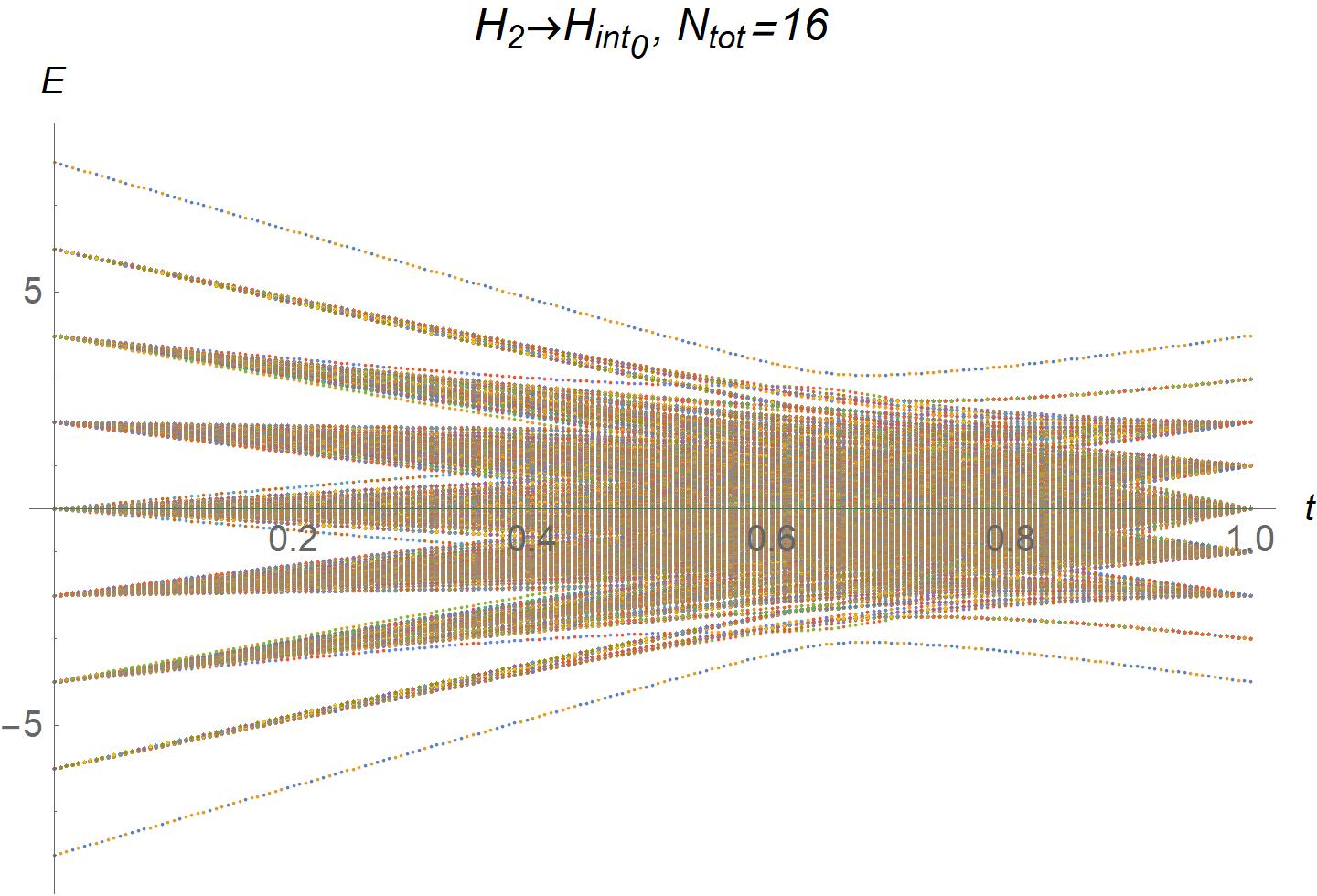}}\hskip 0.5cm
\subfloat[]{\includegraphics[width=0.31\textwidth]{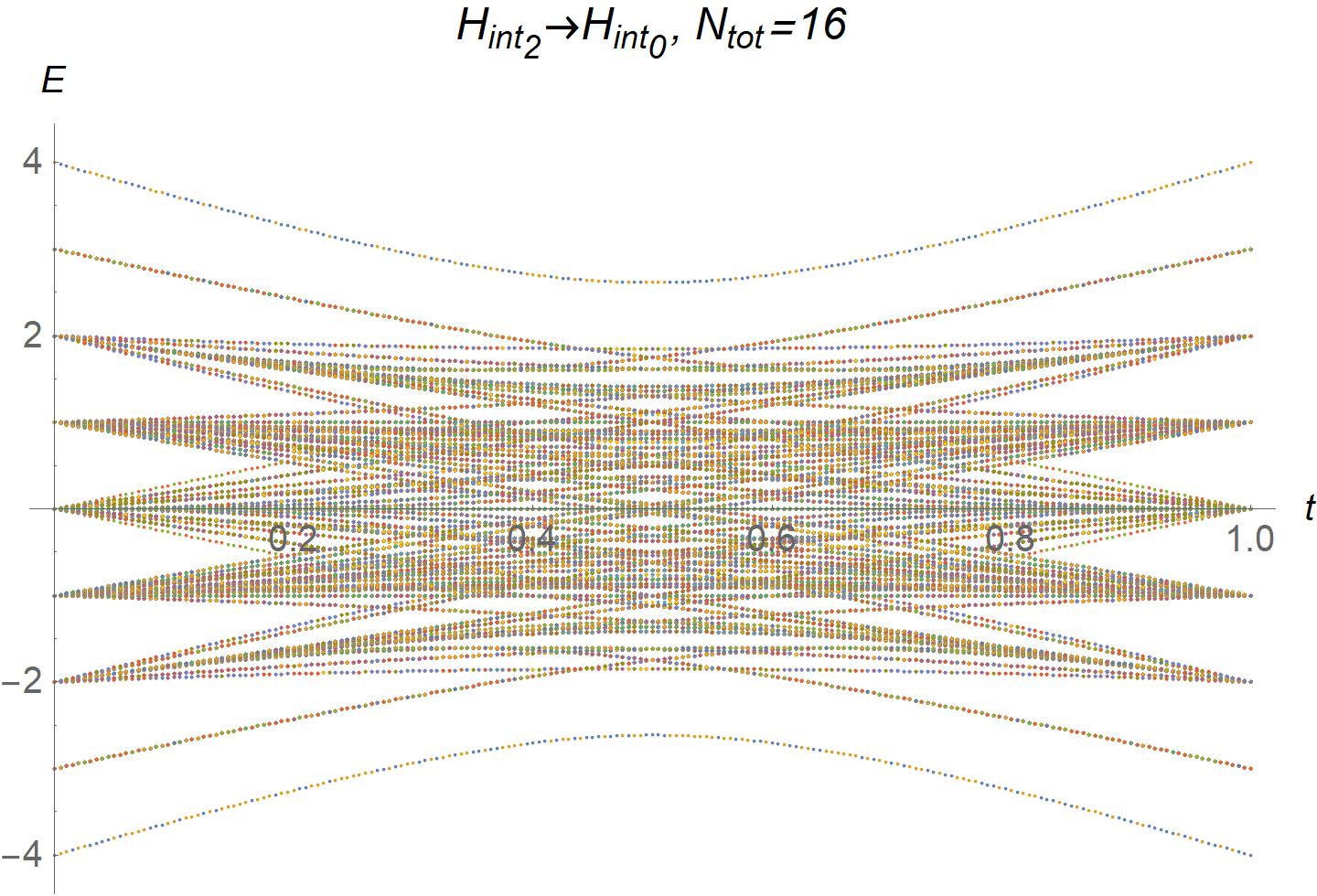}}\\
\caption{The energy spectra of $H_{2}\rightarrow H_{-2}$, $H_{2}\rightarrow H_{int_0}$, and $H_{int_2}\rightarrow H_{int_0}$ with different values of $N_{\text{tot}}$. A level crossing occurs in each deformation for either $N_{\text{tot}}=12$ or $N_{\text{tot}}=16$, indicating a phase transition.}
\label{EN for w2towm2 w2toInt0 In2toInt0 in appendix}
\end{figure}


\begin{figure}[htb!]
\centering
\subfloat[]{\includegraphics[width=0.24\textwidth]{EN_w2toInt2_12.jpg}}\hskip 0.2cm
\subfloat[]{\includegraphics[width=0.24\textwidth]{EN_w2toIntm2_12.jpg}}\hskip 0.2cm
\subfloat[]{\includegraphics[width=0.24\textwidth]{EN_wm2toInt2_12.jpg}}\hskip 0.2cm
\subfloat[]{\includegraphics[width=0.24\textwidth]{EN_wm2toIntm2_12.jpg}}\\
\subfloat[]{\includegraphics[width=0.24\textwidth]{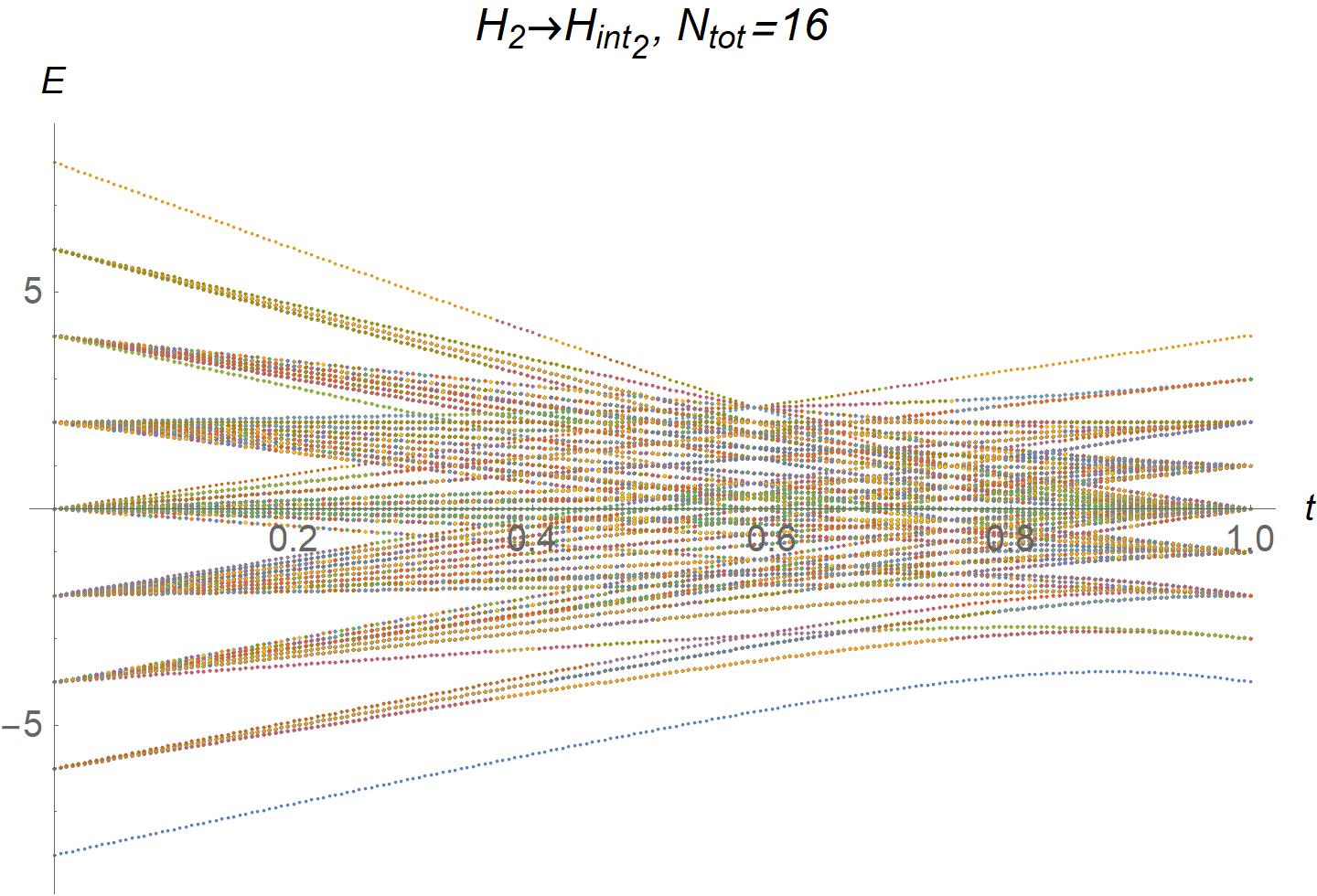}}\hskip 0.2cm
\subfloat[]{\includegraphics[width=0.24\textwidth]{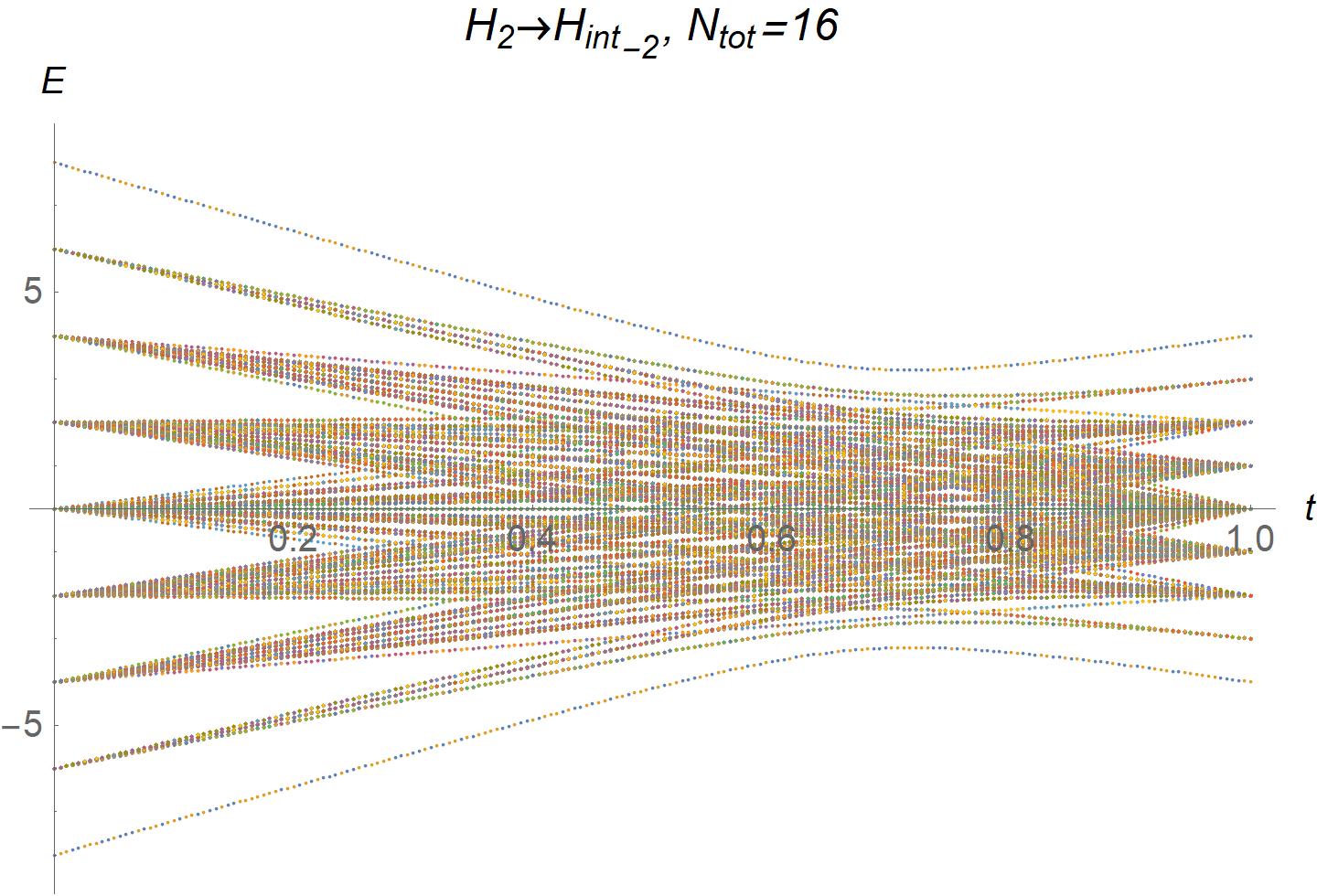}}\hskip 0.2cm
\subfloat[]{\includegraphics[width=0.24\textwidth]{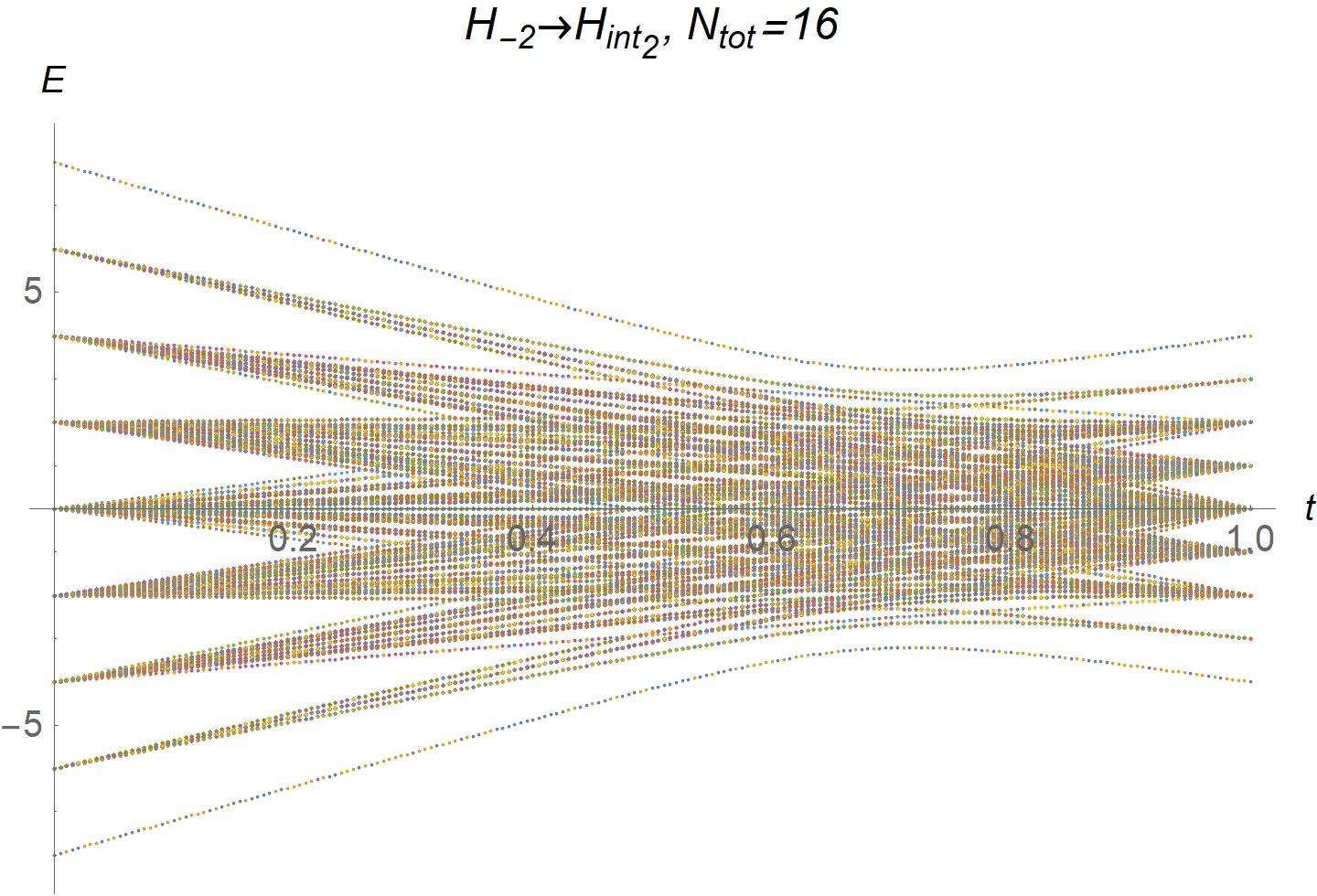}}\hskip 0.2cm
\subfloat[]{\includegraphics[width=0.24\textwidth]{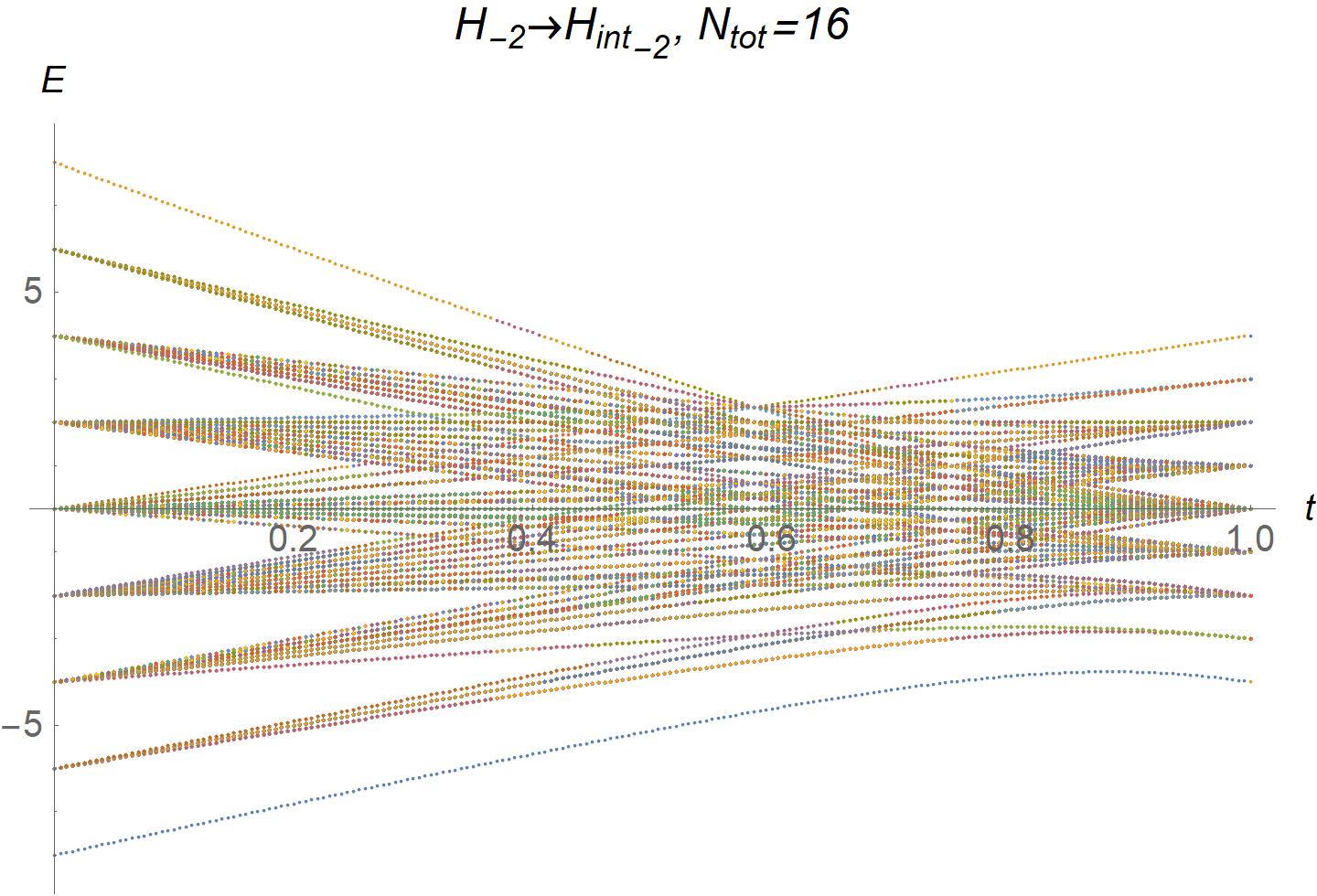}}\\
\caption{Energy spectra of $H_{\pm2}\rightarrow H_{int_{\pm2}}$ with $N_{\text{tot}}=12$ and $N_{\text{tot}}=16$ . Since there are no level crossings in any of the spectra, we can reasonably conclude no phase transition occurs for these deformations.}
\label{EN for Hpm2_to_Intpm2 in appendix}
\end{figure}

In topological band theory, a topological phase transition is typically characterized by the closing of the system's energy gap in the thermodynamic limit. This property can, however, be suppressed by the finite-size effect when considering a system of finite size. For a free fermion system, a finite energy gap can be induced by the finite-size effect, and the behavior of this gap in the thermodynamic limit can be understood precisely from the single-particle spectrum given by the Bloch Hamiltonian.

For a many-body system lacking a Bloch Hamiltonian description, it is generally difficult to predict how finite-size effects will impact the energy spectrum. Moreover, a topological phase transition can occur either when the many-body energy gap vanishes or when the ground states become degenerate while remaining gapped (level crossing) in the thermodynamic limit. To determine whether there is a topological phase transition, it is reasonable to examine the many-body energy spectrum of a finite system with various system sizes. A phase transition is likely present if there exsits a finite $N_{\text{tot}}$ such that a level crossing occurs for the lowest few energy eigenstates at some point in the parameter space. This can be illustrated by the deformed systems discussed in the main part of this article, with the many-body spectra shown in Figures~\ref{ZS EN for w1tow0 in appendix}, \ref{EN for w2towm2 w2toInt0 In2toInt0 in appendix}, and \ref{EN for Hpm2_to_Intpm2 in appendix}.
\section{Atiyah-Hirzebruch spectral sequence} \label{The AHSS in generalized homology}
For the interacting system, the classification of degree $n$ SPT phenomena with crystalline symmetry $G$ can be determined by the generalized homology $h_{n}^{G}(X,Y)$ where $X$ and $Y$ are real space and $Y\subset X$. In general, $h_{n}^{G}(X,Y)$ is quite difficult to compute, so we use the AHSS to simplify the computation \cite{ghomo2}. The way the AHSS is utilized to reach $h_{n}^{G}(X,Y)$ is as follows. We first take the $G$-symmetric cell decomposition of the space manifold $X$ and then define the $p$-skeleton $X_p$ of $X$
\bea\label{skeleton}
X_0=\{\text{0-cells}\}, \qquad X_p=X_{p-1}\cup\{\text{$p$-cells}\}.
\eea
$p$-cell is a $p$-dimensional open cell in the cell decomposition. eq.~\eqref{skeleton} also implies the following relation
\bea\label{skeleton subset}
X_0\subset X_1\subset ...\subset X_d=X,
\eea
where $d$ is the dimension of $X$. To proceed, we introduce the filtration of $h_{n}^{G}(X,Y)$
\bea\label{filtration}
F_ph_n:=\text{Im}[h_{n}^{G}(X_p,X_p\cap Y)\rightarrow h_{n}^{G}(X,Y)].
\eea
$F_ph_n$ can be regarded as the SPT phases on $X$ by embedding the SPT phases on $X_p$ into $X$. Due to eq.~\eqref{skeleton subset}, we have
\bea\label{filtration subset}
0\subset F_0h_n\subset ...\subset F_dh_n=h_{n}^{G}(X,Y).
\eea
Combining eq.~\eqref{filtration subset} and this relation $E_{p,n-p}^{\infty}\cong F_ph_n/F_{p-1}h_n$, we can get a series of short exact sequences:
\bea\label{short exact sequences}
\begin{alignedat}{3}
0\rightarrow & \ \ \, E^{\infty}_{0,n}&&\rightarrow \quad F_{1}h_{n}&&\rightarrow E^{\infty}_{1,n-1}\rightarrow 0,\\
0\rightarrow & \ \ \, F_{1}h_{n}&&\rightarrow \quad F_{2}h_{n}&&\rightarrow E^{\infty}_{2,n-2}\rightarrow 0,\\
&\vdots\\
0\rightarrow &F_{d-1}h_{n}&&\rightarrow h_{n}^{G}(X,Y)&&\rightarrow E^{\infty}_{d,n-d}\rightarrow 0.\\
\end{alignedat}
\eea
$E^{\infty}_{p,-q}$ is called the limiting page, which represents a set of $q$-dimensional SPT phases on $p$-cells that create no anomaly on any adjacent low-dimensional cells and cannot be trivialized by any adjacent high-dimensional cells. Here, if an SPT phase on $p$-cells can be trivialized by the SPT states in an adjacent high-dimensional cell, it means that this SPT phase can be created by adiabatically pumping the SPT states in this adjacent high-dimensional cell. Note that for a $d$-dimensional space manifold, the limiting page $E^{\infty}$ is the same as the converged page $E^{d+1}$. Using the above relation~\eqref{short exact sequences} and given $E^{\infty}$, we can obtain $h_{n}^{G}(X,Y)$ and each $F_ph_n$ by the iterative method.

Now, the last problem is how to get the limiting page $E^{\infty}$. To approach this, we start with the definition of $E^{1}$-page. For each $p$-cell $D^{p}_j$, there is a little group $G_{D^{p}_j} \subset G$ acts on $D^{p}_j$ as on-site symmetry, so the cell decomposition leads to the "local data of SPT phases on $D^{p}_j$", $h_{p-q}^{G_{D^p_j}}(D^{p}_j,\partial D^{p}_j)$. If we consider the collection of these local data, we can define the $E^1$-page
\bea\label{E1-page}
E^1_{p,-q}=\prod _j h_{p-q}^{G_{D^{p}_j}}(D^{p}_j,\partial D^{p}_j),
\eea
where $q$ is the dimension of local SPT phases, and $j$ runs the set of inequivalent p-cells of $X$. Because of the bulk-boundary correspondence \cite{BBC}, it's reasonable to introduce the first differential (boundary map)
\bea\label{1-differential}
d^{1}_{p,-q}:E^{1}_{p,-q}\rightarrow E^{1}_{p-1,-q}.
\eea
Given that $E^{1}_{p,-q}$ can also be interpreted as a set of $(q-1)$-dimensional anomalies over $p$-cells, $d^{1}_{p,-q}$ is a map from the $q$-dimensional SPT phases on $p$-cells to the $(q-1)$-dimensional anomalies on adjacent $(p-1)$-cells. Owing to the fact that the boundary of the boundary is empty, $d^{1}_{p,-q}\circ d^{1}_{p+1,-q}=0$, we can take the homology of $d^1$, which is defined as $E^2$-page
\bea\label{E2-page}
E^{2}_{p,-q}:=\text{Ker}(d^{1}_{p,-q})/\text{Im}(d^{1}_{p+1,-q}).
\eea
Physically, $\text{Ker}(d^{1}_{p,-q})$ is a set of $q$-dimensional SPT phases on $p$-cells that create no anomaly on adjacent $(p-1)$-cells, and $\text{Im}(d^{1}_{p+1,-q})$ represents a set of $q$-dimensional SPT phases on $p$-cells that can be constructed by adiabatically pumping the SPT states in adjacent $(p+1)$-cells. Thus, $E^{2}_{p,-q}$ is a set of $q$-dimensional SPT phases on $p$-cells that can extend to adjacent $(p-1)$-cells without anomaly and cannot be trivialized by the SPT states in adjacent $(p+1)$-cells. Likewise, we can formulate the higher differential and $E^r$-page
\bea\label{r-differential and Er+1-page}
\begin{aligned}
&d^{r}_{p,-q}:E^{r}_{p,-q}\rightarrow E^{r}_{p-r,-q+r-1},\\
&E^{r+1}_{p,-q}:=\text{Ker}(d^{r}_{p,-q})/\text{Im}(d^{r}_{p+r,-q-r+1}).
\end{aligned}
\eea
$d^{r}_{p,-q}$ is called the $r$-th differential. Note that $E^{r+1}_{p,-q}$ is established because the relation $d^{r}_{p,-q}\circ d^{r}_{p+r,-q-r+1}=0$ holds. In general, for a given $E^1$-page, we can get all $E^r$-pages up to the limiting pages by using eq.~\eqref{r-differential and Er+1-page}.

\section{\texorpdfstring{$(Z_f,2\,\text{Arg}[Z^R_{C}]/\pi,2\,\text{Arg}[Z^R_{C'}]/\pi)$}{\textmu} of generators}\label{prove statement}
By considering the restriction \eqref{charges restriction}, eq.~\eqref{classes and ZR}, and the fact that $(N_C,R_C,N_{C'},R_{C'})+n(2,1,-2,1)\sim(N_C,R_C,N_{C'},R_{C'})$ where $n$ is an integer, we can readily verify that the generators~\eqref{generators} satisfy the statement~\eqref{topo statements}. Furthermore, the representing Hamiltonian for these generators can be constructed by localizing charges on the reflection centers $C$ and $C'$. As a reminder, the many-body topological invariants $Z^R_{C}$ and $Z^R_{C'}$ can be determined by $(N_C,R_C)$ and $(N_{C'},R_{C'})$ separately, as shown below
\bea
(\underbrace{N_C,R_C}_{Z^R_{C}},\underbrace{N_{C'},R_{C'}}_{Z^R_{C'}}).
\eea

For $g_{\mathbb{Z}}=(1,0,0,0)$, we can construct the representing Hamiltonian
\bea
H(g_{\mathbb{Z}})=\sum_{j} c^{\dag}_{a,C,j}c_{a,C,j}-c^{\dag}_{b,C,j}c_{b,C,j}
\eea
at half-filling, where $c_{a/b,C,j}$ is the operator for the complex fermion $a/b$ at the reflection center $C$ within subsystem $j$. The ground state of this system is given by $\prod_jc^{\dag}_{b,C,j}\ket{0}$. Since this system is a decomposable system, we can assign the quantum number $(N_C,R_C,N_{C'},R_{C'})$ to it, which is $(1,0,0,0)$. Note that there are two different types of reflection symmetry, $R_{\pm}c^{\dag}_{a/b,C,j}R_{\pm}^{-1}=\pm c^{\dag}_{a/b,C,j}$, and we consider $R_{+}$ here. Using the quantum number, the relation~\eqref{classes and ZR}, and the definition of $Z_f$~\eqref{filling}, we have
\bea
\begin{aligned}
&Z_f(H(ng_{\mathbb{Z}}))=n,\\
&2\,\text{Arg}[Z^R_{C}H(ng_{\mathbb{Z}})]/\pi=n\,\,\text{mod}\,\,4,\\
&2\,\text{Arg}[Z^R_{C'}H(ng_{\mathbb{Z}})]/\pi=0,\\
\end{aligned}
\eea
where $n$ is an integer, and $H(ng)=\bigoplus_{i=1}^{n}H(g)$. The above equation indicates
\bea
(v_f(H(ng_{\mathbb{Z}}),v_{c}(H(ng_{\mathbb{Z}}),v_{c'}(H(ng_{\mathbb{Z}}))=\left(n,n\,\,\text{mod}\,\,4,0\right)\in \mathbb{Z},
\eea
with $(v_f,v_{c},v_{c'})=(Z_f,2\,\text{Arg}[Z^R_{C}]/\pi,$ $2\,\text{Arg}[Z^R_{C'}]/\pi)$. 

For $g_{\mathbb{Z}_2}=(0,1,0,0)$, we can build the Hamiltonian
\bea
H(g_{\mathbb{Z}_2})=\sum_{j} c^{\dag}_{a,C,j}c_{a,C,j}-c^{\dag}_{b,C,j}c_{b,C,j}+h^{\dag}_{a,C,j}h_{a,C,j}-h^{\dag}_{b,C,j}h_{b,C,j},
\eea
where $h$ denotes the operator for a fermion with positive charge $e$. At $Z_f=0$ (i.e., the $filling$ of the particles $c$ is equal to the $filling$ of the particles $h$), the corresponding ground state is $\prod_jc^{\dag}_{b,C,j}h^{\dag}_{b,C,j}\ket{0}$. If we consider $R_{-}c^{\dag}_{a/b,C,j}R_{-}^{-1}=-c^{\dag}_{a/b,C,j}$ and $R_{+}h^{\dag}_{a/b,C,j}R_{+}^{-1}=-h^{\dag}_{a/b,C,j}$, the decomposable system $H(g_{\mathbb{Z}_2})$ is characterized by $(0,1,0,0)$. Therefore, we can get
\bea
\begin{aligned}
&Z_f(H(ng_{\mathbb{Z}_2}))=0,\\
&2\,\text{Arg}[Z^R_{C}H(ng_{\mathbb{Z}_2})]/\pi\sim2\,\text{Arg}[Z^R_{C}H(n(2,0,-2,1))]/\pi=2n\,\,\text{mod}\,\,4,\\
&2\,\text{Arg}[Z^R_{C'}H(ng_{\mathbb{Z}_2})]/\pi=0,\\
\end{aligned}
\eea
which leads to
\bea
(v_f(H(ng_{\mathbb{Z}_2}),v_{c}(H(ng_{\mathbb{Z}_2}),v_{c'}(H(ng_{\mathbb{Z}_2}))=\left(0,n\,\,\text{mod}\,\,2,0\right)\in \mathbb{Z}_2.
\eea

The generator $g_{\mathbb{Z}_4}=(1,0,-1,0)$ can be represented by the Hamiltonian
\bea
H(g_{\mathbb{Z}_4})=\sum_{j} c^{\dag}_{a,C,j}c_{a,C,j}-c^{\dag}_{b,C,j}c_{b,C,j}+h^{\dag}_{a,C',j}h_{a,C',j}-h^{\dag}_{b,C',j}h_{b,C',j}.
\eea
At $Z_f=0$, the ground state is expressed as $\prod_jc^{\dag}_{b,C,j}h^{\dag}_{b,C',j}\ket{0}$. When considering $R_{+}$ for both $c$ and $h$, the associated quantum number is $(1,0,-1,0)$. Consequently, we obtain
\bea
\begin{aligned}
&Z_f(H(ng_{\mathbb{Z}_4}))=0,\\
&2\,\text{Arg}[Z^R_{C}H(ng_{\mathbb{Z}_4})]/\pi=n\,\,\text{mod}\,\,4,\\
&2\,\text{Arg}[Z^R_{C'}H(ng_{\mathbb{Z}_4})]/\pi\sim2\,\text{Arg}[Z^R_{C'}H(n(-3,0,3,0))]/\pi=3n\,\,\text{mod}\,\,4.\\
\end{aligned}
\eea
The equivalence classes of $\left(0,n\,\,\text{mod}\,\,4,3n\,\,\text{mod}\,\,4\right)$ can be written as $[(0,0,0)], [(0,1,3)], [(0,2,2)]$, and $[(0,3,1)]$, so
\bea
(v_f(H(ng_{\mathbb{Z}_4}),v_{c}(H(ng_{\mathbb{Z}_4}),v_{c'}(H(ng_{\mathbb{Z}_4}))=\left(0,n\,\,\text{mod}\,\,4,3n\,\,\text{mod}\,\,4\right)\in \mathbb{Z}_4.
\eea

\section{\texorpdfstring{Deformation between $H_0$ and $H_2$ in the free fermion scope}{\textmu}}
\label{Deformation between $H_0$ and $H_2$ in the free fermion scope}
\begin{figure}[htb!]
\centering
\subfloat[]{\includegraphics[width=0.35\textwidth]{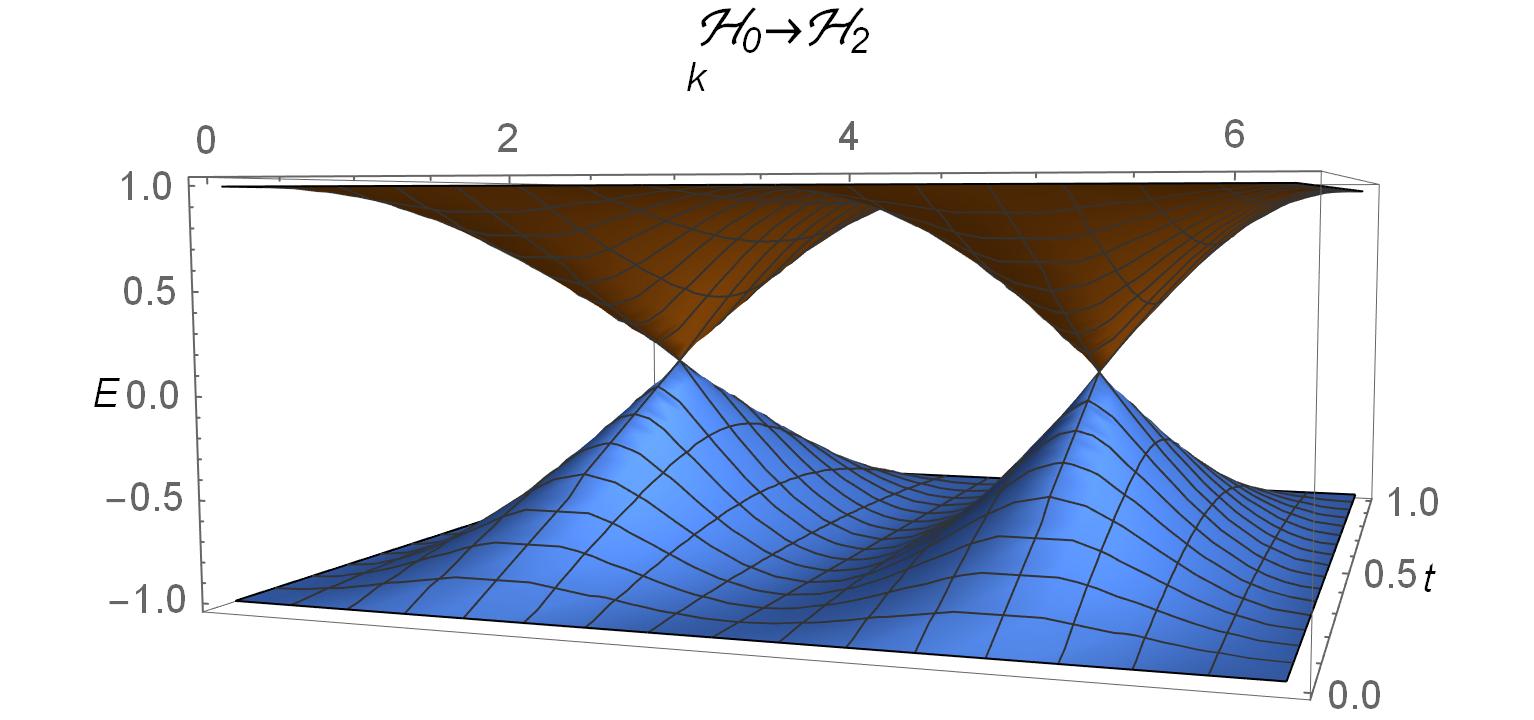}}\hskip 0.5cm
\subfloat[]{\includegraphics[width=0.35\textwidth]{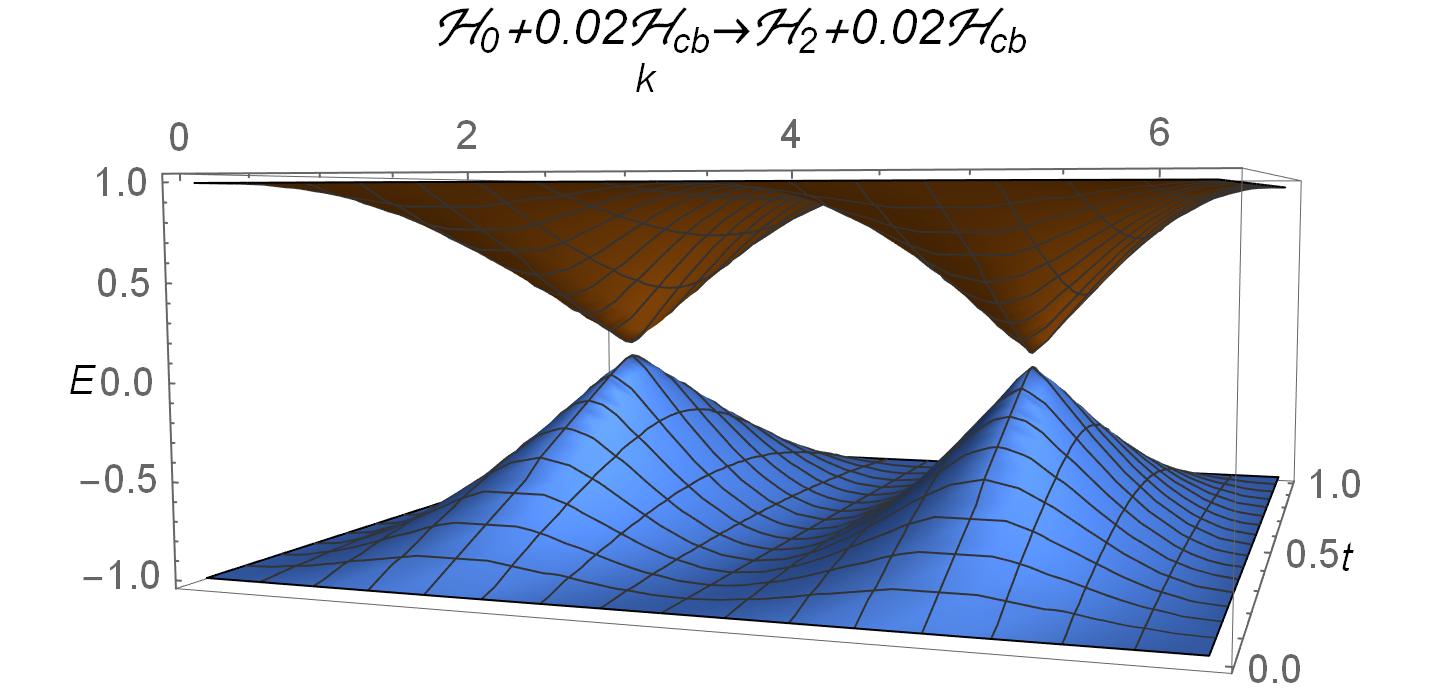}}\\
\caption{The energy spectrum of $\mathscr{H}_{0}\rightarrow \mathscr{H}_{2}$ and $\mathscr{H}_{0}+0.02\mathscr{H}_{cb}\rightarrow \mathscr{H}_{2}+0.02\mathscr{H}_{cb}$ within $k\in[0,2\pi)$.}
\label{Bloch deformation}
\end{figure}

To adiabatically connect $H_0$ and $H_2$ without a phase transition, we consider a non-interacting term that preserves reflection symmetry but breaks chiral symmetry.
\bea
H_{cb}=\sum_{j=1}[-ic_{2j-1}^{\dag}c_{2j+1}+ic_{2j}^{\dag}c_{2j+2}+h.c.].
\eea
Since $H_0$, $H_2$, and $H_{cb}$ can be described in the single-particle basis, we can study their energy spectrum in the thermodynamic limits by considering their Bloch Hamiltonian
\bea
\mathscr{H}_{0}(k)=\left(
\begin{matrix}
0 & 1 \\
1 & 0\\
\end{matrix}
\right),\,\mathscr{H}_{2}(k)=\left(
\begin{matrix}
0 & e^{-2ik} \\
e^{2ik} & 0\\
\end{matrix}
\right),\,\mathscr{H}_{cb}(k)=\left(
\begin{matrix}
2 \sin{k} & 0 \\
0 &  -2\sin{k}\\
\end{matrix}
\right),
\eea
where $\mathscr{H}_{x}(k)$ is the corresponding Bloch Hamiltonian of $H_x$. They respect the reflection symmetry $R\mathscr{H}(k)R^{-1}=\mathscr{H}(-k)$ with
\bea
R=\left(
\begin{matrix}
0 & 1 \\
1 & 0\\
\end{matrix}
\right).
\eea
As shown in Fig.~\ref{Bloch deformation}, gap-closing happens for the deformation $\mathscr{H}_{0}\rightarrow \mathscr{H}_{2}$. However, after introducing the chiral-breaking term as a perturbation, we can see that the deformation $\mathscr{H}_{0}+0.02\mathscr{H}_{cb}\rightarrow \mathscr{H}_{2}+0.02\mathscr{H}_{cb}$ remains gapped.
\end{appendices}
\bibliography{bib}

\begin{thebibliography}{10}

\bibitem{RMPforTI}
Xiao-Liang Qi and Shou-Cheng Zhang.
\newblock Topological insulators and superconductors.
\newblock {\em Rev. Mod. Phys.}, 83:1057--1110, Oct 2011.

\bibitem{SPTphases}
Zheng-Cheng Gu and Xiao-Gang Wen.
\newblock Tensor-entanglement-filtering renormalization approach and
  symmetry-protected topological order.
\newblock {\em Phys. Rev. B}, 80:155131, Oct 2009.

\bibitem{Ktheory}
Alexei Kitaev.
\newblock {Periodic table for topological insulators and superconductors}.
\newblock {\em AIP Conference Proceedings}, 1134(1):22--30, 05 2009.

\bibitem{Chiu:2016aa}
Ching-Kai Chiu, Jeffrey C.~Y. Teo, Andreas~P. Schnyder, and Shinsei Ryu.
\newblock Classification of topological quantum matter with symmetries.
\newblock {\em Reviews of Modern Physics}, 88(3):035005--, 08 2016.

\bibitem{TKtheory1}
Daniel~S. Freed and Gregory~W. Moore.
\newblock {Twisted Equivariant Matter}.
\newblock {\em Annales Henri Poincar{\'e}}, 12 2013.

\bibitem{Shiozaki:2014aa}
Ken Shiozaki and Masatoshi Sato.
\newblock Topology of crystalline insulators and superconductors.
\newblock {\em Physical Review B}, 90(16):165114--, 10 2014.

\bibitem{TKtheory2}
Guo~Chuan Thiang.
\newblock {On the K-Theoretic Classification of Topological Phases of Matter}.
\newblock {\em Annales Henri Poincar{\'e}}, 04 2016.

\bibitem{cohomology1}
Xie Chen, Zheng-Cheng Gu, Zheng-Xin Liu, and Xiao-Gang Wen.
\newblock Symmetry protected topological orders and the group cohomology of
  their symmetry group.
\newblock {\em Phys. Rev. B}, 87:155114, Apr 2013.

\bibitem{cohomology2}
Zheng-Cheng Gu and Xiao-Gang Wen.
\newblock Symmetry-protected topological orders for interacting fermions:
  Fermionic topological nonlinear $\ensuremath{\sigma}$ models and a special
  group supercohomology theory.
\newblock {\em Phys. Rev. B}, 90:115141, Sep 2014.

\bibitem{Song:2017aa}
Hao Song, Sheng-Jie Huang, Liang Fu, and Michael Hermele.
\newblock Topological phases protected by point group symmetry.
\newblock {\em Physical Review X}, 7(1):011020--, 02 2017.

\bibitem{Huang:2017aa}
Sheng-Jie Huang, Hao Song, Yi-Ping Huang, and Michael Hermele.
\newblock Building crystalline topological phases from lower-dimensional
  states.
\newblock {\em Physical Review B}, 96(20):205106--, 11 2017.

\bibitem{Lower-dconstruction3}
Zhida Song, Chen Fang, and Yang Qi.
\newblock Real-space recipes for general topological crystalline states.
\newblock {\em Nature Communications}, 11(1), August 2020.

\bibitem{Lower-dconstruction2}
Meng Cheng and Chenjie Wang.
\newblock Rotation symmetry-protected topological phases of fermions.
\newblock {\em Phys. Rev. B}, 105:195154, May 2022.

\bibitem{Lower-dconstruction1}
Jian-Hao Zhang, Shuo Yang, Yang Qi, and Zheng-Cheng Gu.
\newblock Real-space construction of crystalline topological superconductors
  and insulators in 2d interacting fermionic systems.
\newblock {\em Phys. Rev. Res.}, 4:033081, Jul 2022.

\bibitem{2024realspaceconstruction}
Jian-Hao {Zhang}, Shang-Qiang {Ning}, Yang {Qi}, and Zheng-Cheng {Gu}.
\newblock {Construction and classification of crystalline topological
  superconductor and insulators in three-dimensional interacting fermion
  systems}.
\newblock {\em arXiv e-prints}, page arXiv:2204.13558, April 2022.

\bibitem{ghomo2}
Ken Shiozaki, Charles~Zhaoxi Xiong, and Kiyonori Gomi.
\newblock {Generalized homology and Atiyah-Hirzebruch spectral sequence in
  crystalline symmetry protected topological phenomena}.
\newblock {\em Progress of Theoretical and Experimental Physics}, page ptad086,
  07 2023.

\bibitem{Freed_2021}
Daniel~S Freed and Michael~J Hopkins.
\newblock Reflection positivity and invertible topological phases.
\newblock {\em Geometry {\&} Topology}, 25:1165--1330, may 2021.

\bibitem{cobordism1}
Anton Kapustin.
\newblock Symmetry protected topological phases, anomalies, and cobordisms:
  Beyond group cohomology, 2014.

\bibitem{cobordism2}
Anton Kapustin, Ryan Thorngren, Alex Turzillo, and Zitao Wang.
\newblock {Fermionic symmetry protected topological phases and cobordisms}.
\newblock {\em Journal of High Energy Physics}, 12 2015.

\bibitem{cobordism3}
Kazuya Yonekura.
\newblock {On the Cobordism Classification of Symmetry Protected Topological
  Phases}.
\newblock {\em Communications in Mathematical Physics}, 06 2019.

\bibitem{Thorngren:2018aa}
Ryan Thorngren and Dominic~V. Else.
\newblock Gauging spatial symmetries and the classification of topological
  crystalline phases.
\newblock {\em Physical Review X}, 8(1):011040--, 03 2018.

\bibitem{Hsieh:2014aa}
Chang-Tse Hsieh, Takahiro Morimoto, and Shinsei Ryu.
\newblock Cpt theorem and classification of topological insulators and
  superconductors.
\newblock {\em Physical Review B}, 90(24):245111--, 12 2014.

\bibitem{crystalMTI}
Ken Shiozaki, Hassan Shapourian, and Shinsei Ryu.
\newblock Many-body topological invariants in fermionic symmetry-protected
  topological phases: Cases of point group symmetries.
\newblock {\em Phys. Rev. B}, 95:205139, May 2017.

\bibitem{OnsiteMTI}
Ken Shiozaki, Hassan Shapourian, Kiyonori Gomi, and Shinsei Ryu.
\newblock Many-body topological invariants for fermionic short-range entangled
  topological phases protected by antiunitary symmetries.
\newblock {\em Phys. Rev. B}, 98:035151, Jul 2018.

\bibitem{Pollmann_and_Turner2012}
Frank Pollmann and Ari~M. Turner.
\newblock Detection of symmetry-protected topological phases in one dimension.
\newblock {\em Phys. Rev. B}, 86:125441, Sep 2012.

\bibitem{PRLMTI}
Hassan Shapourian, Ken Shiozaki, and Shinsei Ryu.
\newblock Many-body topological invariants for fermionic symmetry-protected
  topological phases.
\newblock {\em Phys. Rev. Lett.}, 118:216402, May 2017.

\bibitem{Verresen:2017aa}
Ruben Verresen, Roderich Moessner, and Frank Pollmann.
\newblock One-dimensional symmetry protected topological phases and their
  transitions.
\newblock {\em Physical Review B}, 96(16):165124--, 10 2017.

\bibitem{edgedegeneracy1}
Lukasz Fidkowski and Alexei Kitaev.
\newblock Effects of interactions on the topological classification of free
  fermion systems.
\newblock {\em Phys. Rev. B}, 81:134509, Apr 2010.

\bibitem{edgedegeneracy2}
Evelyn Tang and Xiao-Gang Wen.
\newblock Interacting one-dimensional fermionic symmetry-protected topological
  phases.
\newblock {\em Phys. Rev. Lett.}, 109:096403, Aug 2012.

\bibitem{TI}
C.~L. Kane and E.~J. Mele.
\newblock ${Z}_{2}$ topological order and the quantum spin hall effect.
\newblock {\em Phys. Rev. Lett.}, 95:146802, Sep 2005.

\bibitem{BBC1}
Shinsei Ryu and Yasuhiro Hatsugai.
\newblock Topological origin of zero-energy edge states in particle-hole
  symmetric systems.
\newblock {\em Phys. Rev. Lett.}, 89:077002, Jul 2002.

\bibitem{BBC2}
Hui Li and F.~D.~M. Haldane.
\newblock Entanglement spectrum as a generalization of entanglement entropy:
  Identification of topological order in non-abelian fractional quantum hall
  effect states.
\newblock {\em Phys. Rev. Lett.}, 101:010504, Jul 2008.

\bibitem{BBC3}
Y.~Hatsugai.
\newblock Bulk-edge correspondence in graphene with/without magnetic field:
  Chiral symmetry, dirac fermions and edge states.
\newblock {\em Solid State Communications}, 149(27):1061--1067, 2009.

\bibitem{BBC4}
P.~Delplace, D.~Ullmo, and G.~Montambaux.
\newblock Zak phase and the existence of edge states in graphene.
\newblock {\em Phys. Rev. B}, 84:195452, Nov 2011.

\bibitem{BBC5}
Roger S.~K. Mong and Vasudha Shivamoggi.
\newblock Edge states and the bulk-boundary correspondence in dirac
  hamiltonians.
\newblock {\em Phys. Rev. B}, 83:125109, Mar 2011.

\bibitem{BBC6}
Gian~Michele Graf and Marcello Porta.
\newblock Bulk-edge correspondence for two-dimensional topological insulators.
\newblock {\em Communications in Mathematical Physics}, 324(3):851--895,
  October 2013.

\bibitem{BBC7}
J{\'a}nos~K. Asb{\'o}th, L{\'a}szl{\'o} Oroszl{\'a}ny, and Andr{\'a}s
  P{\'a}lyi.
\newblock {\em A Short Course on Topological Insulators}.
\newblock Springer International Publishing, 2016.

\bibitem{BBC8}
Yang Peng, Yimu Bao, and Felix von Oppen.
\newblock Boundary green functions of topological insulators and
  superconductors.
\newblock {\em Phys. Rev. B}, 95:235143, Jun 2017.

\bibitem{BBC9}
Chen-Shen Lee, Iao-Fai Io, and Hsien-chung Kao.
\newblock Winding number and zak phase in multi-band ssh models.
\newblock {\em Chinese Journal of Physics}, 78:96--110, August 2022.

\bibitem{BBCthroK}
Emil Prodan and Hermann Schulz-Baldes.
\newblock {\em Bulk and Boundary Invariants for Complex Topological
  Insulators}.
\newblock Springer International Publishing, 2016.

\bibitem{Greensfunctions}
Andrew~M. Essin and Victor Gurarie.
\newblock Bulk-boundary correspondence of topological insulators from their
  respective green's functions.
\newblock {\em Phys. Rev. B}, 84:125132, Sep 2011.

\bibitem{free_chiral}
Chen-Shen Lee.
\newblock A linear algebra-based approach to understanding the relation between
  the winding number and zero-energy edge states.
\newblock {\em SciPost Phys. Core}, 7:003, 2024.

\bibitem{Turner:2010aa}
Ari~M. Turner, Yi~Zhang, and Ashvin Vishwanath.
\newblock Entanglement and inversion symmetry in topological insulators.
\newblock {\em Physical Review B}, 82(24):241102--, 12 2010.

\bibitem{Hughes:2011aa}
Taylor~L. Hughes, Emil Prodan, and B.~Andrei Bernevig.
\newblock Inversion-symmetric topological insulators.
\newblock {\em Physical Review B}, 83(24):245132--, 06 2011.

\bibitem{d^1PhysRevX.7.011020}
Hao Song, Sheng-Jie Huang, Liang Fu, and Michael Hermele.
\newblock Topological phases protected by point group symmetry.
\newblock {\em Phys. Rev. X}, 7:011020, Feb 2017.

\bibitem{BBC}
Xie Chen, F.~J. Burnell, Ashvin Vishwanath, and Lukasz Fidkowski.
\newblock Anomalous symmetry fractionalization and surface topological order.
\newblock {\em Phys. Rev. X}, 5:041013, Oct 2015.

\end{thebibliography}

\end{document}